\pdfoutput=1

\documentclass[12pt,nofootinbib]{article}
\usepackage{amsmath}
\usepackage{amssymb}
\usepackage{graphicx}
\usepackage[colorlinks=true,linkcolor=blue,citecolor=blue]{hyperref}
\usepackage{amsfonts}
\usepackage{latexsym}
\usepackage{color}
\usepackage{cite}
\usepackage{bm}
\input{colordvi.tex}

\numberwithin{equation}{section}

\renewcommand{\thefootnote}{\#\arabic{footnote}}

\begin{document}

\begin{titlepage}

\setcounter{page}{1} \baselineskip=15.5pt \thispagestyle{empty}

\begin{flushright}
{\footnotesize RESCEU-1/12}
\end{flushright}
\vfil

\bigskip\
\begin{center}
{\LARGE \textbf{Runnings in the Curvaton}}
\vskip 15pt
\end{center}

\vspace{0.5cm}
\begin{center}
{\large 
Takeshi Kobayashi$\,^{1,2,3}$ and Tomo~Takahashi$\,^4$ 
}
\end{center}

\vspace{0.3cm}

\begin{center}

{\em
$^1$Research Center for the Early Universe, School of Science, The
 University of Tokyo,\\ 
7-3-1 Hongo, Bunkyo-ku, Tokyo 113-0033, Japan \\  \vspace{2mm}
$^2$Canadian Institute for Theoretical Astrophysics,
 University of Toronto, \\ 60 St. George Street, Toronto, Ontario M5S
 3H8, Canada \\ \vspace{2mm}
$^3$ Perimeter Institute for Theoretical Physics, \\ 
 31 Caroline St. N, Waterloo, Ontario N2L 2Y5, Canada\\ \vspace{2mm}
$^4$Department of Physics, Saga University, Saga 840-8502, Japan \vspace{5pt} \\
}

\end{center} \vfil

\vspace{0.8cm}

\noindent

We investigate the scale-dependence, or the runnings, of linear and
second order density perturbations generated in various curvaton
scenarios. We argue that the second order perturbations,
i.e. non-Gaussianity, can strongly depend on the scale, even when the
linear perturbations are nearly scale-invariant.  We present analytic
formulae for the runnings from curvatons with general energy
potentials, and clarify the conditions under which $f_{\mathrm{NL}}$
becomes strongly scale-dependent. From the point of view of the
$f_{\mathrm{NL}}$ running, curvaton potentials can be classified into
roughly two categories by whether the potential flattens or steepens
compared to a quadratic one. As such examples, we study
pseudo-Nambu-Goldstone curvatons, and self-interacting curvatons,
respectively.  
The dynamics of non-quadratic curvatons and the
behaviors of the resulting density perturbations are clarified by
analytical methods.
Then we also study models where multiple source can be
responsible for density perturbations such as the multi-curvaton, and
mixed curvaton and inflaton models where the running of
$f_{\mathrm{NL}}$ can also be large due to their multi-source nature.
We make quantitative analysis for each curvaton scenario and discuss
in what cases the scale-dependence, in particular, of
$f_{\mathrm{NL}}$ can be large enough to be probed with future CMB
experiments.

\vfil

\end{titlepage}

\newpage
\tableofcontents

\setcounter{footnote}{0}

\renewcommand{\thepage}{\arabic{page}}
\setcounter{page}{1}
\renewcommand{\thefootnote}{\#\arabic{footnote}}

\section{Introduction}
\label{sec:intro}

The origin of density fluctuations in the Universe is one of the
important issues in cosmology and also gives invaluable information
for high energy physics since they are assumed to be generated in the
very early Universe.  Although quantum fluctuations of the inflaton,
which drives inflation in the very early Universe, has been considered
to be its origin over the years, other mechanisms have also been
discussed. Among possible candidates, the curvaton scenario
\cite{Enqvist:2001zp,Lyth:2001nq,Moroi:2001ct} has been attracting
much attention due to several reasons, one of which is the recent
observational results on primordial non-Gaussianity.  How much the
primordial fluctuations deviate from a Gaussian distribution can be
characterized by the non-linearity parameter $f_{\rm NL}$, whose
constraint from current observations is $-10 < f_{\rm NL} < 74$ (95 \%
C.L.)~\cite{arXiv:1001.4538} for the so-called local type
non-Gaussianity\footnote{
  In this paper we mainly discuss local type non-Gaussianities.  For
  constraints on $f_{\rm NL}$ of other types, see
  \cite{arXiv:1001.4538}.
}. If future data such as those from the on-going Planck satellite
confirms large local type non-Gaussianity of primordial fluctuations
at the level of $f_{\rm NL} \gtrsim \mathcal{O}(10)$, it readily
excludes conventional single-field inflation models predicting $f_{\rm
  NL} \ll \mathcal{O}(1)$ as the origin of the density
perturbations. On the other hand, the curvaton can generate large
non-Gaussianity at this level, which would motivate us to seriously
consider this scenario.  Furthermore, the curvaton may be well fitted
naturally into some particle physics or string theory models (see, for
e.g.,
\cite{Moroi:2002vx,Enqvist:2002rf,Postma:2002et,McDonald:2003xq,Enqvist:2003mr,Kasuya:2003va,Hamaguchi:2003dc,McDonald:2003jk,Enqvist:2003qc,Ikegami:2004ve,Allahverdi:2006dr,Kobayashi:2009cm,Burgess:2010bz}),
thus also in this respect, the curvaton has been the target of intense
study.

In many works on the curvaton model, it is assumed that the potential
for the curvaton has a simple quadratic form and the curvaton is
totally responsible for cosmic density fluctuations.  However, such
assumptions are removed in some microscopic constructions of the
curvaton. Regarding the former assumption, there exist models
realizing potentials that deviate from a purely quadratic one.
The consequences of such non-quadratic curvaton potentials have been
investigated for the self-interacting curvaton
\cite{astro-ph/0508573,arXiv:0807.3069,Huang:2008zj,Enqvist:2009zf,Enqvist:2009eq,Enqvist:2009ww,Fonseca:2011aa}
and the pseudo-Nambu-Goldstone (NG) boson one
\cite{hep-ph/0304050,Kawasaki:2008mc,Chingangbam:2009xi,Kawasaki:2011pd}.
Interestingly, it has been shown that the predictions for
non-Gaussianity of the primordial perturbations significantly depend
on the form of the potential.  In particular, when the potential
deviates from the quadratic form, the non-linearity parameter $f_{\rm
  NL}$ (and also other parameters characterizing non-Gaussianity such
as $\tau_{\rm NL}$ and $g_{\rm NL}$ for the trispectrum) can be
considerably scale-dependent
\cite{arXiv:0911.2780,arXiv:1007.4277,arXiv:1007.5148,arXiv:1008.2641,arXiv:1108.2708},
even when the linear order perturbations are nearly scale-invariant.
On the other hand, the assumption of the curvaton being totally
responsible for the density perturbations is removed for multi-source
scenarios, where the scale dependence of non-Gaussianity can arise
even with quadratic curvaton potentials
\cite{arXiv:0911.2780,arXiv:1007.4277}.  Such a situation can be
realized in the mixed curvaton and inflaton model
\cite{Langlois:2004nn,Moroi:2005kz,Moroi:2005np,Ichikawa:2008iq,Suyama:2010uj}
where fluctuations from the curvaton and inflaton can be both
responsible for cosmic density perturbation today.  One could also
consider a model where there exist multiple curvaton fields and they
can be responsible for density fluctuations. Some authors have studied
a model of two curvatons \cite{Choi:2007fya,Assadullahi:2007uw}.

It should also be noted that in the event of a detection of large
non-Gaussianity, upcoming CMB experiments can set severe constraints
on the scale-dependence of $f_{\mathrm{NL}}$ as well, especially when
combined with large-scale structure surveys, see
e.g. \cite{LoVerde:2007ri,arXiv:0906.0232,Becker:2010hx,Shandera:2010ei}.
Such information beyond $f_{\mathrm{NL}}$ would be a powerful probe of
the physics of the early universe.  In the light of these
considerations, the issue of the scale-dependence or ``running'' of
non-Gaussianity is an interesting and important subject.

In this paper we carry out analytical studies of density perturbations
from curvatons, by applying the method developed in~\cite{Kawasaki:2011pd}. 
We derive generic formulae for the runnings
of the linear and second order density perturbations, which enable us
to go beyond individual case studies and give a systematic treatment of 
the scale-dependence of $f_{\rm NL}$ in the curvaton mechanism. 
Our study not only serves as an analytical counterpart to previous
works that basically relied on numerical computations, but also
clarifies the underlying reason why strongly scale-dependent
non-Gaussianities can be generated from non-quadratic curvatons.
Furthermore, we present conditions for curvatons to
produce large running of $f_{\mathrm{NL}}$, which turn out to take
simple forms (\ref{cond1}) and (\ref{cond2}) for a single curvaton.
In passing, we also investigate the running of the spectral index of the
power spectrum~$\alpha$, which can also give interesting signatures.
We will clarify the direct relation between $\alpha$ and the running of
$f_{\mathrm{NL}}$, which we parametrize as~$n_{f_{\rm NL}}$.
The general study of $n_{f_{\mathrm{NL}}}$ and $\alpha$ would give new insight
into the curvaton mechanism.

Curvaton potentials can roughly be categorized by whether they flatten
or steepen compared to a quadratic one. It will be shown that this
classification is important for discussing $n_{f_{\rm NL}}$ from
curvatons.  
As typical examples of the two cases, we will look into pseudo-Nambu-Goldstone
curvatons and self-interacting curvatons, respectively. For the former
case, the relation between $n_{f_{\rm NL}}$ and $\alpha$ plays an
important role, restricting the running of $f_{\mathrm{NL}}$ from current
observational bounds on~$\alpha$. On the other hand, for the latter
case, strongly scale-dependent $f_{\mathrm{NL}}$ can be produced even
for a suppressed $\alpha$ due to the steepness of the potential.
Regarding $n_{f_{\rm NL}}$ generated in multi-source models, we make a
quantitative study for some models and show in what cases the
scale dependence can be large enough to be probed in future
cosmological observations.

The structure of this paper is as follows: In Section~\ref{sec:GF} we
derive analytic expressions for the scale-dependence of the linear and
second order density perturbations from a curvaton. Then we move on
and apply the generic discussions to pseudo-Nambu-Goldstone curvatons
in Section~\ref{sec:NG}, and self-interacting curvatons in
Section~\ref{sec:SI}.  In Section~\ref{sec:mixed}, we further
investigate the issue for the mixed curvaton and inflaton
scenario. Then in Section~\ref{sec:multi-cur}, we discuss
multi-curvaton model. We present our conclusions in
Section~\ref{sec:conc}.

We give a brief discussion on density perturbations from curvatons
with non-sinusoidal oscillations in Appendix~\ref{app:B}.

\section{Scale-Dependence of $f_{\mathrm{NL}}$ in the Curvaton Mechanism}
\label{sec:GF}

A light curvaton acquires nearly scale-invariant field fluctuations
during inflation, that are converted into the cosmological density
perturbations as the curvaton oscillates and decays in the
post-inflationary era.  In this section we derive generic expressions
for the running of non-Gaussianity from a curvaton, and discuss the
conditions under which the curvaton sources a largely
scale-dependent~$f_{\mathrm{NL}}$.  Our discussions in this section
are based on the work~\cite{Kawasaki:2011pd} which developed analytic
methods for computing density perturbations in the curvaton mechanism.
We extend their results and compute the runnings of the linear and
second order perturbations.  In this Section and also in the following
Sections~\ref{sec:NG} and \ref{sec:SI}, we assume that density
perturbations sourced from the inflaton are neglected. However, we
remove this assumption in Section~\ref{sec:mixed} where we study the
mixed curvaton and inflaton scenario.

\subsection{Density Perturbations}
\label{subsec:DPC}

Let us start by laying out some results of~\cite{Kawasaki:2011pd} for
density perturbations produced by a curvaton~$\sigma$ possessing an
effective potential~$V(\sigma)$.  The potential is assumed to have no
explicit dependence on time, and also that it is well approximated by
a quadratic one around its minimum so that the curvaton oscillations
are sinusoidal\footnote{
  The formulae can be generalized to cases with non-sinusoidal
  oscillations as well, see Appendix~B
  of~\cite{Kawasaki:2011pd}.\label{foot:1}
}.  The curvaton energy density is considered to redshift similarly to
nonrelativistic matter after the onset of the oscillations until when
the curvaton decays into radiation, whereas we suppose the inflaton to
behave as matter from the end of inflation until reheating (=inflaton
decay), after which the inflaton turns into radiation.  The energy
density of the curvaton before the beginning of its oscillation is
assumed to be negligibly tiny compared to the total energy of the
Universe, having little effect on the expansion history.  Furthermore,
the Hubble parameter during inflation is considered to be nearly
constant.  For detailed discussions on the derivations of the
following results, we refer the reader to \cite{Kawasaki:2011pd}.

Using the $\delta \mathcal{N}$-formalism
\cite{229378,astro-ph/9507001,astro-ph/0003278,astro-ph/0411220}, the
density perturbations are obtained by computing quantities such as the
time when the curvaton oscillation starts and the curvaton energy
density as functions of the curvaton field value~$\sigma_*$ (hereafter
the subscript~$*$ denotes values when the CMB scale~$k_*$ exits the
horizon).  The curvaton dynamics prior to the oscillation can be
tracked by the attractor solution
\begin{equation}
 \hat{c} H \dot{\sigma} = -V', \qquad
\mathrm{with} \quad
 \hat{c} = 
 \left\{
   \begin{array}{cl}
     3 & \mbox{(during inflation) } \\
     9/2 & \mbox{(matter domination) } \\
     5 & \mbox{(radiation domination) }, 
   \end{array}
\right.
 \label{CDapp}
\end{equation}
which is a good approximation while $|V'' / \hat{c} H^2 | \ll 1$.
Here, a prime denotes a derivative with respect to~$\sigma$, and an
overdot a time derivative.  Setting the minimum of the potential about
which the curvaton oscillates to $\sigma =0$, we can define the onset
of the oscillation as when the time scale of the curvaton rolling
becomes comparable to the Hubble time, i.e. $|\dot{\sigma} / H \sigma|
= 1$. This, combined with (\ref{CDapp}), gives the Hubble parameter at
the time,
\begin{equation}
 H_{\mathrm{osc}}^2 = \frac{V'(\sigma_{\mathrm{osc}})}{c
  \sigma_{\mathrm{osc}}},
 \label{Hosc}
\end{equation}
where the subscript ``osc'' denotes values at the onset of the
curvaton oscillation, and $c$ is a constant whose value is set by
whether reheating (= inflaton decay, at~$t_{\mathrm{reh}}$) is
earlier/later than the onset of the curvaton oscillation:
\begin{equation}
 c = 
 \left\{
   \begin{array}{cl}
     5 & (t_{\mathrm{reh}} < t_{\mathrm{osc}}) \\
     9/2 &  (t_{\mathrm{reh}} > t_{\mathrm{osc}}).
   \end{array}
\right.
 \label{ciroiro}
\end{equation}

We define the power spectrum~$\mathcal{P}_\zeta$ of the density
perturbations~$\zeta$ as 
\begin{equation}
\langle \zeta_{\boldsymbol{k}} \zeta_{\boldsymbol{k'}} \rangle =
 (2 \pi)^3  \delta^{(3)} (\boldsymbol{k} + \boldsymbol{k'}) 
P_\zeta (k)
\quad \mathrm{with} \quad 
  P_\zeta (k) = \frac{2 \pi^2}{k^3} \mathcal{P}_\zeta (k),
\end{equation}
where $k \equiv |\boldsymbol{k}|$. 
Supposing the curvaton field fluctuations to satisfy
$\mathcal{P}_{\delta \sigma} (k) = (H|_{k = aH } / 2 \pi)^2$ at the time when the
scale~$k$ exits the horizon,
then the linear order density perturbations at the CMB scale 
can be expressed in terms of the curvaton potential as
\begin{equation}
 \mathcal{P}_\zeta (k_*) = \left(\frac{\partial \mathcal{N}}{\partial \sigma_*} 
  \frac{H_*}{2 \pi} \right)^2,
\label{eq-ps}
\end{equation}
with 
\begin{equation}
 \frac{\partial \mathcal{N}}{\partial \sigma_*} = \frac{\hat{r}}{4+3 \hat{r}}
  \left(1 - X(\sigma_{\mathrm{osc}})\right)^{-1}
 \left\{\frac{V'(\sigma_{\mathrm{osc}})}{V(\sigma_{\mathrm{osc}})} -
  \frac{3 X(\sigma_{\mathrm{osc}})}{\sigma_{\mathrm{osc}}}  \right\}
 \frac{V'(\sigma_{\mathrm{osc}})}{V'(\sigma_*)}.
 \label{NfuncX}
\end{equation}
Here, $\hat{r}$ is the energy density ratio between the curvaton and
radiation (which originates from the inflaton) upon curvaton
decay\footnote{
  The following quantity is also used in the literature to express the
  formulae for density perturbations from the curvaton:
\begin{equation}
r_{\rm dec} = \left. \frac{3 \rho_\sigma} {4 \rho_{r} + 3 \rho_\sigma} \right |_{\rm dec},
\end{equation}
which is also evaluated at the curvaton decay. Notice that $\hat{r}$
defined here is a bit different from $r_{\rm dec}$.
}
\begin{equation}
 \hat{r} \equiv \left. \frac{\rho_\sigma }{\rho_r} \right|_{\mathrm{dec}},
\label{def_r}
\end{equation}
while the function~$X$ denotes effects due to the non-uniform onset of
the curvaton oscillations (which are absent for a purely quadratic
curvaton potential), defined as follows:
\begin{equation}
 X(\sigma_{\mathrm{osc}}) \equiv  
 \frac{1}{2(c-3)} \left(\frac{\sigma_{\mathrm{osc}}
	  V''(\sigma_{\mathrm{osc}})}{V'(\sigma_{\mathrm{osc}})} - 1
		  \right).
\label{Xdef}
\end{equation}

From the bispectrum
\begin{equation}
  \langle \zeta_{\boldsymbol{k_1}} \zeta_{\boldsymbol{k_2}}
   \zeta_{\boldsymbol{k_3}} \rangle = 
 (2 \pi)^3  \delta^{(3)} (\boldsymbol{k_1} + \boldsymbol{k_2} + \boldsymbol{k_3})
B_{\zeta} (k_1, k_2, k_3),
\label{eqbi}
\end{equation}
one can generally define the non-linearity parameter~$f_{\mathrm{NL}}$ as
\begin{equation}
 B_\zeta (k_1, k_2, k_3) = \frac{6}{5} f_{\mathrm{NL}}(k_1, k_2, k_3)
 \left[
 P_\zeta (k_1) P_\zeta (k_2) + P_\zeta (k_2) P_\zeta (k_3) + P_\zeta
 (k_3) P_\zeta (k_1) 
 \right].
\end{equation}
However, upon parameterizing the overall amplitude of the local-type bispectra
produced by curvatons,\footnote{It should also be noted that, especially when
$f_{\mathrm{NL}}$ is strongly 
scale-dependent, the bispectrum~(\ref{eqbi}) 
sourced from a curvaton has shapes similar to, but not
exactly of the ``local form''~\cite{Komatsu:2001rj} which is often
given as
 \begin{equation}
 B_{\mathrm{local}}(k_1,k_2,k_3) \propto  \frac{1}{k_1^3 k_2^3} +
 \frac{1}{k_1^3 k_3^3} + \frac{1}{k_2^3 k_3^3}.
\end{equation}
} throughout this paper we discuss
$f_{\mathrm{NL}}$ on the equilateral configuration
\begin{equation}
f_{\mathrm{NL}}(k) \equiv \frac{5}{18}
 B_{\zeta} (k, k, k) 
\left(
 \frac{2 \pi^2}{k^3}   \mathcal{P}_{\zeta}(k)
\right)^{-2}.
\label{fNLdf}
\end{equation}
We suppose that the bispectrum of the curvaton
fluctuations $\langle \delta \sigma_{\boldsymbol{k_1}} \delta
\sigma_{\boldsymbol{k_2}} \delta \sigma_{\boldsymbol{k_3}} \rangle$ with 
$| \boldsymbol{k_1} | \! \!= \! \! |\boldsymbol{k_2} | \! \! = \! \!
| \boldsymbol{k_3} | \! \! = \! \! k$
vanishes when $k$ exits the horizon, and further neglect direct
trispectra of $\delta \sigma$ (i.e.
$\langle \delta \sigma_{\boldsymbol{q_1}} \delta
\sigma_{\boldsymbol{q_2}} \delta \sigma_{\boldsymbol{q_3}}  \delta
\sigma_{\boldsymbol{q_4}} \rangle = 
\langle \delta \sigma_{\boldsymbol{q_1}} \delta
\sigma_{\boldsymbol{q_2}} \rangle \langle \delta \sigma_{\boldsymbol{q_3}} \delta
\sigma_{\boldsymbol{q_4}} \rangle  + (\mathrm{2 \, \, perm.})$).
Then it can be shown that $f_{\mathrm{NL}}$ 
takes the form~\cite{Kawasaki:2011pd}, 
\begin{equation}\label{fNLfuncX}
\begin{split}
 f_{\mathrm{NL}}  & (k_*) =
 \frac{5}{6} \frac{\partial^2 \mathcal{N}}{\partial
  \sigma_*^2} 
 \left( \frac{\partial \mathcal{N}}{\partial \sigma_*} \right)^{-2} \\
 & =
  \frac{40 (1+\hat{r})}{3 \hat{r} (4+3 \hat{r})}  
 + \frac{5 (4+3 \hat{r})}{6 \hat{r}}
 \left\{\frac{V'(\sigma_{\mathrm{osc}})}{V(\sigma_{\mathrm{osc}})}-
 \frac{3 X(\sigma_{\mathrm{osc}})}{\sigma_{\mathrm{osc}}}  \right\}^{-1}  
  \Biggl[(1-X(\sigma_{\mathrm{osc}}))^{-1} X'(\sigma_{\mathrm{osc}})    \\
 & 
  + \left\{\frac{V'(\sigma_{\mathrm{osc}})}{V(\sigma_{\mathrm{osc}})} -
 \frac{3X(\sigma_{\mathrm{osc}})}{\sigma_{\mathrm{osc}}}  \right\}^{-1} 
 \left\{\frac{V''(\sigma_{\mathrm{osc}})}{V(\sigma_{\mathrm{osc}})} -
 \frac{V'(\sigma_{\mathrm{osc}})^2}{V(\sigma_{\mathrm{osc}})^2} -
 \frac{3 X'(\sigma_{\mathrm{osc}})}{\sigma_{\mathrm{osc}}} + \frac{3
 X(\sigma_{\mathrm{osc}})}{\sigma_{\mathrm{osc}}^2} 
 \right\} \\
 & \qquad \qquad \qquad \qquad \qquad \qquad \qquad \qquad \qquad
  + \frac{V''(\sigma_{\mathrm{osc}})}{V'(\sigma_{\mathrm{osc}})} -
 (1-X(\sigma_{\mathrm{osc}}))
 \frac{V''(\sigma_*)}{V'(\sigma_{\mathrm{osc}})} 
\Biggr].
\end{split}
\end{equation}
The scale-dependence of $f_{\mathrm{NL}}$ shows up through~$\sigma_*$,
whereas $\sigma_{\mathrm{osc}}$ and $\hat{r}$ are independent of the
wave number.  Hence for later convenience, let us divide the
expression into terms that explicitly depend on~$\sigma_*$, and the
rest,
\begin{equation}
 f_{\mathrm{NL}} = f_{1}( \sigma_*, \sigma_{\mathrm{osc}}, \hat{r})
 + f_{2}( \sigma_{\mathrm{osc}}, \hat{r}),
\label{f1andf2}
\end{equation}
where we have introduced
\begin{equation}
\begin{split}
 f_{ 1} & \equiv 
- \frac{5 (4+3 \hat{r})}{6 \hat{r}}
 \left\{\frac{V'(\sigma_{\mathrm{osc}})}{V(\sigma_{\mathrm{osc}})}-
 \frac{3 X(\sigma_{\mathrm{osc}})}{\sigma_{\mathrm{osc}}}  \right\}^{-1}  
 (1-X(\sigma_{\mathrm{osc}}))
 \frac{V''(\sigma_*)}{V'(\sigma_{\mathrm{osc}})} \\
 & = -\frac{5}{6} \left(\frac{\partial \mathcal{N}}{\partial \sigma_*}
		\right)^{-1} \frac{V''(\sigma_*)}{V'(\sigma_*)}.
 \label{fNL1}
\end{split}
\end{equation}
Upon obtaining the second line, we have
used~(\ref{NfuncX})\footnote{
  For a curvaton with a quadratic potential $V\propto \sigma^2$, one
  obtains
\begin{equation}
 \frac{\partial \mathcal{N}}{\partial \sigma_*} 
 = \frac{2\hat{r}}{4 + 3\hat{r}}\frac{1}{\sigma_*},
  \qquad
   \frac{\partial^2 \mathcal{N}}{\partial \sigma_*^2}
   = 
\frac{2 \hat{r} (16+8 \hat{r} - 9 \hat{r}^2)}{(4+3 \hat{r})^3} \frac{1
}{\sigma_*^2} .
\end{equation}
Furthermore,  the terms $f_1$ and $f_2$ depend only on~$\hat{r}$,
\begin{equation}
 f_1 = -\frac{5(4+3\hat{r})}{12 \hat{r}}, \qquad
f_2 = \frac{40 (1+\hat{r})}{3 \hat{r} (4+3\hat{r})},
\end{equation}
and thus
\begin{equation}
 f_{\mathrm{NL}} = \frac{5}{12} \left(-3+\frac{4}{\hat{r}} + \frac{8}{4+3 \hat{r}}\right).
\label{eq2.10}
\end{equation}
Large non-Gaussianity is generated for $\hat{r} \ll 1$, under which
\begin{equation}
 f_{\mathrm{NL}} \simeq - f_1 \simeq \frac{1}{2} f_2 \simeq \frac{5}{3\hat{r}}.
\end{equation}
\label{foot:foot4}
}.

The energy density ratio~$\hat{r}$ (\ref{def_r}) is obtained as
\begin{equation}\label{anar}
\begin{split}
 \hat{r} & = \mathrm{Max.}\Biggl[
\frac{V(\sigma_{\mathrm{osc}})}{3 M_p^2 H_{\mathrm{osc}}^{3/2}
 \Gamma_\sigma^{1/2}} \times \mathrm{Min.}\left(
1, \,  \frac{\Gamma_\phi^{1/2}}{H_{\mathrm{osc}}^{1/2}} 
\right),
\\
& \qquad \qquad 
\left\{
\frac{V(\sigma_{\mathrm{osc}})}{3 M_p^2 H_{\mathrm{osc}}^{3/2}
 \Gamma_\sigma^{1/2}} \times \mathrm{Min.}
\left(
1, \,  \frac{\Gamma_\phi^{1/2}}{H_{\mathrm{osc}}^{1/2}} 
\right) 
\right\}^{4/3}
\Biggr],
\end{split}
\end{equation}
where the first and second terms in the Max. parentheses correspond to
the curvaton being subdominant and dominant at its decay,
respectively, while the Min. parentheses are due to whether the onset
of oscillation is after or before reheating.  $\Gamma_\phi$ and
$\Gamma_\sigma$ are constants denoting respectively the decay rates of
the inflaton and the curvaton.  Throughout we adopt the sudden decay
approximation where the scalar fields suddenly decay into radiation
when $H = \Gamma$.

The curvaton field value at the onset of the
oscillations~$\sigma_{\mathrm{osc}}$ is obtained by
integrating~(\ref{CDapp}),
\begin{equation}
 \int^{\sigma_{\mathrm{osc}}}_{\sigma_*} \frac{d\sigma}{V'} =
 -\frac{\mathcal{N}_*}{3 H_{\mathrm{inf}}^2} - \frac{1}{2 c (c-3)
 H_{\mathrm{osc}}^2},  \label{sigmaosc}
\end{equation}
solving which gives $\sigma_{\rm osc}$ as a function
of~$\sigma_*$\footnote{
  When (\ref{sigmaosc}) admits as solutions for $\sigma_{\rm osc}$
  both positive and negative values, one should take the sign of
  $\sigma_{\rm osc}$ to match with that of $\sigma_*$.\label{foot:5}
}.  Here, $\mathcal{N}_*$ is the number of e-folds during inflation
between the horizon exit of the CMB scale and the end of inflation,
and $H_{\mathrm{inf}}$ is the inflationary Hubble scale (we are
assuming a nearly constant Hubble parameter during inflation, thus
$H_{\mathrm{inf}} \simeq H_*$).  Let us also show the derivative of
$\sigma_{\mathrm{osc}}$,
\begin{equation}
 \frac{\partial \sigma_{\mathrm{osc}}}{\partial \sigma_*} = 
\left(1 - X(\sigma_{\mathrm{osc}})\right)^{-1}
\frac{V'(\sigma_{\mathrm{osc}})}{V'(\sigma_*)},
\label{partialsigma_osc}
\end{equation}
which has entered (\ref{NfuncX}) through
$\partial \mathcal{N} / \partial \sigma_* \propto \partial
\sigma_{\mathrm{osc}} / \partial \sigma_*$.

Thus by combining the above expressions, one can compute the resulting
density perturbations from a curvaton with a generic
potential~$V(\sigma)$, given the curvaton field value at the CMB scale
horizon exit~$\sigma_*$, the decay rates of the inflaton~$\Gamma_\phi$
and curvaton~$\Gamma_\sigma$, the inflationary
scale~$H_{\mathrm{inf}}$, and the duration of
inflation~$\mathcal{N}_*$.

\subsection{Scale-Dependence}

Now we extend the above expressions to discuss the scale-dependence of
the linear and second order density perturbations, which are the main
topic of this paper.

Since we are assuming $|\dot{H}/H^2 | \ll 1$ during inflation, the comoving wave
number~$k$ at around the CMB scale satisfies
\begin{equation}
 d \ln k \simeq H_* dt. \label{dlnk}
\end{equation}
Then by using the slow-roll approximation for the curvaton
\begin{equation}
 3 H_* \dot{\sigma}_* \simeq -V'(\sigma_*), \label{3hsigma}
\end{equation}
one obtains the spectral index of the linear order perturbations at the
CMB scale
\begin{equation}
 n_s -1 \equiv \frac{d}{d \ln k}\ln \mathcal{P}_{\zeta} \simeq
 2\frac{\dot{H}_*}{H_*^2} + \frac{2}{3} \frac{V''(\sigma_*)}{H_*^2},
\label{eq:ns}
\end{equation}
as well as its running
\begin{equation}
 \alpha \equiv \frac{d n_s}{d \ln k}
  \simeq 
 2 \frac{\ddot{H}_*}{H_*^3} - 4 \frac{\dot{H}_*^2}{H_*^4} - \frac{4}{3}
 \frac{\dot{H}_*}{H_*^2} \frac{V''(\sigma_*)}{H_*^2} - 
\frac{2}{9} \frac{V'(\sigma_*) V'''(\sigma_*)}{H_*^4 } .
 \label{runningalpha}
\end{equation}
Focusing on the contributions to the scale-dependence that are sourced
purely by the tilt of the curvaton potential, we introduce the
following parameters
\begin{equation}
 \tilde{n}_s -1 \equiv  \frac{2}{3} \frac{V''(\sigma_*)}{H_*^2},
 \label{tildens}
\end{equation}
\begin{equation}
 \tilde{\alpha} \equiv  - 
\frac{2}{9} \frac{V'(\sigma_*) V'''(\sigma_*)}{H_*^4 }.
 \label{tildealpha}
\end{equation}
When the Hubble parameter during inflation is exactly a constant,
these give the spectral index and its running at the leading order,
i.e. $n_s \simeq \tilde{n}_s$ and $\alpha \simeq \tilde{\alpha}$.

\vspace{\baselineskip}

In order to parametrize the scale-dependence of the non-Gaussianity,
we define the spectral index of $f_{\mathrm{NL}}$ as follows:
\begin{equation}
 n_{f_{\mathrm{NL}}} \equiv \frac{d \ln \left| f_{\mathrm{NL}}\right|}{d \ln k} .
\label{nfNLdef}
\end{equation}
We remark that a totally scale-invariant~$f_{\mathrm{NL}}$ corresponds
to $ n_{f_{\mathrm{NL}}} = 0$ (instead of $1$).  Then, in a similar
fashion as above, one arrives at
\begin{equation}
 n_{f_\mathrm{NL}}  \simeq
 \frac{1}{f_{\mathrm{NL}}}
 \frac{5 (4+3\hat{r})}{18 \hat{r}} 
 \left\{\frac{V'(\sigma_{\mathrm{osc}})}{V(\sigma_{\mathrm{osc}})}-
 \frac{3 X(\sigma_{\mathrm{osc}})}{\sigma_{\mathrm{osc}}}  \right\}^{-1}  
 \left( 1 - X(\sigma_{\mathrm{osc}})\right)
 \frac{V'(\sigma_*)}{ V'(\sigma_{\mathrm{osc}})}
 \frac{V'''(\sigma_*) }{ H_*^2 }.
 \label{nfNLfNL}
\end{equation}
Here we note that time derivatives of~$H_*$ do not show up at the
leading order, since~$f_{\mathrm{NL}}$~(\ref{fNLfuncX}) does not
explicitly depend on the Hubble parameter during inflation.  One
clearly sees that a non-vanishing $n_{f_{\mathrm{NL}}}$ (at the leading
order) requires non-zero~$V'''(\sigma_*)$,
and hence $f_{\mathrm{NL}}$ produced by a curvaton with a quadratic potential is
scale-invariant.  This opens
up the possibility that a slight deviation of the curvaton potential
from a quadratic one can be verified through observing the running
of~$f_{\mathrm{NL}}$.

By using~(\ref{NfuncX}), one can simplify (\ref{nfNLfNL}) as
\begin{equation}
 n_{f_{\mathrm{NL}}} \simeq \frac{1}{f_{\mathrm{NL}}}\frac{5}{18} 
 \left( \frac{\partial \mathcal{N}}{\partial \sigma_*}\right)^{-1} 
 \frac{V'''(\sigma_*)}{H_*^2},
\label{2.20}
\end{equation}
which shows that when the power spectrum is fixed to a certain value,
e.g. from the COBE (WMAP) normalization, then the value of the product
$n_{f_{\mathrm{NL}}} f_{\mathrm{NL}}$ is determined only by
information at the CMB scale horizon exit.

Moreover, in terms of the parameters (\ref{fNL1}),
(\ref{tildens}), and (\ref{tildealpha}),  
the expression~(\ref{nfNLfNL}) can be recast into the form of
\begin{equation}
 n_{f_{\mathrm{NL}}} \simeq \frac{\tilde{\alpha}}{\tilde{n}_s - 1}
  \frac{f_1}{f_{\mathrm{NL}}}.
 \label{nfNLalpha}
\end{equation}
This equation makes clear the typical amplitude of the running of
$f_{\mathrm{NL}}$ expected in the curvaton model.  In simple
cases, the running parameter~$\tilde{\alpha}$ (though not necessarily
equivalent to the actual running of the spectral index) has a smaller
size than the spectral index
parameter~$\tilde{n}_s-1$. Furthermore, since~$f_{1}$ explicitly
depends on~$\sigma_*$ while the rest of the terms of~$f_{\mathrm{NL}}$
do not\footnote{
Strictly speaking, $\sigma_{\mathrm{osc}}$ also
  depends on~$\sigma_*$ through (\ref{sigmaosc}), but the point here
  is that $f_{ 1}$ and $f_{ 2}$ depend quite differently
  on~$\sigma_*$.
  },  one may not expect $f_1$ to be much larger than the
sum $f_{\mathrm{NL}} = f_1 + f_2$.  Thus naively one would expect the
running of $f_{\rm NL}$ from curvatons to be highly suppressed.

In other words, $n_{f_{\mathrm{NL}}}$ as large as, say
$|n_{f_{\mathrm{NL}}}| \gtrsim 1$, is realized only when at least one
of the following two conditions are satisfied:
\begin{equation}
 |\tilde{\alpha}| \gtrsim |\tilde{n}_s - 1| , \label{cond1}
\end{equation}
\begin{equation}
 |f_1| \gtrsim |f_{\mathrm{NL}}|.   \label{cond2}
\end{equation}
The former condition~(\ref{cond1}) may be realized by curvaton
potentials possessing inflection points, or more generally, by
potentials that flatten compared to a quadratic as one goes away from
the minimum.  As such an example, in Section~\ref{sec:NG} we study
pseudo-Nambu-Goldstone curvatons with cosine-type potentials. However,
there we will see that the resulting $n_{f_\mathrm{NL}}$ is actually
directly bounded by observational constraints on the running~$\alpha$
of the spectral index of the linear order perturbations.  The latter
condition~(\ref{cond2}) indicates that $f_1$ and $f_2$ cancel each
other, suppressing $f_{\mathrm{NL}}$ compared to $f_1$.  Such
suppression of~$f_{\mathrm{NL}}$ is known to exist for
self-interacting curvatons possessing polynomial terms that are higher
order than quadratic, as was numerically shown in
\cite{astro-ph/0508573,arXiv:0807.3069,Huang:2008zj,Enqvist:2009zf,Enqvist:2009eq,Enqvist:2009ww,Fonseca:2011aa}.
In Section~\ref{sec:SI} we look into self-interacting curvatons and
see that large $n_{f_{\mathrm{NL}}}$ can be obtained even under a
suppressed running~$\alpha$, and further discuss that such behavior
stems from the curvaton potential steepening more rapidly than
quadratic ones.

\vspace{\baselineskip}

The value of $f_{\mathrm{NL}}$ itself needs to be large for its
running to be detectable, and thus experiments are sensitive to the
product $n_{f_{\mathrm{NL}}} f_{\mathrm{NL}}$, instead of
$n_{f_{\mathrm{NL}}}$ alone. Upon discussing example models in the
following sections, we refer to the results of~\cite{arXiv:0906.0232}
where detectability of a scale-dependent non-Gaussianity through CMB
experiments are analyzed\footnote{
We note that the definition of our running
parameter~$n_{f_{\mathrm{NL}}}$~(\ref{nfNLdef}) is not exactly the
same as the $n_{\mathrm{NG}}$ parameter adopted
in~\cite{arXiv:0906.0232}, especially in the sense that
$n_{\mathrm{NG}}$ is set to a constant whereas $n_{f_{\mathrm{NL}}}$ itself
can run. Nevertheless we adopt their results upon discussing
detectability.
}. The running of the local-type~$f_{\mathrm{NL}}$ can
be probed if it is large enough to satisfy
\begin{equation}
 \left| n_{f_{\mathrm{NL}}} \right| > A \times
  \frac{50}{f_{\mathrm{NL}}},
 \label{obs}
\end{equation}
where the right hand side takes $A \simeq 0.68$, $0.10$, and 0.05 for
WMAP\cite{arXiv:1001.4538}, Planck\cite{:2006uk}, and
CMBPol\cite{Baumann:2008aq}, respectively, assuming as fiducial values
$f_{\mathrm{NL}} = 50$, $n_{f_{\mathrm{NL}}} = 0$, and a full-sky
coverage.

\vspace{\baselineskip}

Before ending this section, we should mention about errors in the
analytic formulae. Firstly, the approximation (\ref{dlnk}) contains
error of order~$\dot{H}_*/H_*^2$, and (\ref{3hsigma}) of
order~$V''(\sigma_*) / H_*^2$.  By taking into account such errors,
one can check that they do not modify the above results on $n_s$,
$\alpha$, and $n_{f_{\mathrm{NL}}}$ at the leading order.

However further approximations and simplifications have been carried
out upon obtaining the analytic expressions of
Subsection~\ref{subsec:DPC}, e.g., the approximation~(\ref{CDapp}) on
the curvaton dynamics, which is correct up to order $V''/H^2$.  This
can source errors in the results of order $V''/H^2$, and also of
derivatives of $V''/H^2$ in terms of $\sigma_*$ and/or time~$t$.  It
should be noted that such error with various orders of derivatives can
accumulate and lead to breakdown of the analytic expressions
especially for higher-order correlation functions and running.
Nevertheless, for the explicit examples we study in the following
sections, we will see that the analytic results match well with
results from numerical computations.

\section{Pseudo-Nambu-Goldstone Curvatons}
\label{sec:NG}

Since the parameters $\tilde{n}_s$ and $\tilde{\alpha}$ in the
expression for~$n_{f_{\mathrm{NL}}}$~(\ref{nfNLalpha}) are not
necessarily the actual spectral index and its running under a
non-vanishing~$\dot{H}$ during inflation, one may expect that the
condition~(\ref{cond1}) for a large~$n_{f_{\mathrm{NL}}}$ is easily
satisfied without contradicting with observational constraints on the
flatness of the power spectrum.  One may imagine cases where the
amplitudes of $\tilde{\alpha}$ and $\tilde{n}_s-1$ are much smaller
than unity, though possessing a hierarchy among them as
$|\tilde{\alpha}| \gg |\tilde{n}_s-1|$.

In this section we study the case where the curvaton is realized as a
pseudo-Nambu-Goldstone (NG) boson of a broken U(1) symmetry,
possessing a cosine-type potential
\cite{hep-ph/0304050,Kawasaki:2008mc,Chingangbam:2009xi,Kawasaki:2011pd}.
Given that the curvaton at the CMB scale horizon exit is located close
to the inflection point of the potential, $\tilde{n}_s-1$ vanishes
while~$\tilde{\alpha}$ remains finite.  However, we will see that the
resulting $n_{f_\mathrm{NL}}$ is actually set by the absolute value
of~$\tilde{\alpha}$, thus bounded by observational constraints on the
running of the spectral index (unless $\tilde{\alpha}$ is cancelled
out by the $\dot{H}$ terms in (\ref{runningalpha})).

\vspace{\baselineskip}

We consider the potential of the form
\begin{equation}
 V(\sigma) = \Lambda^4 \left[ 1 - \cos \left(\frac{\sigma}{f} \right)
		       \right],
\end{equation}
where $f$ and $\Lambda$ are mass scales.
It has inflection points at $\sigma / f = ( 1/2 + n) \pi$ with
$n \in \mathbf{Z}$, and 
we focus on the region $0 < \sigma_{\mathrm{osc}} / f ,\,
\sigma_*/f < \pi$ without loss of generality. 
Since
\begin{equation}
 \tilde{n}_s - 1 = \frac{2}{3} \frac{\Lambda^4}{ H_{*}^2 f^2} 
 \cos \left( \frac{\sigma_*}{f} \right),
\label{NGns}
\end{equation}
\begin{equation}
 \tilde{\alpha}  = \frac{2}{9} \frac{\Lambda^8}{ H_{*}^4 f^4} 
 \sin^2 \left( \frac{\sigma_*}{f} \right),
\label{NGalpha}
\end{equation}
one sees that $\tilde{\alpha} / (\tilde{n}_s - 1)$ blows up at $\sigma_*
/ f = \pi/2$. 
However, $n_{f_{\mathrm{NL}}}$ is actually set by the
value itself of~$\tilde{\alpha}$ as we will soon see.

\begin{figure}[htbp]
 \begin{minipage}{.48\linewidth}
  \begin{center}
 \includegraphics[width=\linewidth]{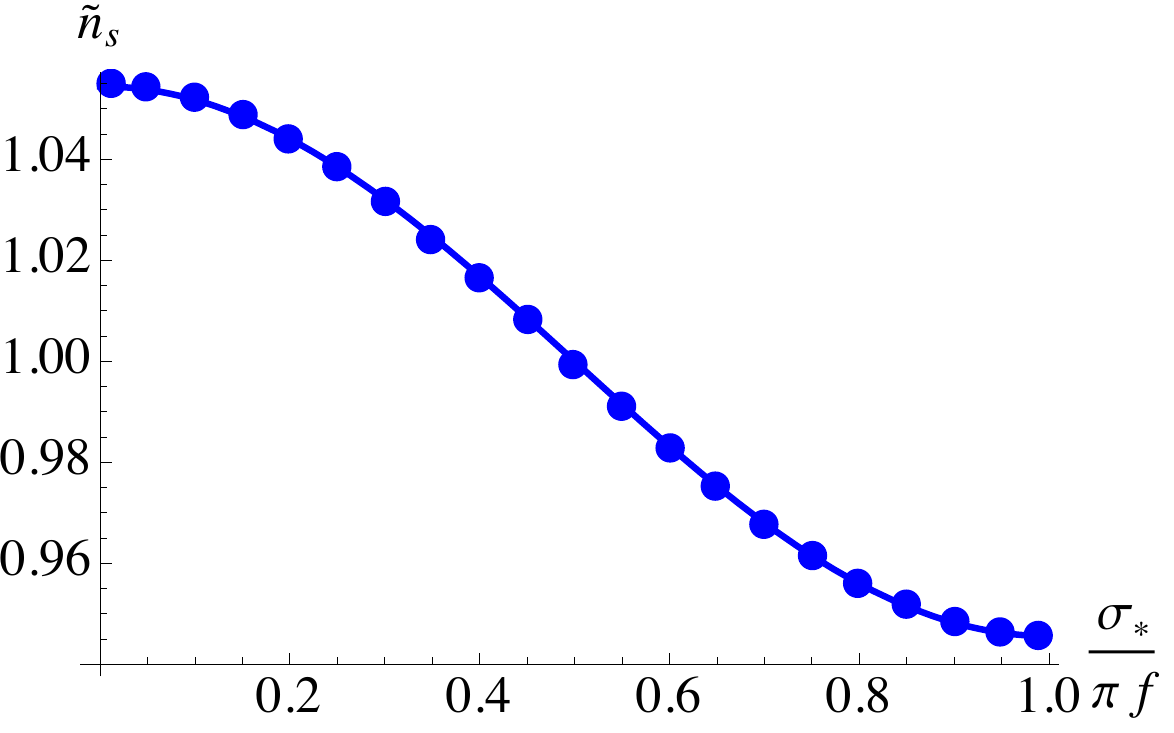}
  \end{center}
  \caption{Curvaton contribution to the spectral index~(\ref{tildens}).}
  \label{fig:NG-ns}
 \end{minipage} 
 \begin{minipage}{0.01\linewidth} 
  \begin{center}
  \end{center}
 \end{minipage} 
 \begin{minipage}{.48\linewidth}
  \begin{center}
 \includegraphics[width=\linewidth]{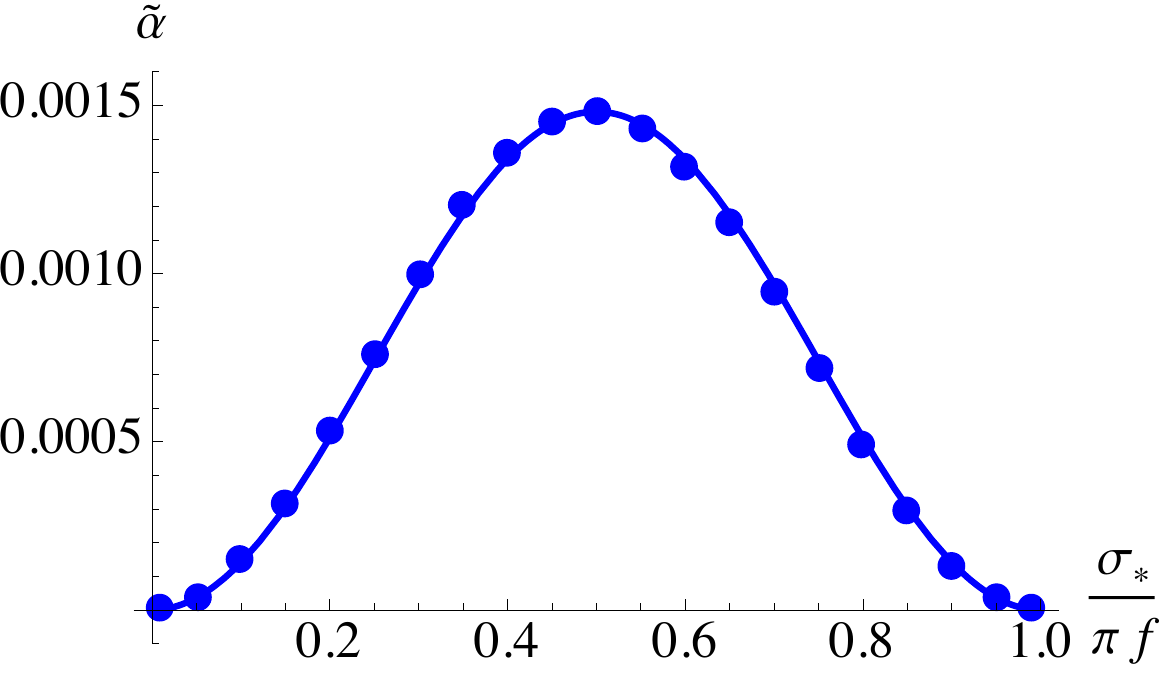}
  \end{center}
  \caption{Curvaton contribution to the running of the spectral
  index~(\ref{tildealpha}).} 
  \label{fig:NG-alpha}
 \end{minipage} 
 \begin{minipage}{.48\linewidth}
  \begin{center}
  \includegraphics[width=\linewidth]{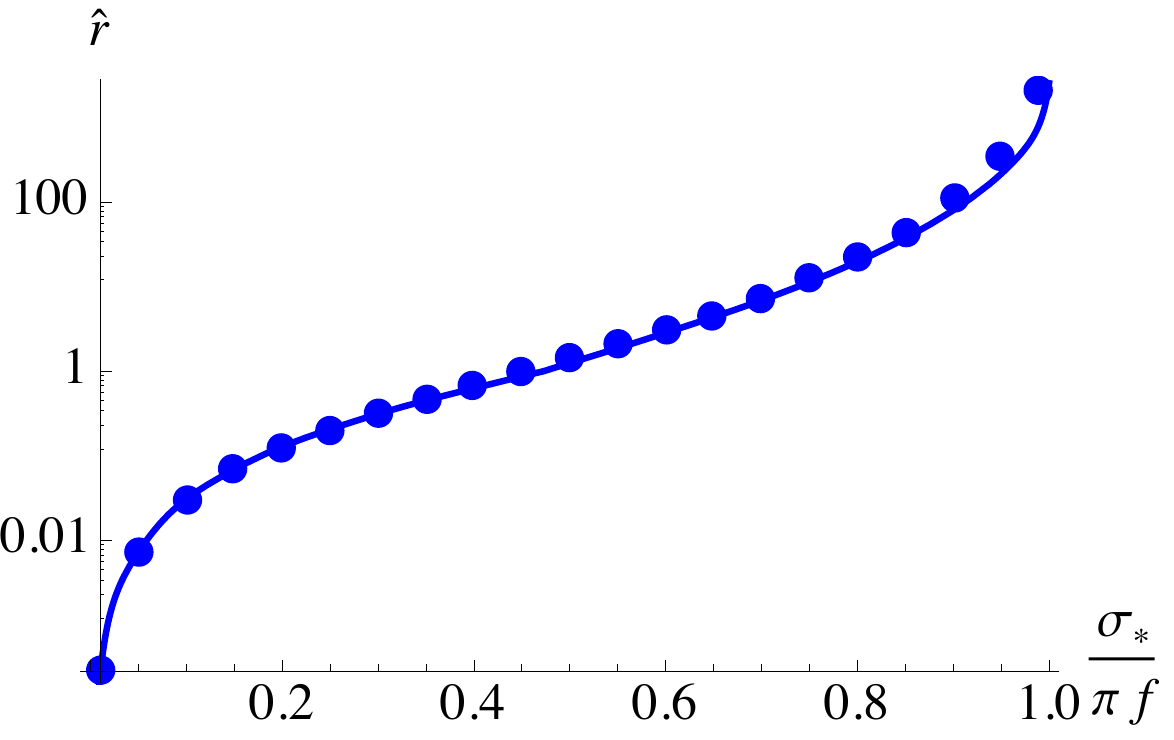}
  \end{center}
  \caption{Energy fraction at decay.}
  \label{fig:NG-r}
 \end{minipage} 
 \begin{minipage}{0.01\linewidth} 
  \begin{center}
  \end{center}
 \end{minipage} 
 \begin{minipage}{.48\linewidth}
  \begin{center}
 \includegraphics[width=\linewidth]{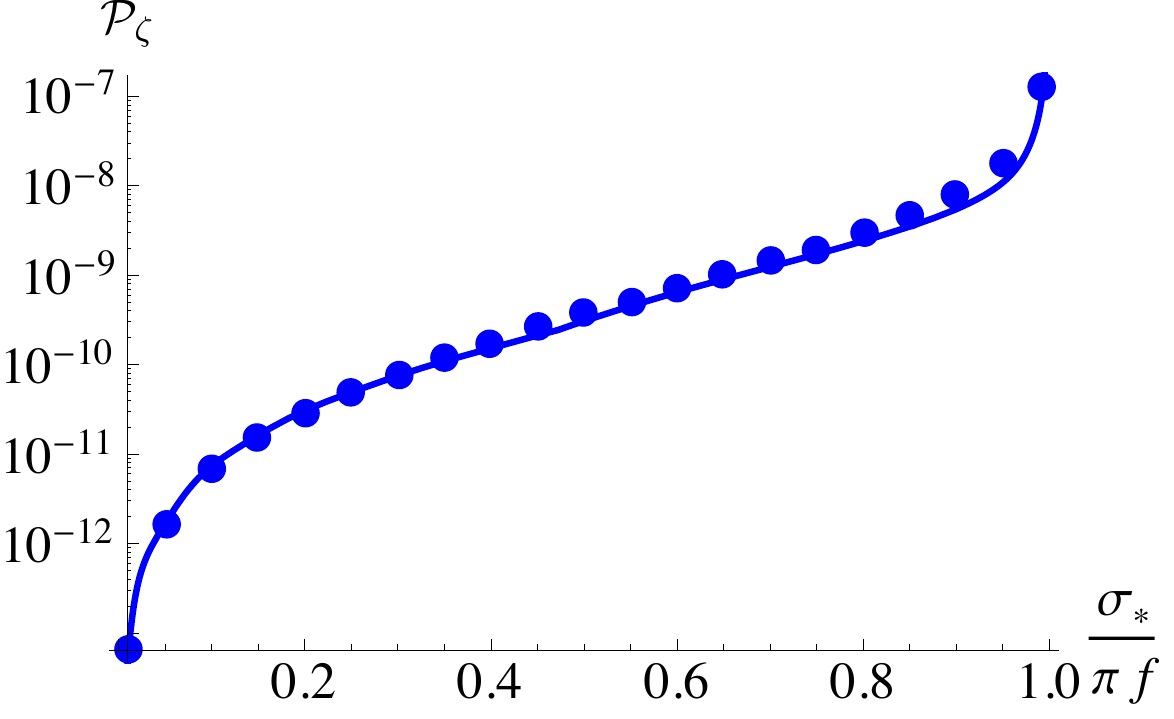}
  \end{center}
  \caption{Linear order perturbations.}
  \label{fig:NG-P}
 \end{minipage} 
 \begin{minipage}{.48\linewidth}
  \begin{center}
  \includegraphics[width=\linewidth]{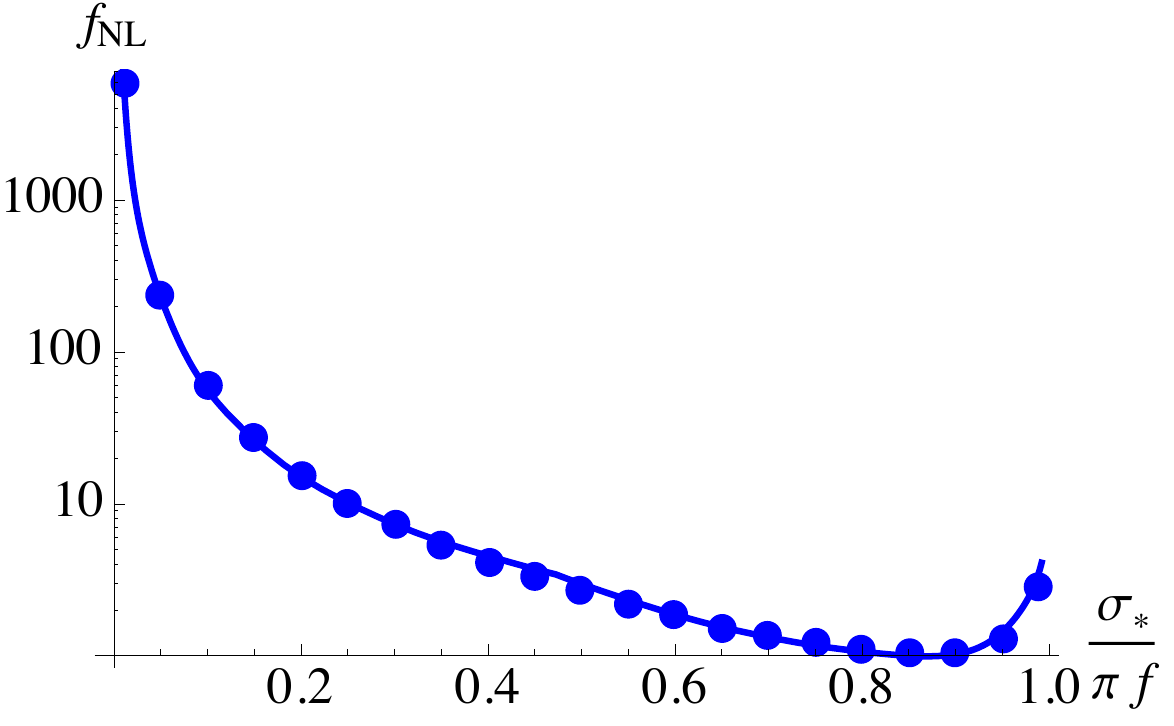}
  \end{center}
  \caption{Non-Gaussianity.}
  \label{fig:NG-fNL}
 \end{minipage} 
 \begin{minipage}{0.01\linewidth} 
  \begin{center}
  \end{center}
 \end{minipage} 
 \begin{minipage}{.48\linewidth}
  \begin{center}
 \includegraphics[width=\linewidth]{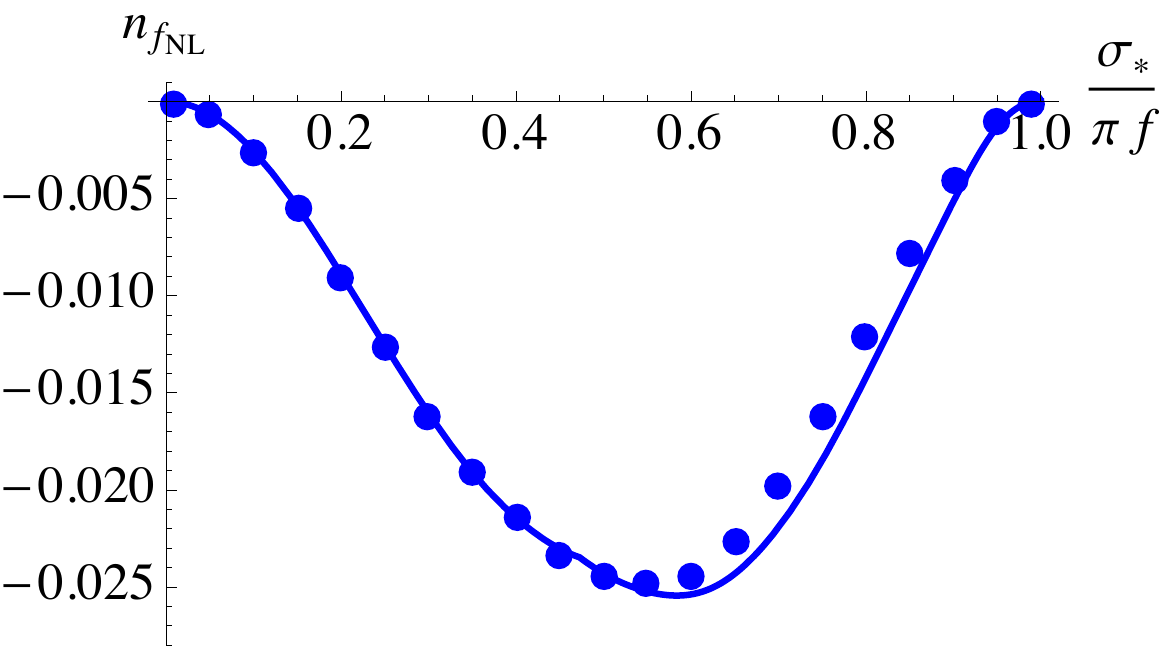}
  \end{center}
  \caption{Running of non-Gaussianity.}
  \label{fig:NG-nfNL}
 \end{minipage} 
\end{figure}

In Figures~\ref{fig:NG-ns} - \ref{fig:NG-nfNL}, we display the density
perturbations plotted as a function of $\sigma_* /f$ under the
parameter set $\Lambda = 10^{15}$ GeV and $f = 10^{17}$ GeV, along
with the inflationary parameters $H_{\mathrm{inf}} = 3.5 \times
10^{13}$ GeV, $\Gamma_\phi = 10^{11}$ GeV (i.e. the
energy density at reheating (= inflaton decay) is
$\rho_{\mathrm{reh}}^{1/4} \approx 6.5 
\times 10^{14}$ GeV), and $\mathcal{N}_* = 50$.\footnote{
  The e-folding number~$\mathcal{N}_*$ is basically determined by
  knowing the scale of inflation and the subsequent expansion history.
  However in order to clarify the dependence of the density
  perturbations on each parameter, we fix $\mathcal{N}_*$ to 50 in
  Sections~\ref{sec:NG} and \ref{sec:SI}.  The direct consequence of a
  larger~(smaller)~$\mathcal{N}_*$ is to
  decrease~(increase)~$\sigma_{\mathrm{osc}}$, hence does not affect
  $n_s$ nor~$\alpha$, while other cosmological observables can be
  affected, especially since $\hat{r}$ is also changed. Nevertheless,
  the overall behavior of the density perturbations are not influenced
  much by the detailed value of~$\mathcal{N}_*$.  Thus the fixing of
  $\mathcal{N}_\ast$ does not affect our discussions greatly.
\label{foot9}
} The curvaton decay rate is set to $\Gamma_\sigma = \frac{1}{16 \pi
}\frac{V''(0)^{3/2}}{f^2} = \frac{1}{16 \pi}\frac{\Lambda^6}{f^5}$,
supposing that the coupling of the NG curvaton with its decay product
is suppressed by the symmetry breaking scale~$f$. This set of
parameters are chosen such that the COBE (WMAP) normalization
value~\cite{arXiv:1001.4538} $\mathcal{P}_\zeta \approx 2.4 \times
10^{-9}$ as well as $\tilde{n}_s \approx 0.96 $ are realized at around
$\sigma_* / \pi f \approx 0.8$.\footnote{The inflationary scale needs to
be rather high when one tries to realize the COBE (WMAP) normalization
and $|\tilde{n}_s -1 | \sim 0.01 $ at $\sigma_*$ values not so close to the
hilltop~\cite{Kawasaki:2011pd}. For such high-scale inflation, depending
on the inflationary mechanism, density perturbations from the inflaton
can also become substantial.}
Moreover, the curvaton starts its oscillation before reheating, thus $c = 9/2$.

\begin{figure}[htbp]
  \begin{center}
 \begin{minipage}{.48\linewidth}
  \begin{center}
  \includegraphics[width=\linewidth]{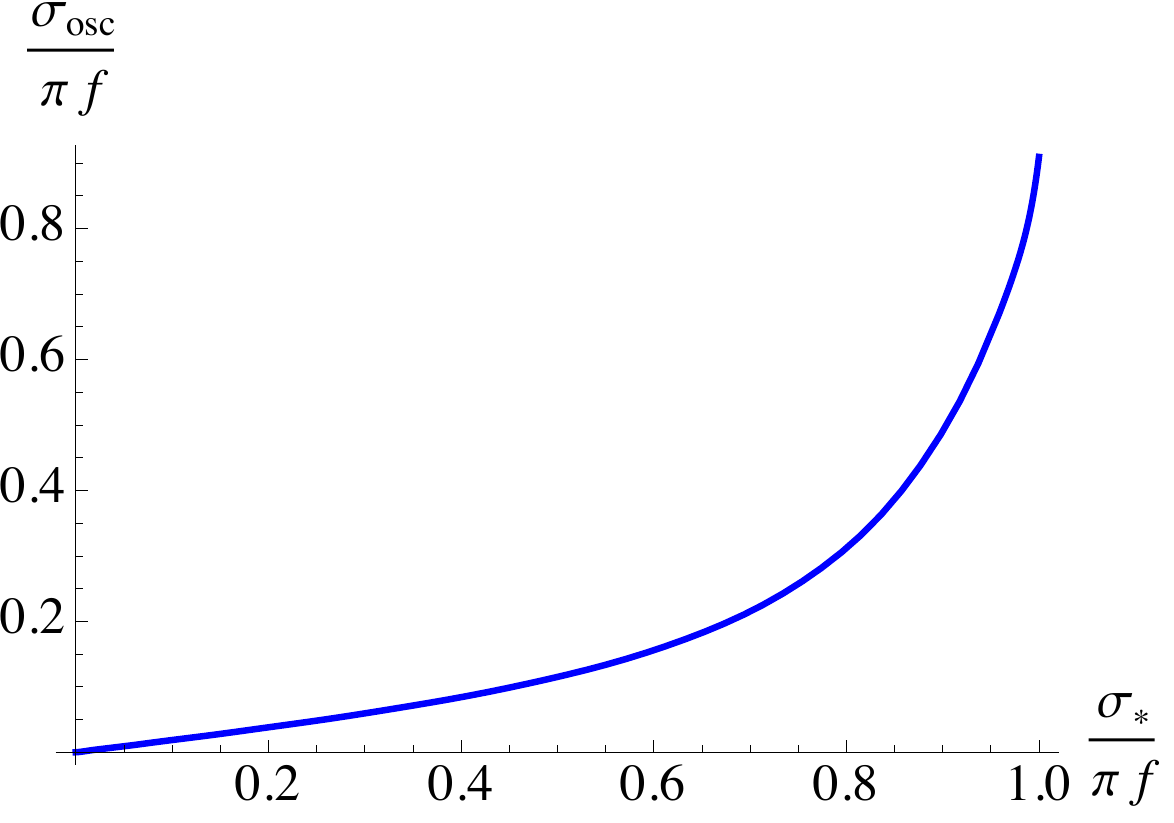}
  \end{center}
  \caption{$\sigma_{\mathrm{osc}}$ as a function of $\sigma_*$.}
  \label{fig:NG-sigma_osc}
 \end{minipage} 
  \end{center}
\end{figure}

$\sigma_{\mathrm{osc}}$ is computed by solving (\ref{sigmaosc}), which
now takes the form
\begin{equation}
 \ln \left[
 \frac{\tan \left(\sigma_{\mathrm{osc}} /2 f\right)}{\tan
 \left(\sigma_* /2 f\right)}
  \right] = 
 - \frac{\mathcal{N}_*}{3 H_{\mathrm{inf}}^2} \frac{\Lambda^4}{f^2} - 
 \frac{1}{2 (c-3)} \frac{\sigma_{\mathrm{osc}}/f }{\sin
 (\sigma_{\mathrm{osc}} /f )}.
 \label{4242}
\end{equation}
By obtaining a fitting function for the solution $\sigma_{\mathrm osc
} (\sigma_*)$ (which is shown in Figure~\ref{fig:NG-sigma_osc}), we
have analytically calculated the density perturbations in terms
of~$\sigma_*$. The analytically estimated results are shown as blue
solid lines in Figures~\ref{fig:NG-ns} - \ref{fig:NG-nfNL} for the
region $ 0.01 \lesssim \sigma_* / \pi f \lesssim 0.99$.

We have also numerically computed the density perturbations, by
solving the curvaton's equation of motion and computing the differences
in the number of
e-foldings obtained from different initial values~$\sigma_*$.  Upon
carrying out the numerical computations, we have set the inflaton
energy to a constant during inflation which lasts for 50 e-foldings
after the CMB scale exits the horizon, then transferred the inflaton
energy to non-relativistic matter redshifting as $\rho \propto
a^{-3}$, and finally to radiation as $\rho \propto
a^{-4}$. Furthermore, we adopted the sudden decay approximation for
the inflaton/curvaton, as well as a sudden end of inflation.  The
numerically computed results are shown as blue dots in the figures.
Since we have fixed the inflaton energy density to a constant during
inflation, the spectral index~(\ref{eq:ns}) and its
running~(\ref{runningalpha}) are expected to match with the parameters
$\tilde{n}_s$~(\ref{tildens}) and $\tilde{\alpha}$~(\ref{tildealpha}),
respectively, at the leading order. Thus we have displayed
$\tilde{n}_s$ and $\tilde{\alpha}$, and the numerically computed $n_s$
and $\alpha$ in the same plots in Figures~\ref{fig:NG-ns} and
\ref{fig:NG-alpha}.  One sees that the analytic estimations are in
good agreement with the results of the numerical calculations in all
Figures~\ref{fig:NG-ns} - \ref{fig:NG-nfNL}.

In most of the displayed $\sigma_* / f$ region the behavior of the
density perturbations from a NG curvaton can be understood similarly
as for a curvaton with a quadratic potential
(cf.~Footnote~\ref{foot:foot4}), except for around the hilltop
$\sigma_* \approx \pi f$ where the linear perturbations and
$f_{\mathrm{NL}}$ are enhanced. (In the figures we have plotted up to
$\sigma_* / \pi f \lesssim 0.99$, however as one goes even closer to
the hilltop, $\mathcal{P}_\zeta$ and $f_{\mathrm{NL}}$ further
increase, cf.~\cite{Kawasaki:2011pd}.)

The amplitude of the running of the
non-Gaussianity~$n_{f_{\mathrm{NL}}}$ is more or less correlated with
the $\tilde{\alpha}$ parameter, which is independent of the energy
fraction~$\hat{r}$.  This is a rather generic feature of density
perturbations from a NG curvaton, which can be understood
from~(\ref{2.20}),
\begin{equation}
 n_{f_{\mathrm{NL}}}  
 \simeq -  \frac{5}{18} 
 \frac{1}{f_{\mathrm{NL}}}
 \left( \frac{\partial \mathcal{N}}{\partial \sigma_*} \right)^{-1}
 \frac{\Lambda^4}{H_*^2 f^3} \sin \left(\frac{\sigma_*}{f}\right)
 = \left( - \frac{6}{5} f_{\mathrm{NL}} \cdot
  \frac{\partial \mathcal{N}}{\partial \sigma_*} \cdot \sigma_*\right)^{-1} 
 \frac{\sigma_*}{f} \sqrt{\frac{\tilde{\alpha}}{2}}.
\label{eq27}
\end{equation}
In the far right hand side, the product inside the parentheses is a
combination whose amplitude is smaller than unity for quadratic
curvatons.\footnote{
  A quadratic curvaton $V \propto \sigma^2$ gives
\begin{equation}
- \frac{6}{5} f_{\mathrm{NL}} \cdot
  \frac{\partial \mathcal{N}}{\partial \sigma_*} \cdot \sigma_*
 = \frac{9\hat{r}^2 - 8\hat{r} - 16}{(3 \hat{r}+4)^2},
\label{eq3.5}
\end{equation}
which monotonically increases with~$\hat{r}$, and approaches $+(-)1$
in the limit $\hat{r} \gg (\ll) 1$.
} Here, recall that the density perturbations from a NG curvaton are
more or less the same as those from a quadratic curvaton, except for
the hilltop region.  Therefore the $()^{-1}$~term in the far right
hand side of~(\ref{eq27}) is roughly of order unity (except for
when~$f_{\mathrm{NL}}$ vanishes), and accordingly,
$|n_{f_{\mathrm{NL}}}|$ is roughly of order~$\sqrt{\tilde{\alpha}}$.
On the other hand, when $|f_{\mathrm{NL}}| \ll 1$, the product inside
the parentheses also becomes much smaller than unity and thus its
inverse blows up (though not seen in Figure~\ref{fig:NG-fNL}, this can
happen depending of the parameter set).  However, in such a case the
product $n_{f_{\mathrm{NL}}} f_{\mathrm{NL}}$ to which experiments are
sensitive is suppressed.  Hence one can conclude that the running
of~$f_{\mathrm{NL}}$ from a NG curvaton is basically set by the
absolute value of~$\tilde{\alpha}$, which is constrained by current
observational bounds on a running spectral index\footnote{The current
  observational 1$\sigma$ limit for the running spectral index from
  WMAP7 is $ \alpha = -0.034 \pm 0.026$ \cite{arXiv:1001.4538}.
} (unless the time-variation of the Hubble parameter cancels out
$\tilde{\alpha}$ from (\ref{runningalpha})).

As mentioned above, the produced perturbations behave quite
differently as one approaches the hilltop, i.e. $\sigma_* / f \to
\pi$.  However there the potential is well approximated by a
quadratic, i.e. $V \simeq V_0 - m^2 \sigma^2$, hence $\tilde{\alpha}$
approaches zero and $n_{f_{\mathrm{NL}}}$ is suppressed.

\vspace{\baselineskip}

\begin{figure}[htbp]
 \begin{minipage}{.48\linewidth}
  \begin{center}
 \includegraphics[width=\linewidth]{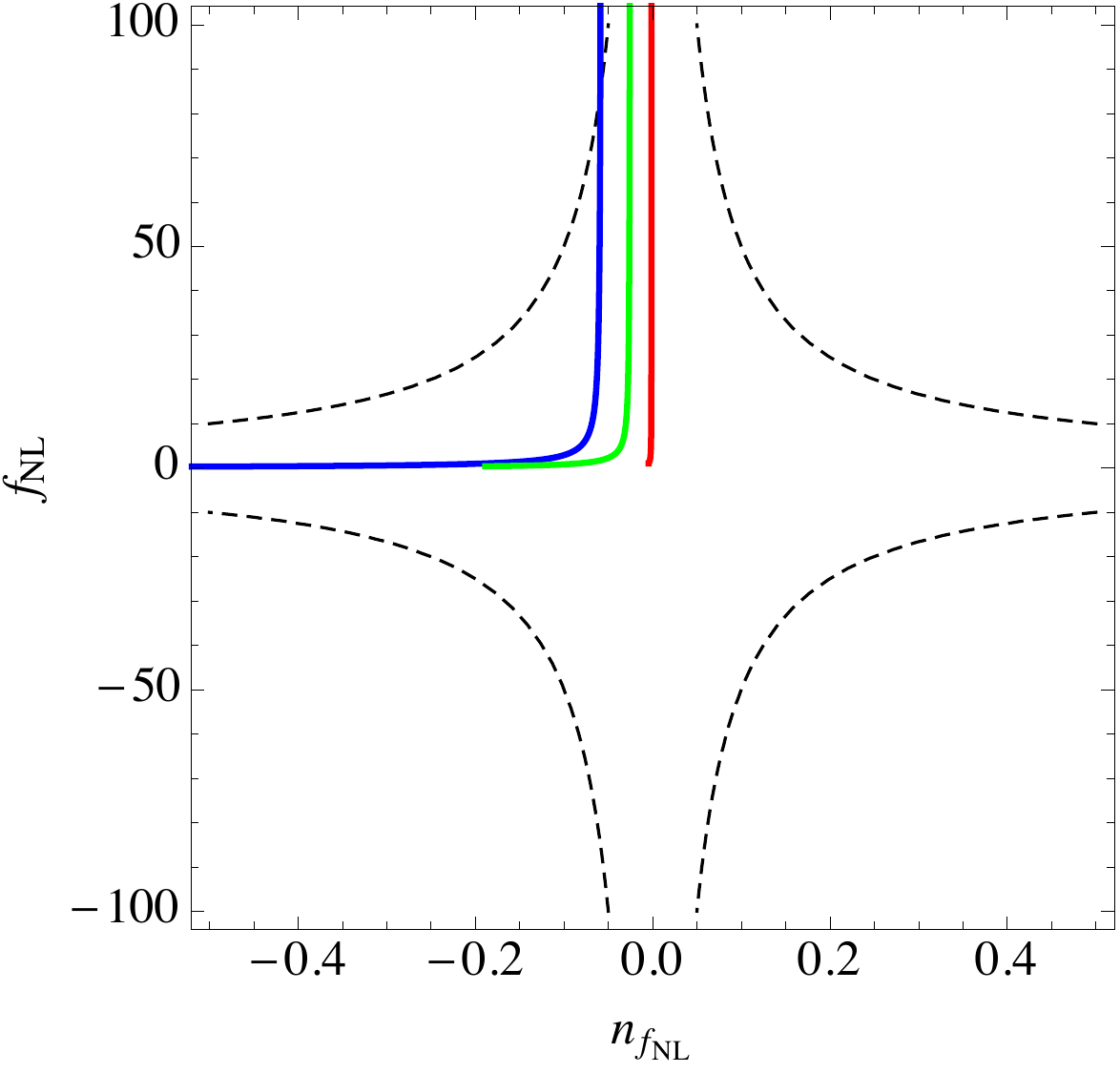}
  \end{center}
  \caption{Varying $\hat{r}$ under $\tilde{n}_s = 0.96$. $\sigma_* /
    \pi f $ is fixed to 0.55 (blue), 0.60 (green), and 0.90 (red),
    corresponding to $\tilde{\alpha} \approx $ 0.03, 0.008, and $8
    \times 10^{-5}$, respectively.  The expected observational
    sensitivity of Planck is also shown (black dashed).}
  \label{fig:NG96}
 \end{minipage} 
 \begin{minipage}{0.01\linewidth} 
  \begin{center}
  \end{center}
 \end{minipage} 
 \begin{minipage}{.48\linewidth}
  \begin{center}
 \includegraphics[width=\linewidth]{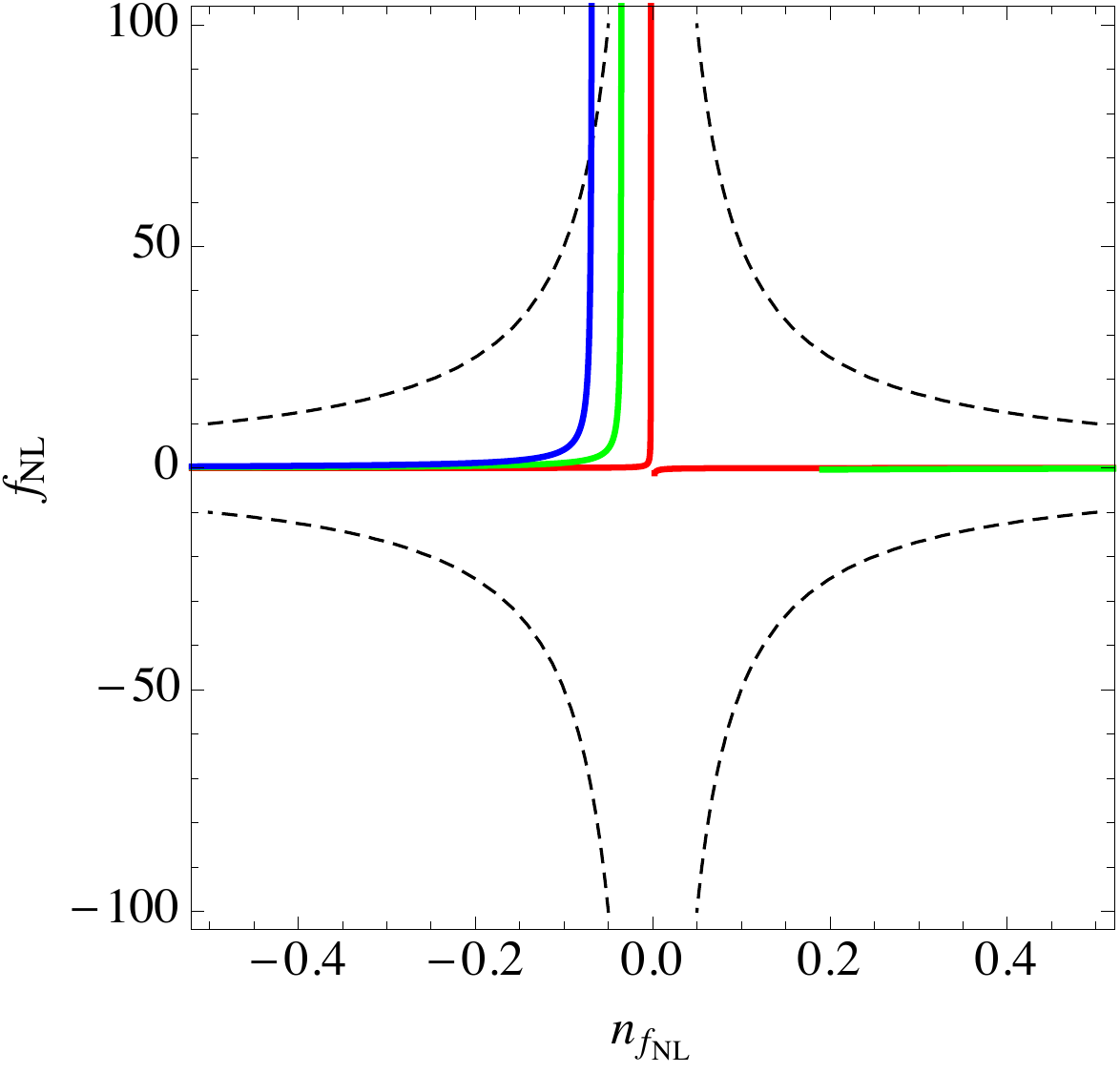}
  \end{center}
  \caption{Varying $\hat{r}$ under $\tilde{n}_s = 1.04$. $\sigma_* / \pi f$ is
  fixed to 0.45 (blue), 0.40 
  (green), and 0.10 (red), corresponding to $\tilde{\alpha} \approx $
  0.03, 0.008, and $8 \times 10^{-5}$, respectively.
  } \vspace{10mm}
  \label{fig:NG04}
 \end{minipage} 
\end{figure}

We show contours in the $n_{f_{\mathrm{NL}}}$ - $f_{\mathrm{NL}}$
planes from a NG curvaton in Figures~\ref{fig:NG96} and
\ref{fig:NG04}.  Instead of choosing a certain parameter set for
e.g. $\Lambda$ and $f$ as in the previous figures, here we have fixed
$\tilde{n}_s $ to 0.96 and 1.04, respectively, and plotted the
contours in the following way: For a given~$\tilde{n}_s$, by
using~(\ref{NGns}) one can rewrite $\Lambda^4 / H_{\mathrm{inf}}^2
f^2$ in terms of $\sigma_* / f$.  Hence $f_{\mathrm{NL}}$ as well as
$n_{f_{\mathrm{NL}}}$ are uniquely determined by $\sigma_* / f$,
$\sigma_{\mathrm{osc}} / f$, $\hat{r}$, $\tilde{n}_s$, and $c$.
Furthermore, $\sigma_{\mathrm{osc}} / f$ is given as a function of
$\sigma_* / f$ after solving (\ref{4242}) (note that $\Lambda^4 /
H_{\mathrm{inf}}^2 f^2$ is now determined by $\tilde{n}_s$ and
$\sigma_* / f$), where we took $\mathcal{N}_* = 50$ and $c = 9/2$,
i.e. $t_{\mathrm{reh}} > t_{\mathrm{osc}}$ (taking instead $c = 5$,
i.e. $t_{\mathrm{reh}} < t_{\mathrm{osc}}$, makes little difference
for the resulting contours).  Each colored solid line in the figures
is plotted for a fixed value of $\sigma_* / \pi f$, under a
varying~$\hat{r}$. Different values for $\sigma_* / \pi f$ give
different values for $\tilde{\alpha}$: blue, green, and red lines
correspond to $\tilde{\alpha} \approx $ 0.05, 0.008, and $8 \times
10^{-5}$, respectively.  We note that this procedure for plotting the
contours does not fix the linear perturbation amplitude, i.e., the
individual values for $\Lambda$, $f$, and $H_{\mathrm{inf}}$ can be
varied to change the perturbation amplitude for a fixed $\Lambda^4
/H_{\mathrm{inf}}^2 f^2 $.  As reference values for detectability, we
have also shown $n_{f_{\mathrm{NL}}} f_{\mathrm{NL}} / 50 = 0.10 $
contours as black dashed lines. Outside these contours corresponds to
the region detectable by Planck, cf.~(\ref{obs}).

Taking smaller~$\hat{r}$ ($\ll 1$) corresponds to moving upwards along
the colored contour lines.  It is clearly seen that a large $
n_{f_{\mathrm{NL}}} f_{\mathrm{NL}} $ is realized for larger values of
$\tilde{\alpha}$, as was discussed below (\ref{eq27}).  Since the
contours in the two figures are chosen to take the same values
for~$\tilde{\alpha}$, they are more or less the same in the $\hat{r}
\ll 1$ regime.

On the other hand, as $\hat{r}$ is increased, the contour lines
eventually turns left and $n_{f_{\rm NL}}$ decreases.  There the blue
contours realize $|n_{f_{\mathrm{NL}}}|$ much larger than $|\alpha|$,
due to the suppressed $|f_{\mathrm{NL}}| \ll 1$.  In some cases
$f_{\mathrm{NL}}$ can even cross zero (as it happens for quadratic
curvatons, cf.~(\ref{eq3.5})), which is accompanied by
$n_{f_{\mathrm{NL}}}$ blowing up and changing sign. This is seen in
Figure~\ref{fig:NG04} as the contour lines extending towards
$n_{f_{\rm NL}} \to - \infty$, and then coming back from the right.
Although such behavior is absent in Figure~\ref{fig:NG96}, it can
happen for $\sigma_* / f > \pi / 2$ cases as well, depending on the
parameter values.  Since the blowing up of $n_{f_{\mathrm{NL}}}$
happens when $f_{\mathrm{NL}}$ is tiny, its observational detection
would be challenging, even if it happened.  In the $\hat{r} \to \infty
$ limit, $n_{f_{\mathrm{NL}}}$ approaches a certain value, which can
be seen as the end points of the contour lines.  (The end points of
the blue lines in both figures are outside the displayed region.)

Focusing on a contour line with a fixed value for $\sigma_* / \pi f$,
then taking a larger value for $|\tilde{n}_s - 1|$ is equivalent to
increasing~$\tilde{\alpha}$, hence the contour would shift towards
larger~$|n_{f_{\mathrm{NL}}}|$.  In summary, for NG curvatons, large
$|n_{f_{\mathrm{NL}}} f_{\mathrm{NL}}|$ is produced together with a
large~$\tilde{\alpha}$, therefore it is already strictly constrained
by current observational bounds on running spectral index, unless
$\tilde{\alpha}$ is cancelled out from the expression for~$\alpha$ due
to a varying Hubble parameter during inflation.

\section{Self-Interacting Curvatons}
\label{sec:SI}

As a simple example of curvaton potentials that steepen more rapidly
than a quadratic, in this section we explore curvatons possessing a
mass term and an additional higher-order polynomial
term\cite{astro-ph/0508573,arXiv:0807.3069,Huang:2008zj,Enqvist:2009zf,Enqvist:2009eq,Enqvist:2009ww,Fonseca:2011aa}:
\begin{equation}
 V(\sigma) = \Lambda^4 
\left[
\left(\frac{\sigma}{f}\right)^2 + \left(\frac{\sigma}{f}\right)^m
\right],
\label{SIpot}
\end{equation}
where $\Lambda$ and $f$ are positive constants with mass dimension,
and $m$ is an even integer with $m > 2$. $f$ denotes where the
higher-order term becomes important, while $\Lambda$ sets the overall
scale of the curvaton potential. Furthermore, the curvaton mass at the
potential minimum is given by $M_\sigma^2 = 2 \Lambda^4 / f^2$.  We
especially focus on the region $|\sigma | \lesssim f$, where the
analytic expressions give good estimations of the resulting curvature
perturbations and a large running of~$f_{\mathrm{NL}}$ can show
up. The validity of the analytic expressions at large $\sigma$ values
are discussed towards the end of this section.

The potential~(\ref{SIpot}) gives
\begin{equation}
 \tilde{n}_s - 1 = \frac{4}{3} \frac{\Lambda^4}{H_*^2 f^2}
 \left\{1 + \frac{m (m-1)}{2} \left(\frac{\sigma_*}{f}\right)^{m-2}\right\},
\end{equation}
\begin{equation}
 \tilde{\alpha} = -\frac{4  m (m-1) (m-2)}{9} \frac{\Lambda^8}{H_*^4 f^4}
 \left(\frac{\sigma_*}{f}\right)^{m-2} 
 \left\{ 1 +\frac{m}{2} \left(\frac{\sigma_*}{f}\right)^{m-2} \right\},
\end{equation}
where one can see that the resulting $n_s$ and $\alpha$ are positive
and negative, respectively, for $\sigma_* \neq 0$.  We also note that
$ - \frac{\tilde{\alpha}}{(\tilde{n}_s-1)^2} $ takes its maximum value
$\frac{m-1}{8}$ at $( \frac{ \sigma_*} {f})^{m-2} = \frac{2}{m
  (m-3)}$.  Hence especially for $m \lesssim 10$, then
$-\tilde{\alpha} \lesssim (\tilde{n}_s - 1)^2$.

\begin{figure}[htb]
 \begin{minipage}{.44\linewidth}
  \begin{center}
 \includegraphics[width=\linewidth]{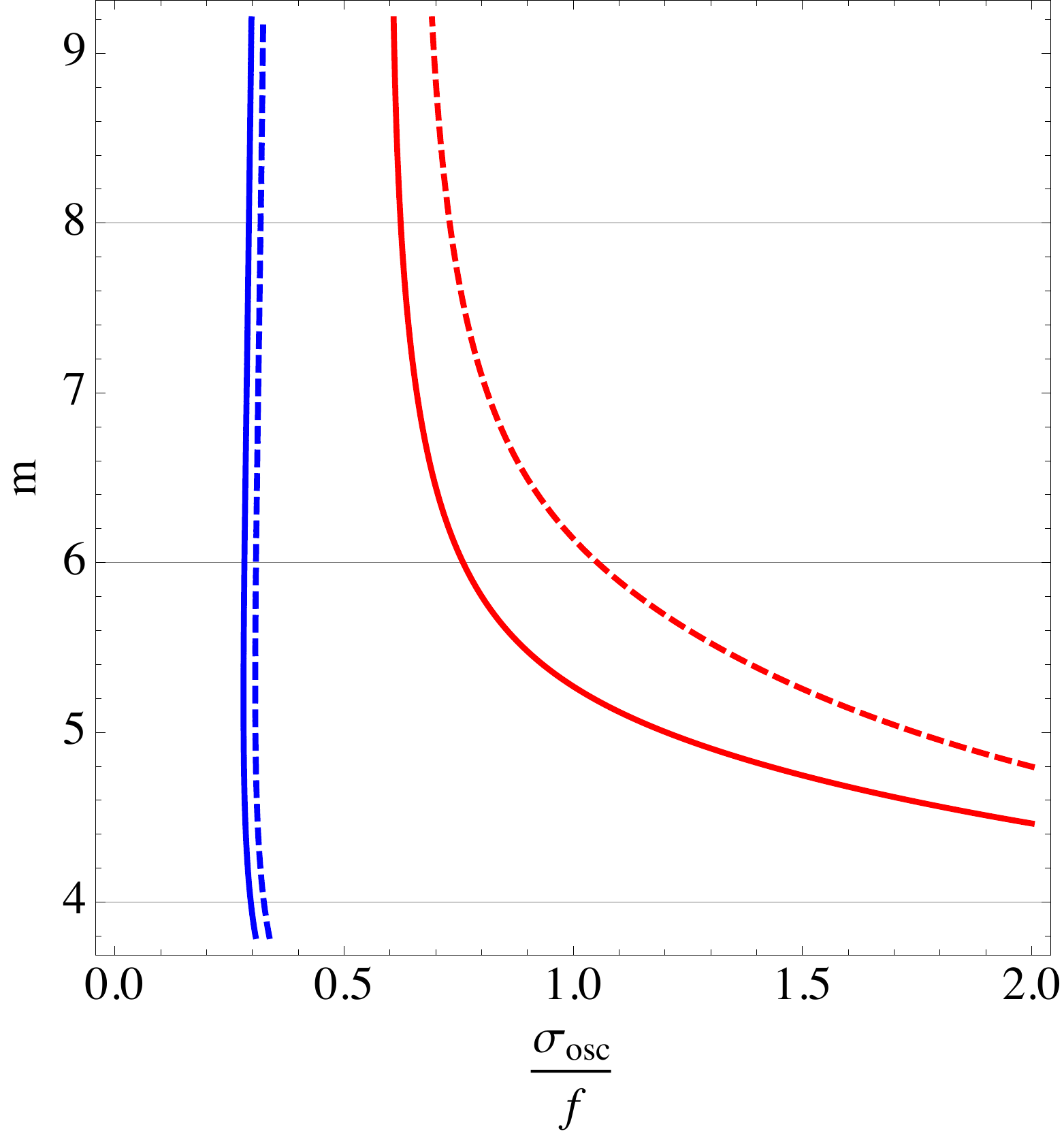}
  \end{center}
  \caption{The right hand side of (\ref{eq4.6}) becomes negative on
    the right sides of the lines. Blue and red lines denote
    $\mathcal{N}_* \Lambda^4 / H_{\mathrm{inf}}^2 f^2 = 1$ and
    $10^{-2}$, while solid and dashed lines denote $c = 9/2$ and 5,
    respectively.}
  \label{fig:SI-sigmaosc_m}
 \end{minipage} 
 \begin{minipage}{0.01\linewidth} 
  \begin{center}
  \end{center}
 \end{minipage} 
 \begin{minipage}{.48\linewidth}
  \begin{center}
 \includegraphics[width=\linewidth]{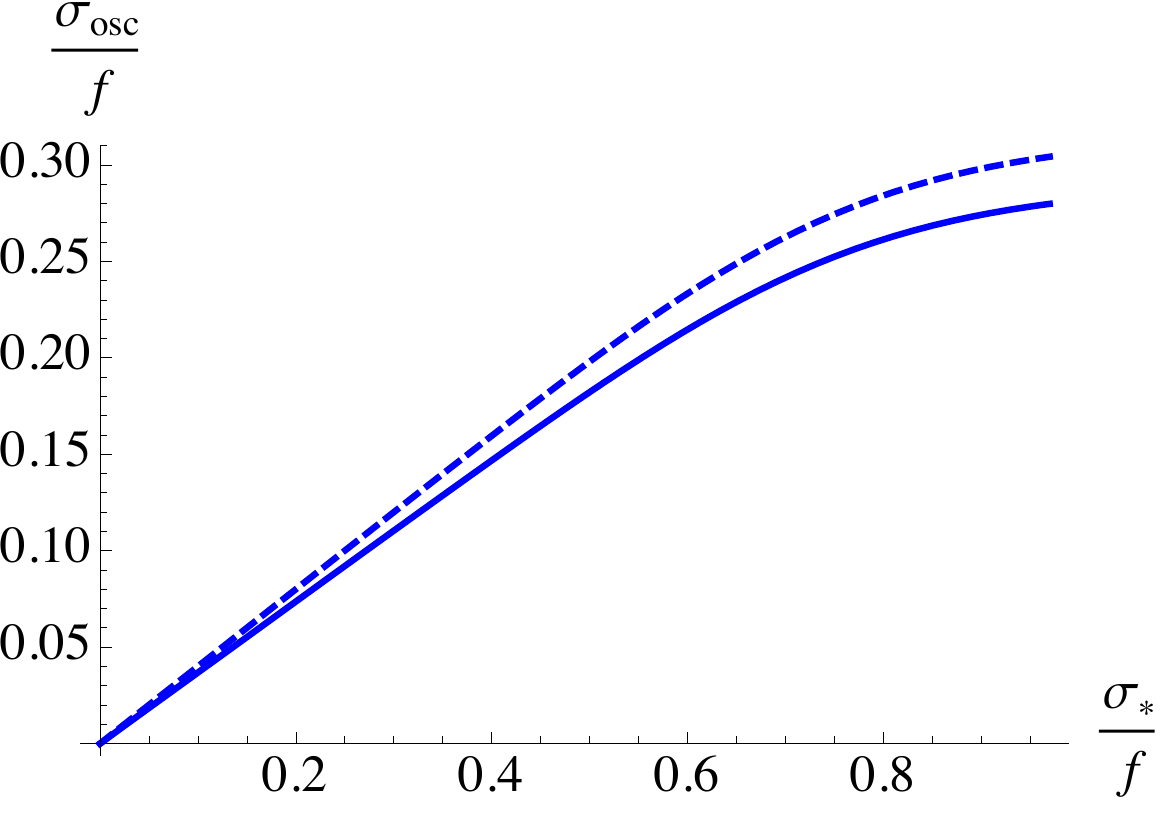}
  \end{center}
  \caption{$\sigma_{\mathrm{osc}}$ as a function of $\sigma_*$, for
    $m=8$ and $ \mathcal{N}_* \Lambda^4 / H_{\mathrm{inf}}^2 f^2 =
    1$. Solid and dashed lines denote $c = 9/2$ and 5, respectively.}
  \label{fig:SI-sigma_osc}
 \end{minipage} 
\end{figure}

Let us now examine $\sigma_{\mathrm{osc}}$ since understanding its
behavior in terms of $\sigma_*$ is essential for discussing largely
scale-dependent~$f_{\mathrm{NL}}$ produced from self-interacting
curvatons. The Hubble parameter at the onset of
oscillation~(\ref{Hosc}) is
\begin{equation}
 H_{\mathrm{osc}}^2 = \frac{1}{c} \frac{\Lambda^4}{f^2} 
 \left(2 + m \left(\frac{\sigma_{\mathrm{osc}}}{f}\right)^{m-2}  \right),
\end{equation}
thus the relation~(\ref{sigmaosc}) between $\sigma_*$ and
$\sigma_{\mathrm{osc}}$ becomes
\begin{equation}
 \ln \left[
 \frac{m + 2 \left(\sigma_*/f\right)^{-(m-2)}}{m + 2
 \left(\sigma_{\mathrm{osc}}/f\right)^{-(m-2)}}  
\right]
=
 -(m-2) \left\{
\frac{2 }{3 }\mathcal{N}_*\frac{\Lambda^4}{H_{\mathrm{inf}}^2 f^2} + 
 \frac{1}{c-3} \left(2 + m
\left(\frac{\sigma_{\mathrm{osc}}}{f}\right)^{m-2}\right)^{-1} 
\right\}.
\label{eq4.5}
\end{equation}
By solving this equation, $\sigma_{\mathrm{osc}}$ is obtained as a
function of~$\sigma_*$. (Here $\sigma_{\rm osc}$ should take the same
sign as $\sigma_*$, cf. Footnote~\ref{foot:5}.) We do this
numerically, as shown in Figure~\ref{fig:SI-sigma_osc}, but the
behavior of $\sigma_{\mathrm{osc}}$ can simply be understood by
rewriting (\ref{eq4.5}) as\footnote{
  Here we exponentiate both sides of the equation in order to make
  clear the behavior of its solution, but let us note that since
  (\ref{sigmaosc}) itself is an approximate relation, one should in
  general be careful about the size of the errors when exponentiating
  the equation.
  }
\begin{multline}
 2 \left(\frac{\sigma_*}{f} \right)^{-(m-2)} = -m
 + \left( m +  2 \left(\frac{\sigma_{\mathrm{osc}}}{f} \right)^{-(m-2)}
   \right) \times \\
 \exp \left[
 -(m-2) \left\{
\frac{2 }{3 }\mathcal{N}_*\frac{\Lambda^4}{H_{\mathrm{inf}}^2 f^2} + 
 \frac{1}{c-3} \left(2 + m
\left(\frac{\sigma_{\mathrm{osc}}}{f}\right)^{m-2}\right)^{-1} 
\right\}
\right].
\label{eq4.6}
\end{multline}
Since the left hand side of this equation is positive,
$\sigma_{\mathrm{osc}}$ should not take values that make the right
hand side negative, which can happen for e.g., large $m$,
$|\sigma_{\mathrm{osc}} / f|$, and $\mathcal{N}_* \Lambda^4 /
H_{\mathrm{inf}}^2 f^2$.  Such values of $\sigma_{\mathrm{osc}}$ are
displayed in Figure~\ref{fig:SI-sigmaosc_m}, where regions of the
$\sigma_{\mathrm{osc}} / f$ - $m$ plane on the right sides of the
lines give negative values to the right hand side of (\ref{eq4.6}).
For example, when $ m = 8$ and $\mathcal{N}_* \Lambda^4 /
H_{\mathrm{inf}}^2 f^2 = 1$, then $\sigma_{\mathrm{osc}} \gtrsim 0.3
f$ cannot be a solution of (\ref{eq4.6}), i.e. the curvaton rolls down
before it starts its oscillations
to values smaller than about $0.3 f$ even if $\sigma$ during inflation
takes much larger field values\footnote{
  Of course, discussions here are basically limited to curvaton field
  values that satisfy the approximation (\ref{CDapp}) until the onset
  of the oscillations.
}.  Such behavior is in contrast to a quadratic curvaton, for which
one can check that $\sigma_{\rm osc} \propto \sigma_*$.  In
Figure~\ref{fig:SI-sigma_osc} we plot $\sigma_{\mathrm{osc}}$ as a
function of $\sigma_*$ by numerically solving (\ref{eq4.5}) for $m=8$
and $ \mathcal{N}_* \Lambda^4 / H_{\mathrm{inf}}^2 f^2 = 1$.  One sees
that the growing rate of $\sigma_{\mathrm{osc}}$ is suppressed at
$\sigma_* \sim f$ where the curvaton potential steepens due to the
higher-order self-interaction, and that $\sigma_{\mathrm{osc}}$ is
limited to values smaller than $\sim 0.3 f$ as is indicated in
Figure~\ref{fig:SI-sigmaosc_m}.  Such flattening of
$\sigma_{\mathrm{osc}}$ is more significant for larger values of $m$
and $\Lambda^4/H_{\mathrm{inf}}^2 f^2$ which make the potential
steeper, and for larger~$\mathcal{N}_*$ providing a longer period for
the curvaton to roll down during inflation.

\begin{figure}[htbp]
 \begin{minipage}{.48\linewidth}
  \begin{center}
 \includegraphics[width=\linewidth]{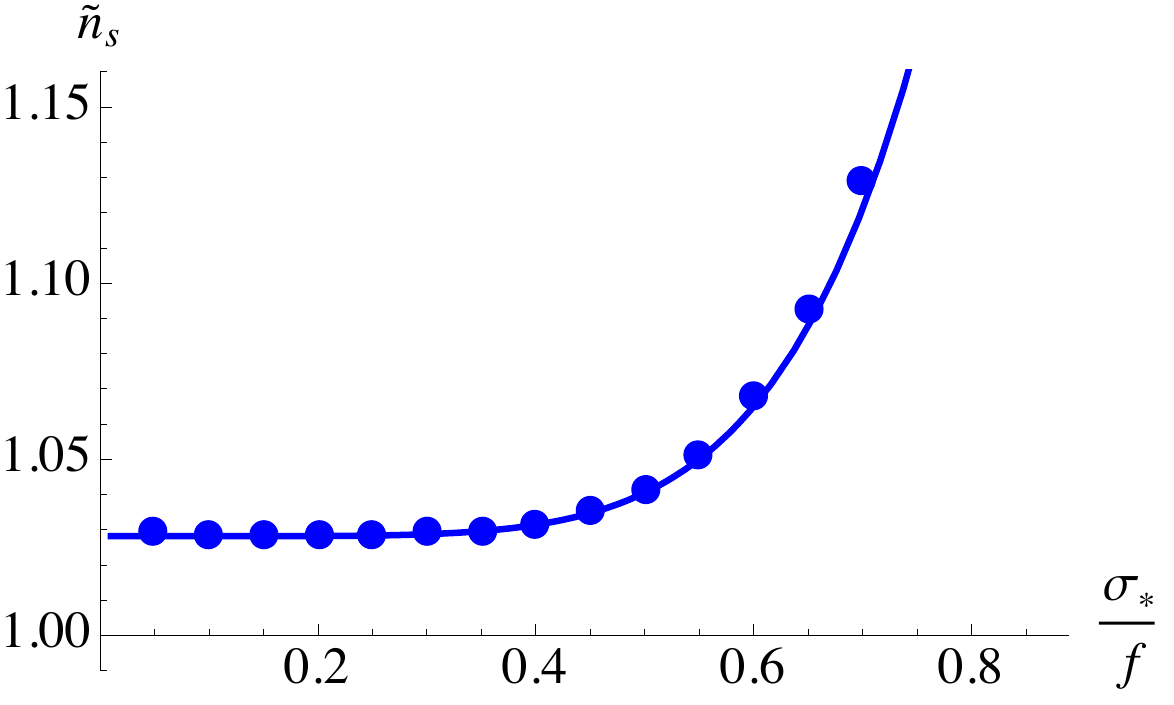}
  \end{center}
  \caption{Curvaton contribution to the spectral index~(\ref{tildens}).}
  \label{fig:SI-ns}
 \end{minipage} 
 \begin{minipage}{0.01\linewidth} 
  \begin{center}
  \end{center}
 \end{minipage} 
 \begin{minipage}{.48\linewidth}
  \begin{center}
 \includegraphics[width=\linewidth]{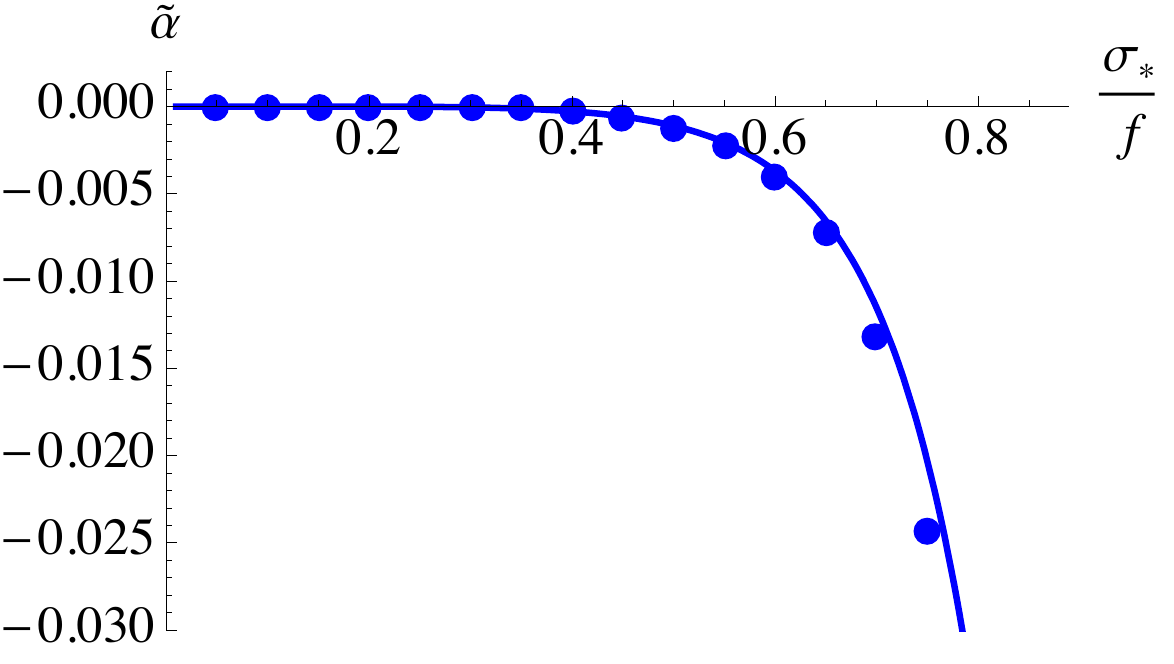}
  \end{center}
  \caption{Curvaton contribution to the running of the spectral
  index~(\ref{tildealpha}).}
  \label{fig:SI-alpha}
 \end{minipage} 
 \begin{minipage}{.48\linewidth}
  \begin{center}
  \includegraphics[width=\linewidth]{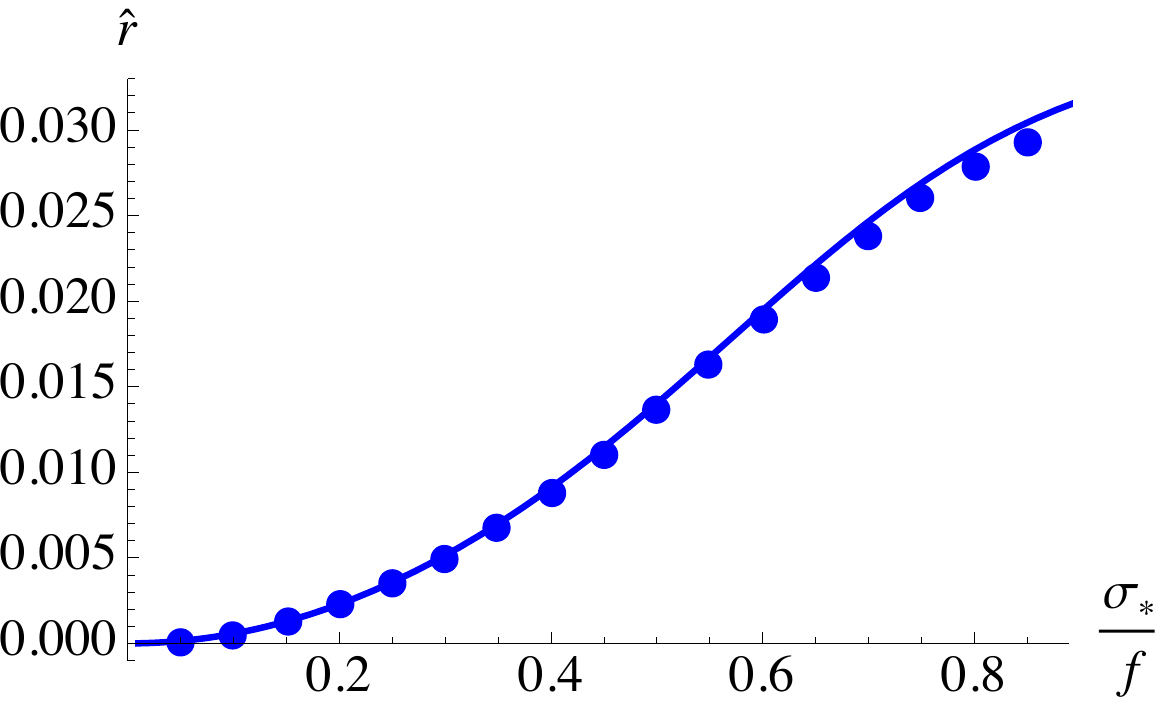}
  \end{center}
  \caption{Energy fraction at decay.}
  \label{fig:SI-r}
 \end{minipage} 
 \begin{minipage}{0.01\linewidth} 
  \begin{center}
  \end{center}
 \end{minipage} 
 \begin{minipage}{.48\linewidth}
  \begin{center}
 \includegraphics[width=\linewidth]{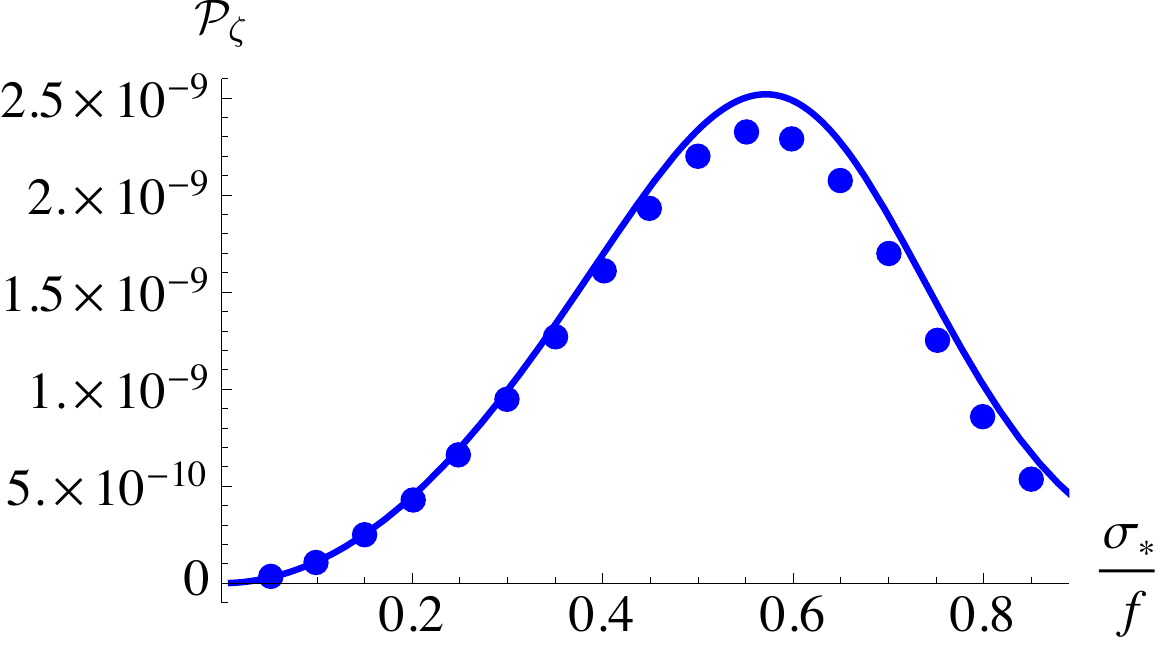}
  \end{center}
  \caption{Linear order perturbations.}
  \label{fig:SI-P}
 \end{minipage} 
 \begin{minipage}{.48\linewidth}
  \begin{center}
  \includegraphics[width=\linewidth]{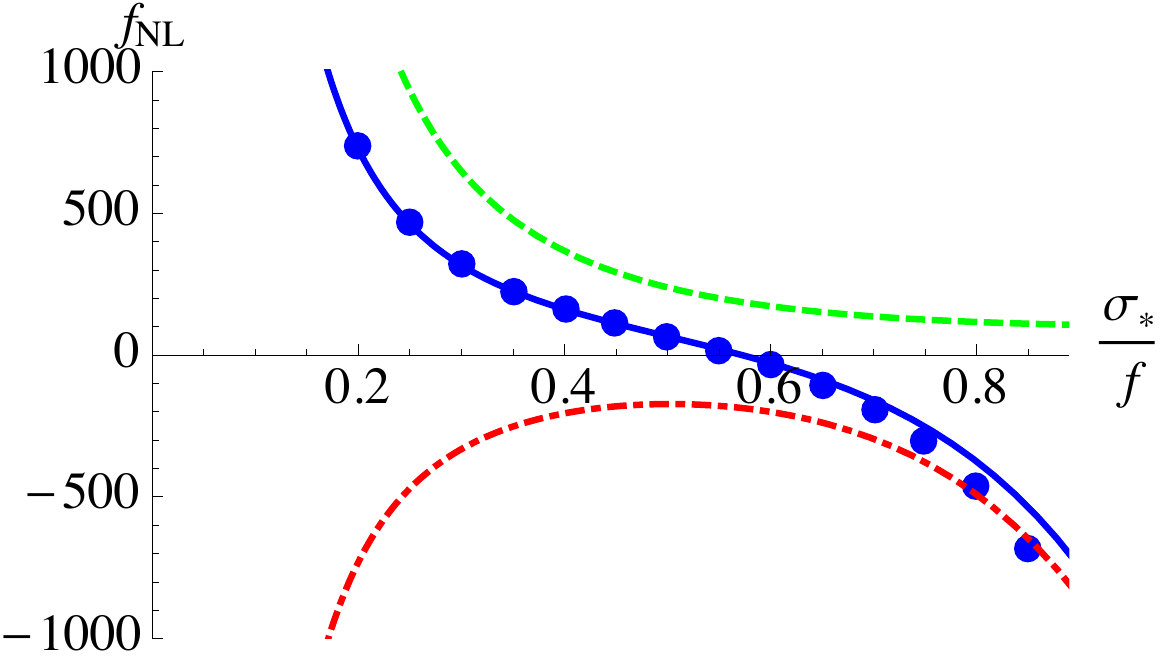}
  \end{center}
  \caption{Non-Gaussianity. $f_{\mathrm{NL}}$: blue solid, $f_1$: red
  dot-dashed, $f_2$: green dashed.}
  \label{fig:SI-fNL}
 \end{minipage} 
 \begin{minipage}{0.01\linewidth} 
  \begin{center}
  \end{center}
 \end{minipage} 
 \begin{minipage}{.48\linewidth}
  \begin{center}
 \includegraphics[width=\linewidth]{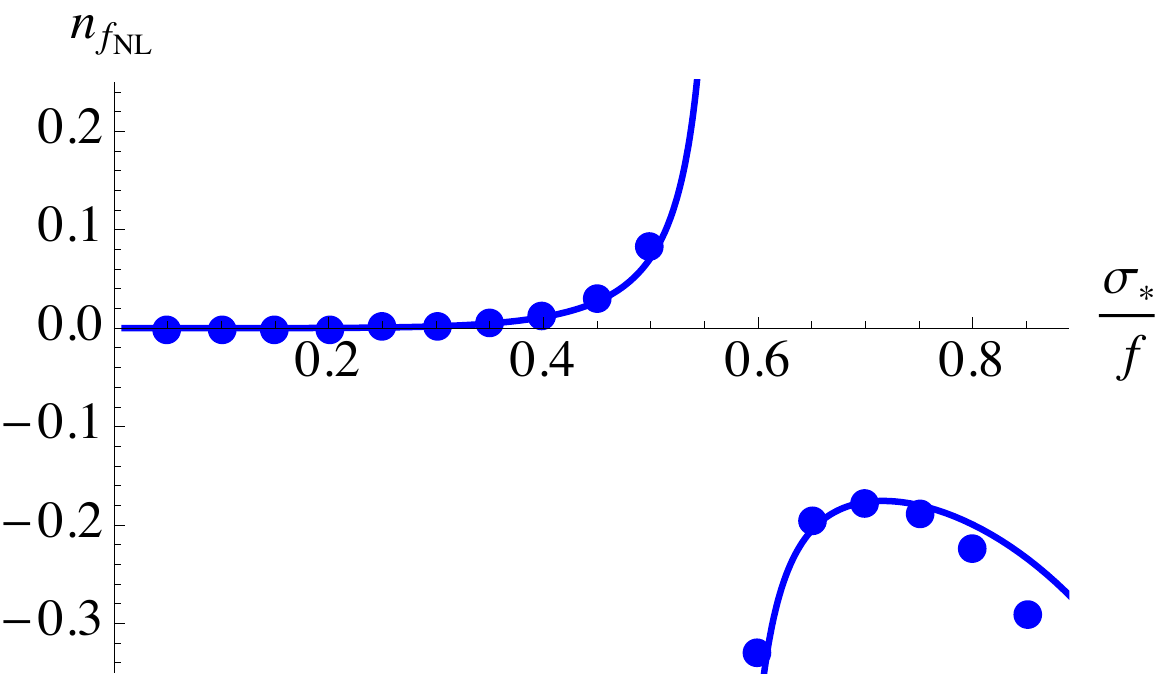}
  \end{center}
  \caption{Running of non-Gaussianity.}
  \label{fig:SI-nfNL}
 \end{minipage} 
\end{figure}

\begin{figure}[htb]
 \begin{center}
  \begin{center}
  \includegraphics[width=.48\linewidth]{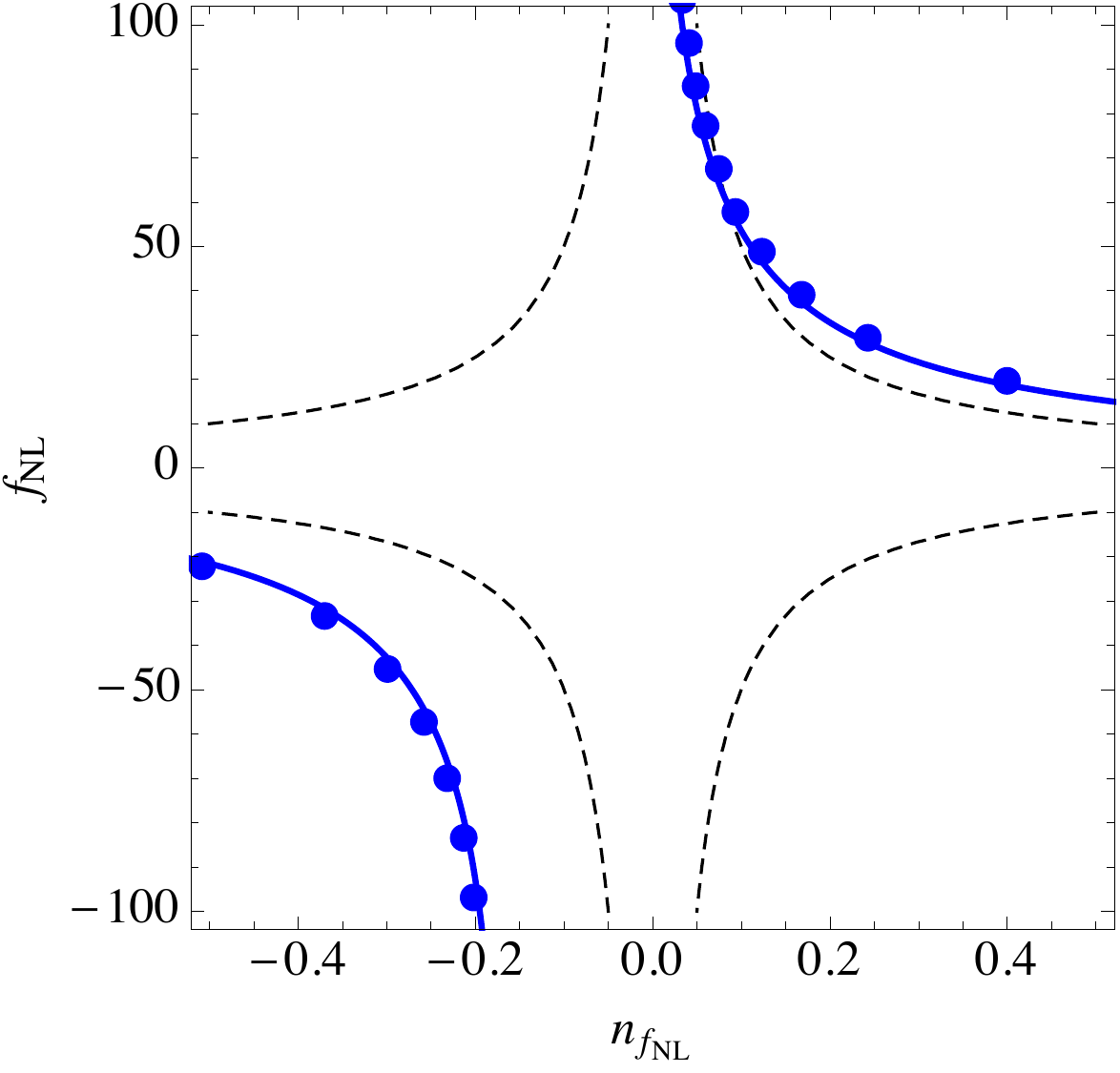}
  \end{center}
  \caption{Plot in the $n_{f_{\mathrm{NL}}}$ - $f_{\mathrm{NL}}$
    plane. The region $ 0.45 \lesssim \sigma_* / f \lesssim 0.65 $ is
    shown in the displayed area.  The expected observational
    sensitivity of Planck is also shown (black dashed).}
  \label{fig:SI-fNL_nfNL}
\end{center}
\end{figure}

The rolling of the curvaton due to the steep potential makes
$\sigma_{\mathrm{osc}}$ substantially smaller than $\sigma_*$, and as
we will soon see, this leads to a running non-Gaussianity.  Let us now
show the density perturbations generated by a self-interacting
curvaton with the parameter set $m=8$, $\Lambda = 2.5 \times
10^{12}$~GeV, $f = 4.3 \times 10^{13}$~GeV, and $\Gamma_\sigma =
10^{-15}$~GeV (i.e. the energy density of the universe at the curvaton
decay is $\rho_{\mathrm{dec}}^{1/4} \approx 65 $~GeV), along with the
inflationary parameters $H_{\mathrm{inf}} = 10^{12}$ GeV, $\Gamma_\phi
= 3.8 \times 10^3$ GeV (i.e. the energy density at reheating (=
inflaton decay) is $\rho_{\mathrm{reh}}^{1/4} \approx 1.3 \times
10^{11}$ GeV), and $\mathcal{N}_* = 50$ (see also
Footnote~\ref{foot9}).  This set of parameters gives $\Lambda^4 /
H_{\mathrm{inf}}^2 f^2 \approx 0.02$ (which determines the magnitude
of $\tilde{n}_s -1$), and realizes the COBE (WMAP) normalization value
as well as $\hat{r} \sim 10^{-2}$ at around $\sigma_* / f \approx
0.6$, where $n_{f_{\mathrm{NL}}}$ blows up.  The curvaton starts its
oscillation before reheating, thus $c = 9/2$.  The value of
$\sigma_{\mathrm{osc}}$ in terms of $\sigma_*$ for this parameter set
is illustrated by the blue solid line in
Figure~\ref{fig:SI-sigma_osc}.  Along with the analytic calculations,
we also carried out numerical computations for the density
perturbations in a similar fashion as we have done for NG curvatons in
Section~\ref{sec:NG}. The results are shown as functions of $\sigma_*
/ f$ in Figures~\ref{fig:SI-ns} - \ref{fig:SI-fNL_nfNL}, where the
blue solid lines denote the analytic calculations and the blue dots
the numerically computed results. Upon the numerical computations we
have fixed the inflaton energy density to a constant during inflation,
thus plotted $\tilde{n}_s$ and $\tilde{\alpha}$, and the numerically
computed $n_s$ and $\alpha$ in the same figures. In the presence of
nonzero $\dot{H}_*$ or $\ddot{H}_*$, the actual spectral index and its
running obtain offsets as shown in (\ref{eq:ns}) and
(\ref{runningalpha}). For example, a $\phi^6$ type chaotic inflation
with $\mathcal{N}_* = 50$ realizes $n_s = \tilde{n}_s -
0.06$\footnote{
  Although the $\phi^6$ chaotic inflation model is excluded due to too
  large tensor-to-scalar ratio and red-tilted spectral index when the
  inflaton is totally responsible for the primordial density
  perturbation, in the curvaton framework, chaotic inflation models
  with high order polynomials are still viable.
}, and thus red-tilts the resulting perturbation spectrum at
$\sigma_* /  f \approx 0.6 $ 
where $n_{f_{\mathrm{NL}}}$ blows up.

We also note that for the above parameter set, the curvaton's classical
rolling in a Hubble time is larger than its quantum
fluctuations during inflation, i.e. $V' / 3 H_{\mathrm{inf}}^2 >
H_{\mathrm{inf}} / 2 \pi$, at around $\sigma_* / f \approx 0.6$ and
larger.  Although this is not necessarily the case for smaller
$\sigma_*$ (for $\sigma_* / f \lesssim 0.54$ the quantum fluctuations
become dominant over the classical rolling by the end of
inflation), we have plotted in the figures down to small $\sigma_*$ 
regions supposing that the curvaton dynamics during inflation can be
treated as classical, in order to see the full (classical) $\sigma_*$-dependence
of the density perturbations\footnote{
  We also remark that for the example parameters we have adopted, the
  classical rolling is not so large compared to the quantum
  fluctuations at around $\sigma_* / f\approx 0.6$, hence random
  effects from quantum fluctuations may also need to be taken into
  account for a more rigorous treatment.
}.

The $\sigma_* \ll f$ region is well-approximated by the familiar
quadratic curvaton, but as one goes towards larger~$\sigma_*$, the
system starts to behave quite differently.  The power spectrum of the
linear order perturbations~(\ref{eq-ps}) in Figure~\ref{fig:SI-P}
increases in the small field regime with~$\sigma_*$ due to the
increase of~$\hat{r}$. However for larger field values, the power
spectrum starts to decrease as a function of~$\sigma_*$.  This is
attributed to the behavior of $\sigma_{\mathrm{osc}}$ shown in
Figure~\ref{fig:SI-sigma_osc}:
When $\sigma_* / f$ approaches unity,
$\sigma_{\mathrm{osc}}$ ceases to grow as rapidly as~$\sigma_*$, hence
the terms in the expression~(\ref{NfuncX}) are insensitive to the
value of~$\sigma_*$, except through the explicit dependence $\partial
\mathcal{N} / \partial \sigma_* \propto \partial \sigma_{\mathrm{osc}}
/
\partial \sigma_* \propto 1 / V'(\sigma_*)$. As a consequence, the
power spectrum becomes a decreasing function of~$\sigma_*$ for large
field regime.  In other words, in the large~$\sigma_*$ regime the
steep potential forces the curvaton to roll down to take similar
values for~$\sigma_{\mathrm{osc}}$ almost independently of~$\sigma_*$,
therefore $\partial \mathcal{N}_* / \partial \sigma_*$ is suppressed.
The peak of the power spectrum is located in the intermediate region,
i.e. at $\sigma_* \approx 0.6 f$, and this roughly matches with the
asymptotic value of $\sigma_{\mathrm{osc}}$ that can be read off from
Figure~\ref{fig:SI-sigma_osc} (or \ref{fig:SI-sigmaosc_m}).  When
starting from large $\sigma_*$, the initial field fluctuations are
reduced as the curvaton rolls down to field values below the
asymptotic value of $\sigma_{\rm osc}$, and thus the linear
perturbations are suppressed.

The existence of the peak in the power spectrum indicates that the
non-linearity parameter $f_{\mathrm{NL}}$ crosses zero and its
running~$n_{f_{\mathrm{NL}}}$ blows up. This is understood also in
terms of $f_1$ and $f_2$ (\ref{f1andf2}), which are plotted in
addition to~$f_{\mathrm{NL}}$ in Figure~\ref{fig:SI-fNL}.  At
$\sigma_* \ll f$ the amplitudes of $f_1$ and $f_2$ are comparable and
their sum results in a positive $f_{\mathrm{NL}}$ as is the case for
quadratic curvatons (cf.~Footnote~\ref{foot:foot4}). However as one
takes larger $\sigma_*$, $f_2$ which is a function of
$\sigma_{\mathrm{osc}}$ and $\hat{r}$ becomes insensitive
to~$\sigma_*$, while $f_1$ rapidly decreases on the negative side
through its explicit dependence on $\sigma_*$, resulting in a negative
$f_{\mathrm{NL}}$.  One clearly sees that the condition~(\ref{cond2})
is satisfied at around $\sigma_* / f \approx 0.6$ where
$f_{\mathrm{NL}}$ crosses zero, being able to produce a
large~$n_{f_{\mathrm{NL}}}$ even under a suppressed~$\tilde{\alpha}$.
Furthermore, since $\hat{r} \ll 1$, the amplitude of $f_{\mathrm{NL}}$
itself is large (except for the very vicinity of its vanishing point)
while satisfying (\ref{cond2}), i.e.  $ 1 \ll |f_{\mathrm{NL}}| \ll
|f_1|$.  (This is in contrast to quadratic or NG curvatons, for which
$f_{\mathrm{NL}}$ vanishes at rather large~$\hat{r}$. This was why the
product $n_{f_{\mathrm{NL}}} f_{\mathrm{NL}}$ from NG curvatons was
suppressed even when $n_{f_{\mathrm{NL}}}$ blew up,
cf.~Figure~\ref{fig:NG04}.)  Therefore a large and strongly
scale-dependent $f_{\mathrm{NL}}$ can be produced, whose running is in
the detectable range by upcoming CMB observations. Contour lines in
the $n_{f_{\mathrm{NL}}}$ - $f_{\mathrm{NL}}$ plane are shown in
Figure~\ref{fig:SI-fNL_nfNL}, where the black dashed line denotes the
Planck detection limit $n_{f_{\mathrm{NL}}} f_{\mathrm{NL}} / 50 =
0.10 $, cf.~(\ref{obs}).  Compared to the Figures~\ref{fig:SI-ns} -
\ref{fig:SI-nfNL}, in Figure~\ref{fig:SI-fNL_nfNL} we have added more
numerically computed points, which are equally spaced in terms of
$\Delta \sigma_*$.

\vspace{\baselineskip}

Let us also comment on the validity of the analytic estimations.  In
most of the displayed region, the results from the analytic
calculations are in good agreement with those from the numerical
computations.  However, as one goes towards larger $\sigma_*$,
i.e. larger $\sigma_{\mathrm{osc}}$, a period of non-sinusoidal
oscillations along the $\sigma^m$ potential needs to be taken into
account.  Thus the deviations between the analytic and numerical
results which one can already see in, e.g. Figure~\ref{fig:SI-fNL},
becomes even larger for $\sigma_*$ beyond the region displayed in the
figures.  In Appendix~\ref{app:B} we extend the expressions in
Section~\ref{sec:GF} to incorporate a period of non-sinusoidal
oscillations, which becomes important for large
$\sigma_{\mathrm{osc}}$ values.  We also note that, 
with the choice of the parameter set here, the curvaton's
effective mass during inflation is comparable to $H_{\mathrm{inf}}$
for $\sigma_* \sim f$, which violates the slow-roll approximation and
sources additional errors to the analytic estimations.

For self-interacting curvatons with even larger
$\sigma_{\mathrm{osc}}$, we should remark that our approximations for
the analytic expressions can break down.  This is because, especially
for cases with large~$m$ and $\sigma_{\mathrm{osc}} /f$, the potential
curvature~$V''$ quickly decreases after oscillation starts (which we
have defined by when $|\dot{\sigma} / H \sigma| = 1$) as the curvaton
rolls down to smaller field values. This forces the curvaton to
recover (though not completely) the attractor dynamics~(\ref{CDapp})
for a short time after~$t_{\mathrm{osc}}$.  As a consequence, the
simple picture of the curvaton suddenly starting its oscillations at
$t = t_{\mathrm{osc}}$ is no longer a good approximation. Such
breakdown of the approximations happen especially for large $m$ or
$\sigma_* /f$, or small $\Lambda^4 / H_{\mathrm{inf}}^2 f^2$, which
realize large values for $\sigma_{\mathrm{osc}} $.  We also note that
large $m$ and $\sigma_*$ can lead to the breakdown of the attractor
solution~(\ref{CDapp}) well before the onset of the oscillations,
which can also spoil the analytic estimations.  Density perturbations
from a self-interacting curvaton with large $\sigma_*$ values have
been worked out numerically in~\cite{arXiv:1108.2708}.


\vspace{\baselineskip}

\begin{figure}[htb]
 \begin{minipage}{.48\linewidth}
 \begin{center}
  \includegraphics[width=\linewidth]{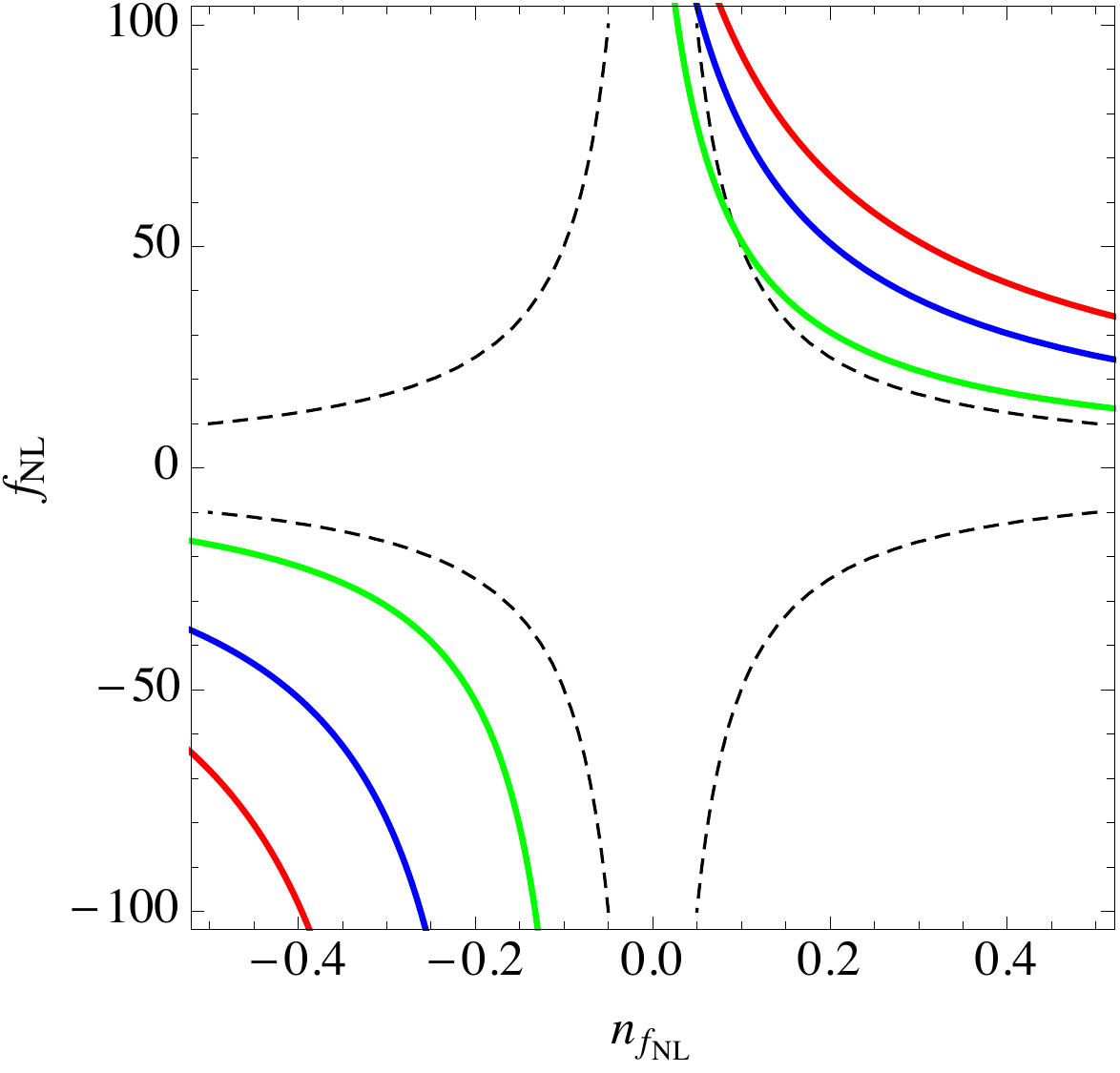}
  \end{center}
  \caption{Varying $\sigma_* / f$ for $\Lambda^4 /H_{\mathrm{inf}}^2
    f^2 = 0.03 $ (red), $0.02$ (blue), $0.01$ (green), with fixed $m =
    8$ and $\hat{r} = 10^{-2}$.  The expected observational
    sensitivity of Planck is also shown (black dashed).}
  \label{fig:var_mass}
 \end{minipage} 
 \begin{minipage}{0.01\linewidth} 
  \begin{center}
  \end{center}
 \end{minipage} 
 \begin{minipage}{.48\linewidth}
  \begin{center}
 \includegraphics[width=\linewidth]{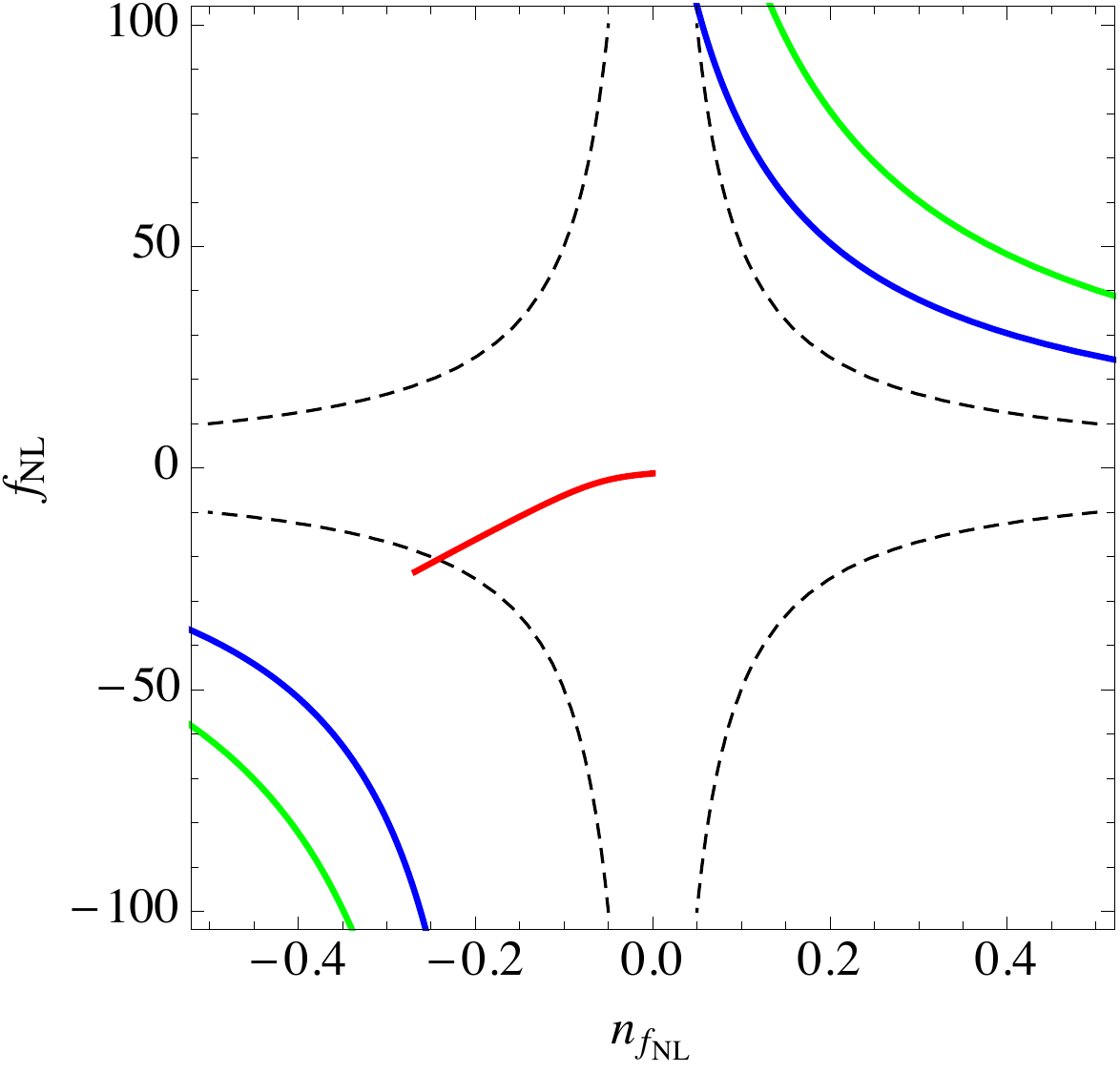}
  \end{center}
  \caption{Varying $\sigma_* / f$ for $\hat{r} = 10^{3}$ (red),
    $10^{-2}$ (blue), $10^{-2.2}$ (green), with fixed $m = 8$ and
    $\Lambda^4 /H_{\mathrm{inf}}^2 f^2 = 0.02$.}
  \vspace{10mm}
  \label{fig:var_r}
 \end{minipage} 
\end{figure}

In Figures~\ref{fig:var_mass} and \ref{fig:var_r} we further show
contours in the $n_{f_{\mathrm{NL}}}$ - $f_{\mathrm{NL}}$ planes for a
varying $\sigma_*$ under fixed values of $m$, $\Lambda^4 /
H_{\mathrm{inf}}^2 f^2$, and $\hat{r}$ (instead of fixing individual
parameters). We have taken $\mathcal{N}_* = 50$ and $t_{\mathrm{osc}}
< t_{\mathrm{reh}}$, i.e. $c = 9/2$, but we remark that the results do
not change significantly for $t_{\mathrm{osc}} > t_{\mathrm{reh}}$,
i.e. $c = 5$.  (It should also be noted that in order to compute
$f_{\mathrm{NL}}$ and $n_{f_{\mathrm{NL}}}$, we only need to specify
$m$, $\Lambda^4 / H_{\mathrm{inf}}^2 f^2 $, $\hat{r}$, $\sigma_* / f$,
$\mathcal{N}_*$, and $c$. Note also that in order to determine the
liner perturbation amplitude~$\mathcal{P}_\zeta$, we further need to
fix the ratio $\Lambda / f$.)  Here we are fixing~$\hat{r}$, which can
be considered as varying $\Gamma_\sigma$ along with $\sigma_* / f$,
hence the resulting $f_{\mathrm{NL}}$ and $n_{f_{\mathrm{NL}}}$ can
behave somewhat differently compared to Figures~\ref{fig:SI-fNL} -
\ref{fig:SI-fNL_nfNL}, especially in the region $|\sigma_* | \ll f$.

Steeper potentials, i.e., larger $\Lambda^4 / H_{\mathrm{inf}}^2 f^2$
(or larger $m$), realize flatter functions of $\sigma_{\mathrm{osc}}$
in terms of $\sigma_*$, which tend to produce $f_{\mathrm{NL}}$ with
stronger scale-dependence as shown in Figure~\ref{fig:var_mass}.
Moreover, smaller~$\hat{r}$ basically leads to larger
$|f_{\mathrm{NL}}|$, pushing $n_{f_{\mathrm{NL}}}$ towards the
detectable regions as one sees in Figure~\ref{fig:var_r}.  A somewhat
different behavior is seen for the $\hat{r} \gg 1$ case shown as the
red contour line in Figure~\ref{fig:var_r}, where $f_{\mathrm{NL}}$ is
always negative and the $\sigma_* \to 0 $ limit corresponds to the
endpoint of the line at $n_{f_{\mathrm{NL}}} = 0$.  This is understood
as the $\hat{r} \gg 1 $ behavior of quadratic curvatons,
cf.~(\ref{eq2.10}).  Moreover, we have plotted the red line up to
$\sigma_* / f \approx 0.9$, beyond which $\tilde{n}_s $ becomes larger
than $1.5$ and the curvaton no longer slow-rolls during
inflation. This corresponds to the other endpoint of the red line,
where rather large $|n_{f_{\mathrm{NL}}}|$ (together with a largely
negative~$f_{\mathrm{NL}}$) is realized due the large $\tilde{n}_s -1
$ and $|\tilde{\alpha}|$, coming close to satisfy the condition
(\ref{cond1}).

\vspace{1\baselineskip}

In summary, the steep potential forces the self-interacting curvaton
to roll down to small field values by the onset of the curvaton
oscillation. This greatly diminishes the initial differences in the
curvaton field values during inflation $\sigma_*$, thus suppresses the
resulting linear order density perturbations.  When fixing all the
parameters of the system except for $\sigma_*$, then a maximally large
linear perturbation amplitude is obtained from $\sigma_*$ that is
close to the asymptotic value of $\sigma_{\mathrm{osc}}$, and around
this $\sigma_*$ value is where a strongly scale-dependent
$f_{\mathrm{NL}}$ is produced.  We expect such behavior of a
self-interacting curvaton to be a rather generic feature of curvatons
whose potentials are approximated by quadratic around their minimum,
but steepens more rapidly than a quadratic away from the minimum.

\clearpage

\section{Mixed Curvaton and Inflaton}
\label{sec:mixed}

In the preceding two sections, we have assumed that fluctuations from
the inflaton are negligible and those from the curvaton are only
responsible for cosmic density fluctuations. But now in this section,
we further consider the case where the inflaton~$\phi$ also
contributes to the density perturbations as well as those from the
curvaton, which is called ``mixed curvaton and inflaton model'' in the
literature
\cite{Langlois:2004nn,Moroi:2005kz,Moroi:2005np,Ichikawa:2008iq,Suyama:2010uj}.

\vspace{\baselineskip}

We start with giving the expression of the curvature perturbations in
terms of the $\delta\mathcal{N}$-formalism, assuming that the field
fluctuations of the inflaton~$\delta\phi$ and the curvaton~$\delta
\sigma$ obey Gaussian distributions (to be more precise, in the sense
discussed below (\ref{fNLdf})),
with $\mathcal{P}_{\delta \phi} (k) = \mathcal{P}_{\delta \sigma}
(k) = (H|_{k = aH} /2 \pi)^2$ when $k$ exits the horizon.
We further suppose that there are no direct couplings between the
curvaton and inflation, thus no
correlations between their fluctuations $\delta \phi$ and $\delta
\sigma$.  Then, considering a homogeneous and isotropic universe
background, the power spectrum $\mathcal{P}_{\zeta}$ and the
non-linearity parameter~$f_{\mathrm{NL}}$ at the CMB scale in this
model can be written as~\cite{Ichikawa:2008iq}
\begin{equation}
 \mathcal{P_{\zeta}} = 
\left( \mathcal{N}_\phi^2 + \mathcal{N}_\sigma^2 \right)
 \left(\frac{H_*}{2 \pi}\right)^2,
\end{equation}
\begin{equation}
 f_{\mathrm{NL}}  = \frac{5}{6}
\frac{
\mathcal{N}_\phi^2 \mathcal{N}_{\phi \phi}  +
\mathcal{N}_\sigma^2 \mathcal{N}_{\sigma \sigma} +
2 \mathcal{N}_\sigma \mathcal{N}_\phi \mathcal{N}_{\phi \sigma} 
}{\left(\mathcal{N}_\phi^2 + \mathcal{N}_\sigma^2 \right)^2},
\end{equation}
at the leading order in terms of the field fluctuations.  Here the
subscripts $\phi$ and $\sigma$ denote partial derivatives with respect
to the fields, i.e. $\mathcal{N}_\phi \equiv \partial \mathcal{N}
/ \partial \phi_* $, $\mathcal{N}_\sigma \equiv \partial \mathcal{N}
/ \partial \sigma_* $, etc.  

Given that the end of inflation is independent of $\phi_*$ (so that
quantities such as the energy density of the universe and the inflaton
field value at the end of inflation are independent of~$\phi_*$), one
can find that
\begin{equation}
 \mathcal{N}_{\phi} = -   \frac{H_*}{\dot{\phi}_*} ,
\label{leadingNphi}
\end{equation}
where we have assumed that $H $ and $ \dot{\phi}$ during inflation are
determined merely by~$\phi$ (e.g., as in slow-roll inflation).  Here
we should remark that there are actually further contributions to
$\mathcal{N}_\phi$ since fluctuations in the duration of inflation due
to $\delta \phi$ lead to curvaton field fluctuations in the
post-inflationary era, which can be seen as
\begin{equation}
 \frac{\partial \sigma_{\mathrm{osc}}}{\partial \phi_*} = 
\left(1 - X(\sigma_{\mathrm{osc}})\right)^{-1}
\frac{V'(\sigma_{\mathrm{osc}})}{3 H_* \dot{\phi}_*}.
\label{sigma_osc_phi}
\end{equation}
This can be derived in a similar fashion as (\ref{partialsigma_osc}),
considering a curvaton potential~$V(\sigma)$ that is a function only
of~$\sigma$. In this sense, the inflaton can further produce density
perturbations after inflation by fluctuating the curvaton.  Moreover,
we note that such effect sources~$\mathcal{N}_{\phi \sigma}$, which
otherwise vanishes since the inflaton and curvaton only affect the
e-folding numbers during and after inflation, respectively.  However,
in this section, we ignore effects due to~(\ref{sigma_osc_phi}) since
we would like to show that $f_{\mathrm{NL}}$ can become largely
scale-dependent simply by having multiple sources (even in the absence
of the cross term~$\mathcal{N}_{\phi \sigma}$), and also because their
effects are negligible for the case of a quadratic curvaton and a
slow-rolling inflaton which we will study later in this section\footnote{
  To be precise, for the mixed case of a quadratic curvaton (with
  mass~$m$) and a slow-rolling inflaton, effects from
  (\ref{sigma_osc_phi}) on the power spectrum~$\mathcal{P}_\zeta$ are
  suppressed by a factor~$m^2 / H_*^2$ compared to the
  contribution~(\ref{leadingNphi}). As for~$f_{\mathrm{NL}}$, unless
  $\hat{r}$ takes a specific value $\hat{r} \approx 1.85$ which
  vanishes $\mathcal{N}_{\sigma\sigma}/\mathcal{N}_\sigma^2$,
  contributions from (\ref{sigma_osc_phi}) on~$f_{\mathrm{NL}}$ are
  either sufficiently smaller than unity, or subleading compared to
  other terms.  }.

For convenience, we introduce a parameter which represents the
fraction of the curvaton contribution to the total power spectrum:
\begin{equation}
 q \equiv \frac{\mathcal{N}_{\sigma}^2}{\mathcal{N}_{\phi}^2 +
  \mathcal{N}_{\sigma}^2 }.
\end{equation}
This quantity satisfies $0 \leq q \leq 1$ by definition. Then using
this quantity, the power spectrum and the non-linearity parameter
$f_{\rm NL}$ can be rewritten as
\begin{equation}
 \mathcal{P}_{\zeta} = \frac{1}{q} \mathcal{N}_{\sigma}^2 
 \left( \frac{H_*}{2 \pi}\right)^2,
\end{equation}
\begin{equation}
 f_{\mathrm{NL}} = \frac{5}{6} \left[
(1-q)^2 \frac{\mathcal{N}_{\phi \phi}}{\mathcal{N}_{\phi}^2} + 
 q^2 \frac{\mathcal{N}_{\sigma \sigma}}{\mathcal{N}_{\sigma}^2} 
\right].
\end{equation}

\vspace{\baselineskip}

Now let us write down the expressions in terms of the effective
potentials.  We assume that the inflaton and the curvaton are
decoupled from each other so that the total potential can be divided
into the inflaton potential~$U (\phi)$ which drives inflation as well
as determines the inflaton dynamics, and the curvaton potential~$V
(\sigma)$ which governs the curvaton dynamics. Both potentials are
considered to have no explicit dependence on time.  Furthermore, for
the inflaton, we assume a slow-roll inflation, i.e.
\begin{equation}
 3 H \dot{\phi} = - U', \quad
 3 H^2 M_p^2 = U,
\end{equation}
where a prime on~$U$ denotes differentiation with respect to~$\phi$. 
These approximations give
\begin{equation}
 \mathcal{N}_\phi = \left. \frac{U}{M_p^2 U'} \right|_*, \quad
 \mathcal{N}_{\phi \phi} = \frac{1}{M_p^2}
 \left. \left( 1 - \frac{U U''}{U'^2} \right) \right|_*.
\end{equation}
Introducing the slow-roll parameters
\begin{equation}
 \epsilon \equiv \frac{M_p^2}{2}\left( \frac{U'}{U} \right)^2, \quad
 \eta \equiv M_p^2 \frac{U''}{U},
\end{equation}
one arrives at
\begin{equation}
 \frac{\mathcal{N}_{\phi \phi}}{\mathcal{N}_\phi^2} = 2 \epsilon_* - \eta_*.
\end{equation}

The curvaton is also assumed to slow-roll during inflation, i.e.  $ 3
H \dot{\sigma} = -V'$, where a prime here for $V$ denotes a derivative
in terms of~$\sigma$.  For the $\delta \mathcal{N}$ from the curvaton
we adopt the formulae used in Section~\ref{sec:GF}:
$\mathcal{N}_\sigma$ is shown in (\ref{NfuncX}), and
$\mathcal{N}_{\sigma \sigma}$ can be read off from (\ref{fNLfuncX}).

Then, at the leading order in terms of the slow-roll
parameters~$\epsilon_*$ and $\eta_*$, we obtain expressions for the
spectral index $n_s-1$, its running~$\alpha$, and the product
$n_{f_{\mathrm{NL}}} f_{\mathrm{NL}}$ as follows:
\begin{equation}
 n_s - 1 \simeq 
(1-q) (-6 \epsilon_* + 2 \eta_*) + q \left( 
 -2    \epsilon_* + 2
  \frac{V''(\sigma_*)}{ 3 H_*^2}\right) ,
\label{mixedns}
\end{equation}
\begin{equation}
\begin{split}
 \alpha \simeq  & (1-q) \left(-24 \epsilon_*^2 + 16 \epsilon_* \eta_* - 2
		      \xi_* \right) \\
  & + 4 q (1-q) \left( 2  \epsilon_* - \eta_* 
+ \frac{V''(\sigma_*)}{3H_*^2}
\right)^2 
 \\
  & + q \left(  - 8 \epsilon_*^2 + 4 \epsilon_* \eta_*  
+   4 \epsilon_* \frac{V''(\sigma_*)}{3 H_*^2}
-2 \frac{V'(\sigma_*) V'''(\sigma_*)}{9 H_*^4}  
\right),
\end{split}
\label{mixedalpha}
\end{equation}
\begin{equation}
\begin{split}
 n_{f_{\mathrm{NL}}} f_{\mathrm{NL}} \simeq
 \frac{5}{6} \biggl[
 & (1-q)^2 (8 \epsilon_*^2 - 6 \epsilon_* \eta_* + \xi_*) \\
  & + 4 q (1-q) \left( 2 \epsilon_* - \eta_* +
	      \frac{V''(\sigma_*)}{3 H_*^2} \right)
 \left(
 (1-q) (-2 \epsilon_* + \eta_*) + q \frac{\mathcal{N}_{\sigma
 \sigma}}{ \mathcal{N}_{\sigma}^2} 
\right) \\
 & + \frac{q^2}{\mathcal{N}_{\sigma}} \frac{V'''(\sigma_*)}{3 H_*^2} 
\biggr],
\label{mixednfNL}
\end{split}
\end{equation}
where we have further introduced another slow-roll parameter for the inflaton:
\begin{equation}
 \xi \equiv M_p^4 \frac{U' U'''}{U^2}.
\end{equation}

Let us stress that upon obtaining the above results, we have made use
of the slow-roll approximations for $\phi$ and $\sigma$, and further
used the analytic formulae for $\mathcal{N}_\sigma$ and
$\mathcal{N}_{\sigma \sigma}$ which also relies on some approximations
(including the sudden decay) as discussed in Section~\ref{sec:GF}.  It
should be noted that errors contained in the approximations can
accumulate, especially as one goes to higher order derivatives.

\begin{figure}[htb]
 \begin{minipage}{.48\linewidth}
 \begin{center}
  \includegraphics[width=\linewidth]{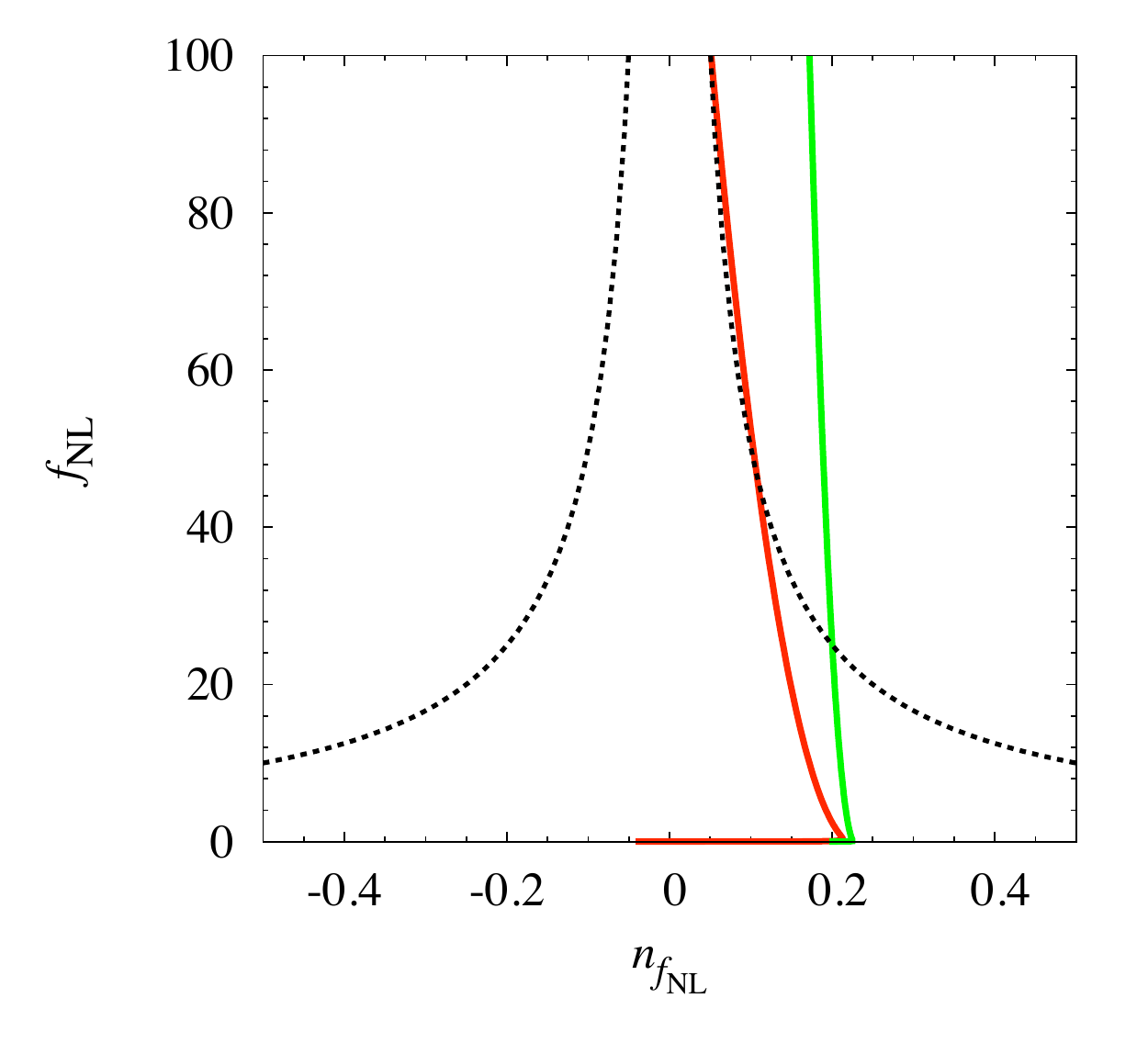}
  \end{center}
  \caption{Contours for $\hat{r}=10^{-2}$ (red line) and $10^{-3}$
    (green line) with $q$ being varied in the $n_{f_{\rm
        NL}}$--$f_{\rm NL}$ plane. Here we assume that
    $m^2/(3H_*^2)=0.05$.  The slow-roll parameters for the inflaton
    sector are taken as $\epsilon_\ast = 0.016, \eta_\ast = 0.025$,
    and $\xi_\ast = 0.0004$.  The expected observational sensitivity of
    Planck is also shown (black dashed).}
  \label{fig:mixed}
 \end{minipage} 
 \begin{minipage}{0.01\linewidth} 
  \begin{center}
  \end{center}
 \end{minipage} 
 \begin{minipage}{.48\linewidth}
  \begin{center}
 \includegraphics[width=\linewidth]{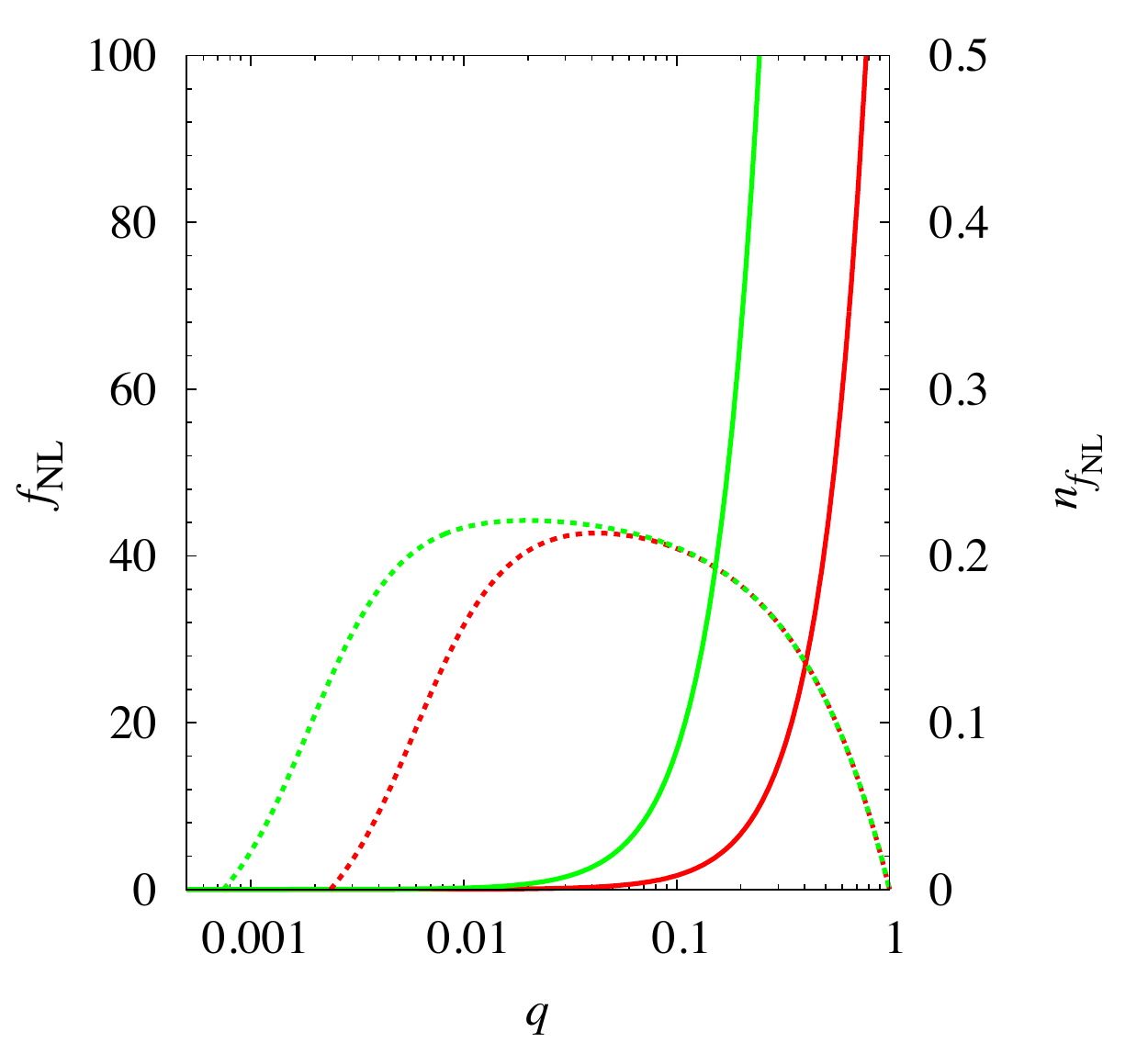}
  \end{center}
  \caption{The same cases as Figure~\ref{fig:mixed} are shown as a
    function of $q$.  $f_{\rm NL}$ is plotted for $\hat{r}=10^{-2}$
    (red) and $10^{-3}$ (green).  $n_{f_{\rm NL}}$ is for
    $\hat{r}=10^{-2}$ (red dashed) and $\hat{r}=10^{-3}$ (green
    dashed).  Notice that the value of $n_{f_{\rm NL}}$ becomes
    insensitive to $\hat{r}$ as $q$ approaches to unity.}
  \label{fig:mixed_q}
 \end{minipage} 
\end{figure}

\begin{figure}[htb]
 \begin{minipage}{.48\linewidth}
 \begin{center}
 \hspace{-2.5cm}
  \includegraphics[width=1.3\linewidth]{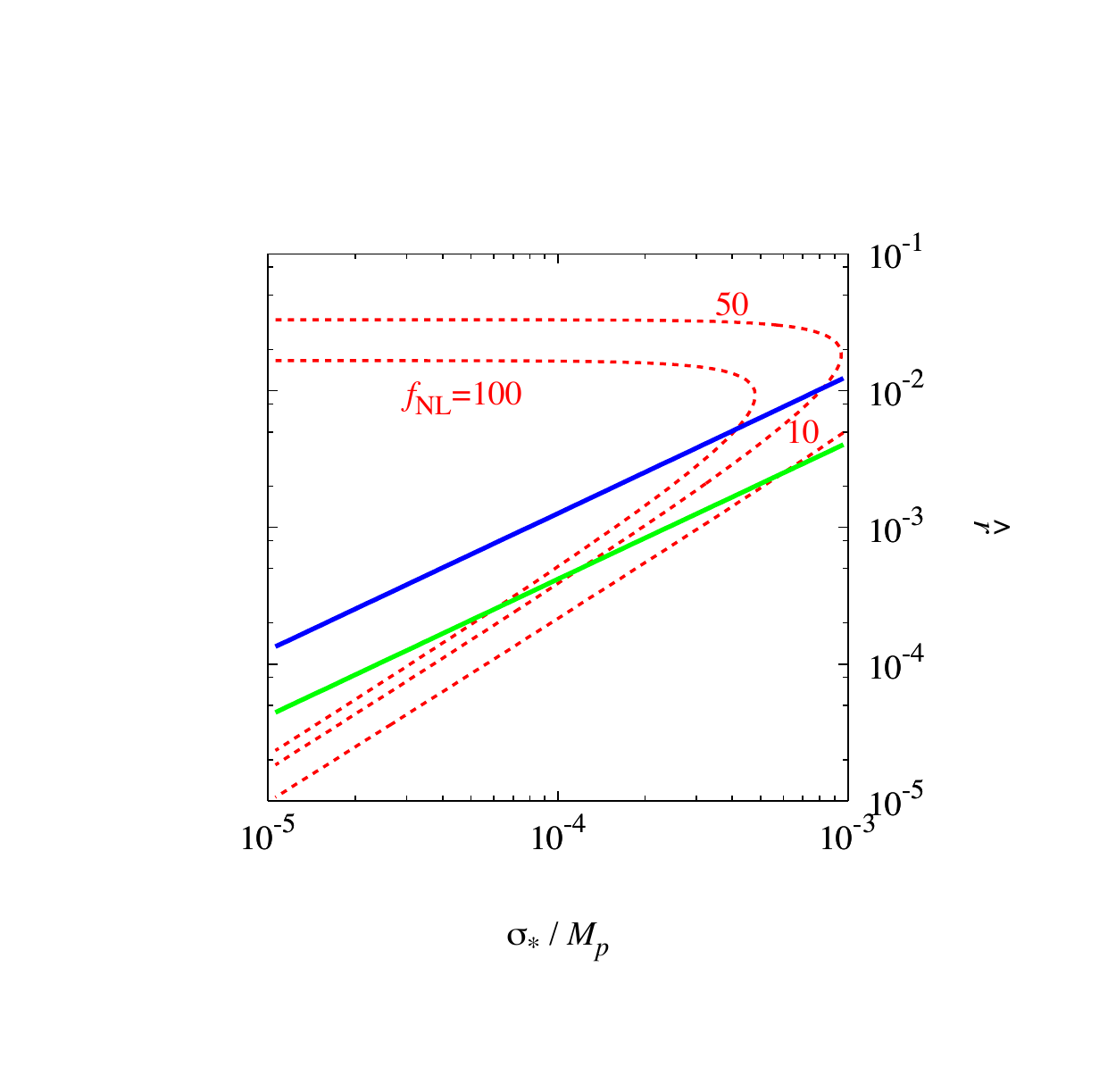}
  \end{center}
  \caption{Contours of $f_{\rm NL}$ and $n_{f_{\rm NL}}$ in
    $\sigma_\ast/M_p$--$\hat{r}$ plane.  Here we show the
    contours of $n_{f_{\rm NL}}=0.2$ (green) and $0.1$ (blue).  The
    values of $f_{\rm NL}$ are shown in the figure.  Other parameters
    are assumed as $m^2/(3H_*^2)=0.05$, $\epsilon_\ast = 0.016,
    \eta_\ast = 0.025$, and $\xi_\ast = 0.0004$.  }
  \label{fig:mixed_sig_r}
 \end{minipage} 
 \begin{minipage}{0.01\linewidth} 
  \begin{center}
  \end{center}
 \end{minipage} 
 \begin{minipage}{.48\linewidth}
  \begin{center}
 \includegraphics[width=1.3\linewidth]{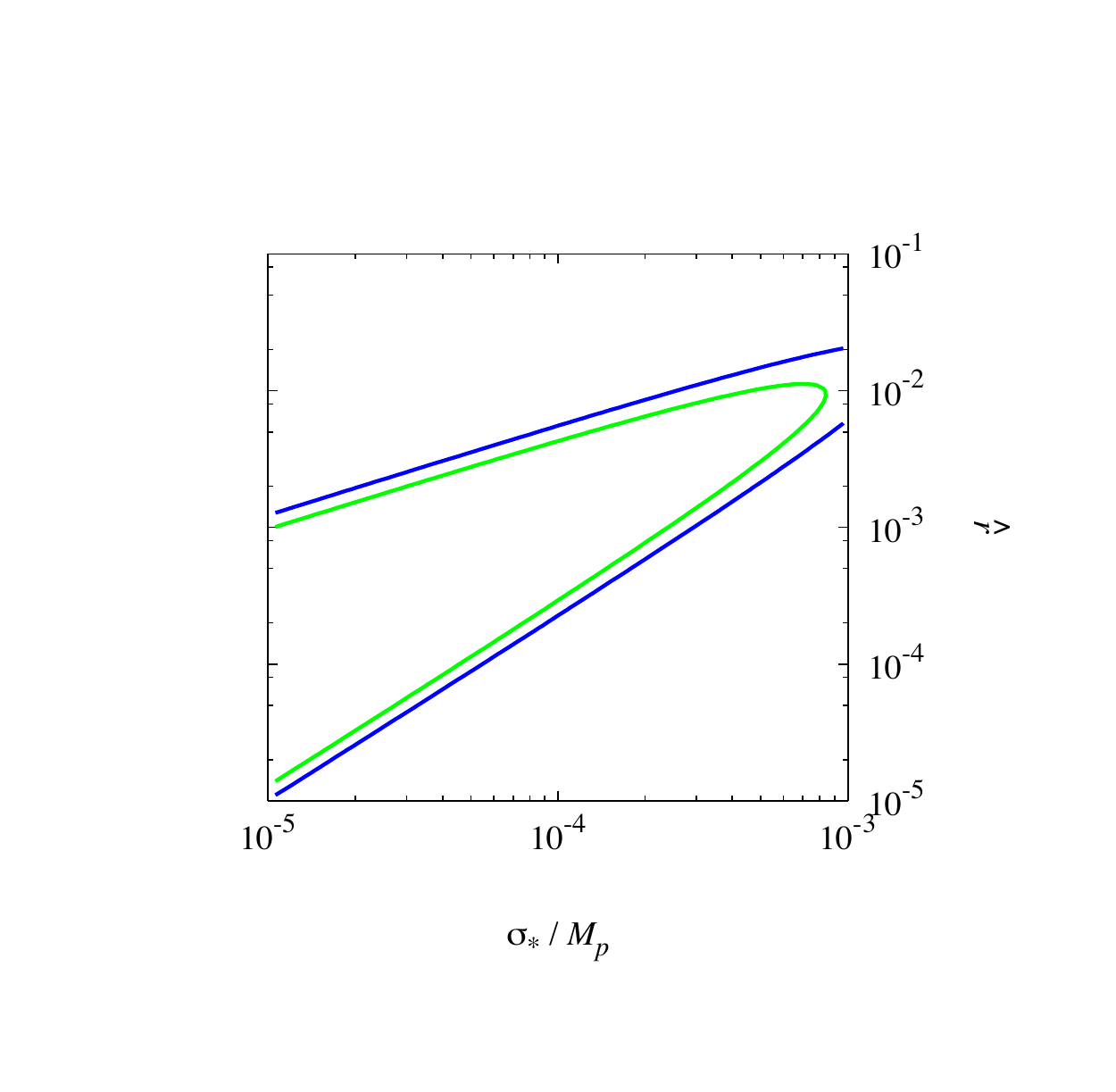}
  \end{center}
  \caption{Shown are contours of $n_{f_{\rm NL}} f_{\rm NL} =5$
    (green) and $2.5$ (blue), which correspond to the detection limit
    values for Planck and CMBPol, respectively.  Other parameters are
    taken as the same as in Figure~\ref{fig:mixed_sig_r}.  }
  \label{fig:mixed_sig_r_2}
 \end{minipage} 
\end{figure}

\vspace{\baselineskip}

The running of $f_{\mathrm{NL}}$ receives additional contributions in
the mixed case from the second line of (\ref{mixednfNL}), which is
absent for a pure curvaton or inflaton case.  In particular, large
$f_{\rm NL}$ and $n_{f_{\rm NL}}$ can be obtained when the curvaton
mainly contributes to the (running) non-Gaussianity, while the
inflaton mainly generates the liner order perturbations.  In fact, the
running $\alpha$ also has the contribution specific to the mixed model
such as the second line of \eqref{mixedalpha}. However, it is
suppressed as $(\mathcal{O}(\epsilon_\ast, \eta_\ast, \eta_\sigma))^2$
with $\eta_\sigma = V^{''}/(3H_\ast^2)$ corresponding to a slow-roll
parameter for $\sigma$, hence $\alpha$ is small in this model.

To see to what extent the running of non-Gaussianity can be large in
this kind of model, as an example, let us consider a quadratic
curvaton $V(\sigma) = \frac{1}{2} m^2 \sigma^2 $ for which
$n_{f_{\mathrm{NL}}}$ vanishes for the curvaton-only case, i.e. $q =
1$. (Note that for quadratic curvatons, $f_{\mathrm{NL}}$ and
$n_{f_{\mathrm{NL}}}$ are independent of $\sigma_*$ and
$\sigma_{\mathrm{osc}}$.) By doing this, we can see $n_{f_{\rm NL}}$
generated from the multi-source nature of this scenario.

Given that $\epsilon_*$, $|\eta_*|$, $|\xi_*|$, $m^2 / H_*^2 \ll 1$,
then one can check that $|f_{\mathrm{NL}}| \gg 1$ along with
$|n_{f_{\mathrm{NL}}} f_{\mathrm{NL}} | \gtrsim 1$ are realized if and
only if
\begin{equation}
 \hat{r} \lesssim q^2 (1-q) \left|
 2 \epsilon_* - \eta_* + \frac{m^2}{3 H_*^2} 
\right|
\end{equation}
is satisfied.
In such a case, $f_{\mathrm{NL}}$ and its running are approximated by
\begin{equation}
\label{eq:nfNL_2}
 f_{\mathrm{NL}} \simeq \frac{5 q^2}{3\hat{r}}, \qquad
 n_{f_{\mathrm{NL}}} \simeq 4 (1-q) \left(
 2 \epsilon_* - \eta_* + \frac{m^2}{3 H_*^2}
\right).
\end{equation}

In Figure~\ref{fig:mixed}, plots in the
$n_{f_{\mathrm{NL}}}$--$f_{\mathrm{NL}}$ plane are shown for cases
with $\hat{r} = 10^{-2}$ and $10^{-3}$ with $q$ being varied.  For the
curvaton mass, we assume $m^2 /(3H_\ast^2) = 0.05$. The slow-roll
parameters for the inflaton sector are taken as $\epsilon_\ast =
0.016, \eta_\ast = 0.025$, and $\xi_\ast = 0.0004$, which correspond to
the values for $\phi^4$ chaotic inflation model with $\mathcal{N}_\ast
= 60$.  Since non-Gaussianity mainly comes from the curvaton sector,
$f_{\rm NL}$ becomes larger as $q$ approaches unity as read off from
\eqref{eq:nfNL_2}. But on the other hand, $n_{f_{\rm NL}}$ becomes
small in this limit because the running of the non-Gaussianity arises
due to the multi-source nature of the model and the case with $q
\rightarrow 1$ corresponds to the single-source limit.

Regarding the scale dependence of $f_{\rm NL}$, $n_{f_{\rm NL}}$ takes
its maximum value at intermediate range of $q$ since one has to have
non-zero value of $q(1-q)$ to have sizable $n_{f_{\rm NL}}$ in this
kind of a mixed model.  It should be noted here that although the
value of $f_{\rm NL}$ becomes very small as $q$ goes to 0, however, if
we assume a very small value $\hat{r}$ by choosing some appropriate
parameters for the curvaton sector, the smallness of $q$ is somewhat
compensated and $f_{\rm NL}$ can be sizable.  Thus for some values of
$q$, both $f_{\rm NL}$ and $n_{f_{\rm NL}}$ can be large enough to be
detected by Planck satellite as seen from Figure~\ref{fig:mixed}.  To
see how $f_{\rm NL}$ and $n_{f_{\rm NL}}$ depend on $q$, we also plot
the values of these quantities as a function of $q$ in
Figure~\ref{fig:mixed_q}, from which we can clearly see how the size
of $q$ affects these quantities.

Since $q$ is determined once we fix $\epsilon_\ast, \sigma_\ast$ and
$\hat{r}$, we also show the prediction of the model in the
$\sigma_\ast$--$\hat{r}$ plane in Figure~\ref{fig:mixed_sig_r} where
contours of $f_{\rm NL}$ and $n_{f_{\rm NL}}$ are shown.  As already
mentioned in the previous section, the detectability of $n_{f_{\rm
    NL}}$ rather depends on the product of $n_{f_{\rm NL}} f_{\rm NL}
$, hence we further plot contours of this quantity in
Figure~\ref{fig:mixed_sig_r_2}.  In the figure, the contours
corresponding to the detection sensitivity limit for Planck and CMBPol
are depicted. From the figure, we can see that, in some parameter
region, $n_{f_{\rm NL}}$ can be detectable in near future
observations.

\section{Multi-curvaton}
\label{sec:multi-cur}

As a final example model related to the curvaton, we in this section
consider the multi-curvaton scenario where multiple curvaton fields
can be responsible for cosmic density perturbations.  Models of this
kind have been investigated in the literature
\cite{Choi:2007fya,Assadullahi:2007uw} and some explicit examples in
particle physics have also been discussed \cite{Lee:2011dy}.  In this
section, we again neglect the contribution from the inflaton
fluctuations to the total curvature perturbations.  Although there may
generally exist many curvaton fields, here we consider a two curvaton
case.

As in the previous sections, first we write down the expressions for
the spectral index $n_s$, the running $\alpha$, non-linearity
parameter $f_{\rm NL}$, and its running $n_{f_{\rm NL}}$.  In general,
the curvature perturbations in the model can be written as, up to the
second order of the field fluctuations,
\begin{equation}
\zeta 
= 
\mathcal{N}_a  \delta a_\ast + \mathcal{N}_b  \delta b_\ast
+ \frac12 \left( 
\mathcal{N}_{aa}  (\delta a_\ast)^2 
+
\mathcal{N}_{bb}  (\delta b_\ast)^2 
+
2\mathcal{N}_{ab}  \delta a_\ast \delta b_\ast
\right),
\end{equation}
where $\delta a_\ast$ and $\delta b_\ast $ are fluctuations of the
curvaton fields $``a"$ and $``b"$, respectively, and $\mathcal{N}_a
= \partial \mathcal{N} / \partial a_\ast$ and so on.  Throughout this
section, we denote $``a"$ and $``b"$ curvatons as the ones that decay
first and later, respectively.

The explicit forms of $\mathcal{N}_a$, $\mathcal{N}_b$, etc., can be
expressed in terms of the curvaton potentials by extending the
discussions in Section~\ref{sec:GF} to the multiple curvaton case.
However, since the explicit expressions are in general very
complicated and we are interested in the scale-dependence which comes
from the multi-source nature of this scenario, here we limit ourselves
to the case where the curvaton potentials are quadratic, i.e.,
\begin{equation}
 V(a, b) = \frac{1}{2} m_a^2 a^2 + \frac{1}{2} m_b^2 b^2.
\end{equation}
We do not specify which of the masses $m_a$ and $m_b$ is larger.
We further suppose that the curvatons start to oscillate after the
inflaton decay, and that the energy density of the curvatons are
negligibly tiny compared to the total energy of the universe until
both curvatons start their oscillations. Then, one can find that
$\mathcal{N}_a$, $\mathcal{N}_{aa}$, $\cdots$ are of the form
\begin{equation}
 \mathcal{N}_a \propto \frac{1}{a_\ast}, \quad \mathcal{N}_{aa} \propto
  \frac{1}{a_\ast^2 },
\quad \mathcal{N}_{ab} \propto \frac{1}{a_\ast b_\ast} ,
\label{NaNaaNab}
\end{equation}
(the same for replacing $a$ with $b$), where the constants of
proportionality are given as functions of the ratios between the
energy densities of $a$, $b$, and radiation upon decays of $a$ and $b$
(the full expressions of the coefficients up to the third order are
given in~\cite{Suyama:2010uj}, where the density perturbations are
computed in a different approach from Section~\ref{sec:GF}):
\begin{equation}
 \hat{r}_{a1} \equiv \left. \frac{\rho_a}{\rho_r} \right|_{t=t_{a \rm
  dec}},
\quad
\hat{r}_{b1} \equiv 
\left. \frac{\rho_b}{\rho_r} \right|_{t=t_{a \rm dec}},
\quad
\hat{r}_{b2} \equiv \left. \frac{\rho_b}{\rho_r} \right|_{t=t_{b \rm
dec}}.
\label{ra1rb1rb2}
\end{equation}
Here we have defined $\hat{r}_{a1}$ and $\hat{r}_{b2}$ similarly to
$\hat{r}$ defined in~\eqref{def_r}.  $\rho_a, \rho_b$ and $\rho_r$ are
energy densities of $a$ curvaton, $b$ curvaton and radiation,
respectively, while $t_{a \rm dec}$ and $t_{b \rm dec}$ are the times
at the decay of $a$ and $b$ curvatons. Notice that $\hat{r}_{a1}$ and
$\hat{r}_{b2}$ are defined at different times, thus we also put the
subscripts $``1"$ and $``2"$ in addition to $``a"$ and $``b"$ in the
definitions.  We also remark that the $\rho_r$ in the denominators of
$\hat{r}_{a1}$ and $\hat{r}_{b1}$ come from the inflaton decay, while
$\rho_r$ in $\hat{r}_{b2}$ also includes the decay products from the
$a$~curvaton.

Since the proportionality factors in (\ref{NaNaaNab}) are determined
by the three ratios (\ref{ra1rb1rb2}), the scale-dependence of the
density perturbations are sourced through $a_*$ and $b_*$.  Hence the
spectral index $n_s$ and its running $\alpha$ for the power spectrum
are (assuming Gaussian field fluctuations in the sense explained below
(\ref{fNLdf}), with $\mathcal{P}_{\delta
a}(k) = \mathcal{P}_{\delta b} (k) = (H|_{k = aH} / 2 \pi)^2$ when $k$
exits the horizon, and no
correlations between $\delta a$ and $\delta b$)
\begin{equation}
\label{eq:two_cur_ns}
n_s - 1 
\simeq
2 \frac{\dot{H}_*}{H_*^2} + 2 K_a \eta_a + 2 K_b \eta_b,
\end{equation}
\begin{equation}
\label{eq:two_cur_alpha}
\alpha
\simeq
2 \frac{\ddot{H}_*}{H_*^3} 
-4 \frac{\dot{H}_*^2}{H_*^4} 
- 4\frac{\dot{H}_*}{H_*^2} 
\left( K_a \eta_a  +  K_b \eta_b \right)
+ 4  K_a K_b \left(  \eta_a - \eta_b \right)^2 ,
\end{equation}
where   $\eta_a$ and $\eta_b$ are defined as
\begin{equation}
\eta_a \equiv \frac{ m_a^2}{ 3 H_\ast^2},
\qquad
\eta_b \equiv \frac{ m_b^2}{ 3 H_\ast^2}.
\end{equation}
We have also introduced the parameters $K_a$ and $K_b$ representing
the fractional contribution of the curvatons $a$ and $b$ to the total
power spectrum:
\begin{equation}
K_a \equiv  \frac{\mathcal{N}_a^2}{\mathcal{N}_a^2 + \mathcal{N}_b^2},
\qquad
K_b \equiv  \frac{\mathcal{N}_b^2}{\mathcal{N}_a^2 + \mathcal{N}_b^2},
\end{equation}
which clearly satisfies $K_a + K_b = 1$.

Regarding the non-Gaussianity, the non-linearity parameter~$f_{\rm
  NL}$ is generally given by
\begin{equation}
\label{eq:two_cur_fNL}
f_{\rm NL} = \frac56 \left( 
K_a^2 \frac{\mathcal{N}_{aa}}{\mathcal{N}_a^2} 
+ K_b^2 \frac{\mathcal{N}_{bb}}{\mathcal{N}_b^2} 
+ 2 K_a K_b \frac{\mathcal{N}_{ab}}{\mathcal{N}_a \mathcal{N}_b} 
\right),
\end{equation}
and its running $n_{f_{\rm NL}}$ can be computed as
\begin{eqnarray}
\label{eq:two_cur_nfNL}
n_{f_{\rm NL}} f_{\rm NL} 
\simeq
\frac{10}{3} K_a K_b 
\left(  \eta_a - \eta_b \right)
\left[
K_a \frac{\mathcal{N}_{aa}}{\mathcal{N}_a^2} -K_b \frac{\mathcal{N}_{bb}}{\mathcal{N}_b^2}
- \frac{\mathcal{N}_{ab}}{\mathcal{N}_a \mathcal{N}_b} (K_a - K_b) 
\right].
\end{eqnarray}

In the limit of either one of these curvatons being solely responsible
for density fluctuations (i.e., in the limit of $K_a (K_b) \rightarrow
1$ and $K_b (K_a) \rightarrow 0$), the above formulae reduce to a
single curvaton case.  What is specific to this model is the terms
with $K_a K_b$ which disappear in a single curvaton scenario. In
particular, when the masses of two curvatons are different (i.e.,
$\eta_a \ne \eta_b$), the running $n_{f_{\rm NL}}$ can be sizable,
which has been pointed out in \cite{arXiv:0911.2780,arXiv:1007.4277}.
In fact, $\alpha$ also receives the contribution which originates from
the mass difference of the curvatons. However, as seen from
\eqref{eq:two_cur_alpha}, such a term is suppressed with $(\eta_a -
\eta_b)^2$, thus $\alpha$ would be small in this model.

In fact, even if we assume a quadratic potential for the curvatons,
general expressions for the coefficients such as $\mathcal{N}_a,
\mathcal{N}_{aa}$ and so on are still very complicated.  Thus in the
following we discuss some limiting cases where both curvatons are {\it
  subdominant} and {\it dominant} at the time of their decays, paying
particular attention to the non-linearity parameter $f_{\rm NL}$ and
its running~$n_{f_{\rm NL}}$.

\subsection{Both curvaton subdominant at their decays}

First we consider the case where the energy densities of both
curvatons at their decays are subdominant, i.e.,
\begin{eqnarray}
\hat{r}_{a1}, \, 
\hat{r}_{b2} \ll 1.
\end{eqnarray}
Since we adopt a purely quadratic potential for the curvatons, we can
treat the curvatons as nonrelativistic fluids once they start to
oscillate.  By using the sudden decay approximation, one can show that
the explicit forms of (\ref{NaNaaNab}) in this case
become~\cite{Suyama:2010uj}\footnote{To be precise, here we are also
  assuming that the decays of the curvatons $a$ and $b$ are well
  separated along the expansion timeline.}
\begin{equation}
 N_a \simeq \frac{\hat{r}_{a1}}{2} \frac{1}{ a_*}, \quad
N_b \simeq \frac{\hat{r}_{b2}}{2 } \frac{1}{b_*}, \quad
N_{aa} \simeq \frac{\hat{r}_{a1}}{2} \frac{1}{ a_*^2}, \quad
N_{ab} \simeq -\frac{3 \hat{r}_{a1} \hat{r}_{b2}}{4} \frac{1}{ a_* b_*}, 
\quad 
N_{bb} \simeq \frac{\hat{r}_{b2}}{2 } \frac{1}{b_*^2},
\label{6.12}
\end{equation}
which give $n_s, \alpha, f_{\rm NL}$, and $n_{f_{\rm NL}}$ through the
formulae given in~\eqref{eq:two_cur_ns}--\eqref{eq:two_cur_nfNL}.  For
convenience, we also introduce a parameter representing the ratio of
the amplitude of the $a$ and $b$ curvatons:
\begin{equation}
\lambda \equiv  \frac{b_\ast}{a_\ast}. 
\end{equation}

\begin{figure}[htbp]
 \begin{center}
  \resizebox{180mm}{!}{
  \includegraphics{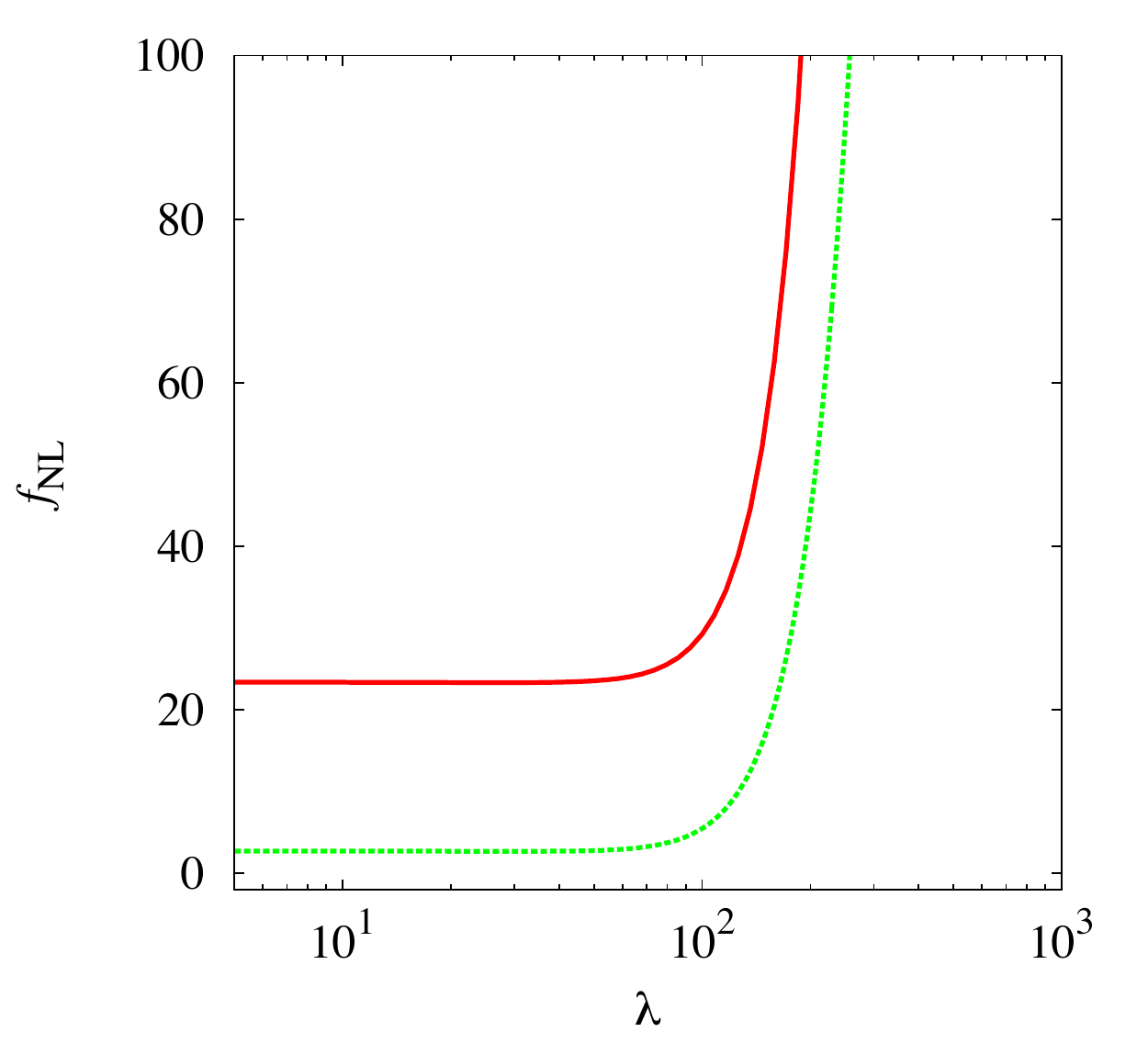}
   \includegraphics{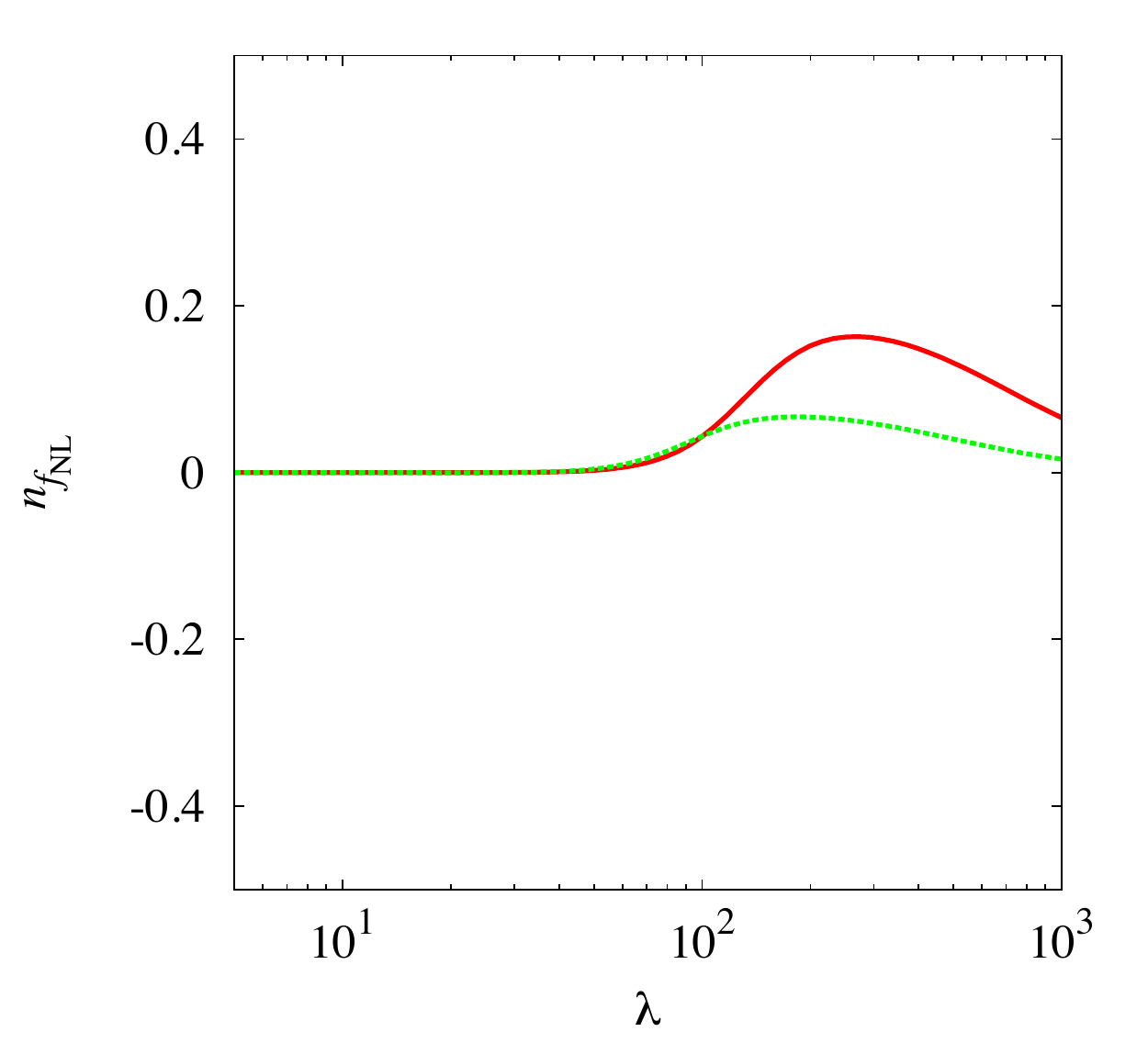}
   }
   \caption{Plots of $f_{\rm NL}$ (left panel) and $n_{f_{\rm NL}}$
     (right panel) as a function of $\lambda$ for the cases with both
     curvatons being {\it subdominant} at their decays.  Here we
       fix the parameters as $(\hat{r}_{a1},\hat{r}_{b1},
       \hat{r}_{b2}, \eta_a)=(10^{-4},10^{-5}, 0.07, 0.05)$ (red) and
       $(10^{-3},10^{-4}, 0.5, 0.02)$ (green), while $ \eta_b$ takes
       values much smaller than~$\eta_a$
       (cf. Footnote~\ref{foot18}).}
    \label{fig:subdom_two_lam}
  \end{center}
\end{figure}

In Figure~\ref{fig:subdom_two_lam}, we plot $f_{\rm NL}$ and
$n_{f_{\rm NL}}$ as functions of $\lambda$ for several parameter sets
of $(\hat{r}_{a1}$, $\hat{r}_{b1}$, $\hat{r}_{b2}$, $ \eta_a$,
$\eta_b)$\footnote{
  For completeness, we also give the value of $\hat{r}_{b1}$ used in
  plotting the figures since we use a full expression provided in
  \cite{Suyama:2010uj} for the 
  computations.  However, we remark that its actual
  value is not important in this limit as can be noticed from
  (\ref{6.12}) where an explicit dependence on $\hat{r}_{b1}$
  disappears.
\label{foot:r_b1}
}.  Actually, these parameters are not independent, as will be
  discussed around~(\ref{6.17}).  Here we take different values for
$\eta_a$ and $\eta_b$\footnote{
As can be seen from (\ref{6.17}), the parameters are not
independent and $\eta_{b}$ in the figure takes values much smaller
than $\eta_a$.  However, when $\eta_b \ll \eta_a$ is satisfied,
the explicit value of~$\eta_{b}$ is not important for obtaining
$f_{\rm NL}$ and $n_{f_{\rm NL}}$.
\label{foot18} 
}, which can give a sizable value for $n_{f_{\rm NL}}$ as noted above.
When fixing $\hat{r}_{a1}$ and $\hat{r}_{b2}$, the fraction
  $K_a$ increases together with $\lambda$.   From
\eqref{eq:two_cur_nfNL}, one can see that large values of $ n_{f_{\rm
    NL}} f_{\rm NL}$ would be obtained when $K_a \sim K_b$
(corresponding to $\lambda \sim \hat{r}_{b2}/\hat{r}_{a1} \sim
\mathcal{O}(10^2)$ for the parameter sets assumed in
Figure~\ref{fig:subdom_two_lam}), otherwise the factor $K_a K_b$
becomes very small.  Furthermore, in the limit of both curvatons {\it
  subdominant} we are considering here, one has
\begin{equation}
\frac{\mathcal{N}_{aa}}{ \mathcal{N}_a^2} \simeq \frac{2}{ \hat{r}_{a1}}, 
\qquad
\frac{\mathcal{N}_{bb} }{ \mathcal{N}_b^2} \simeq \frac{2}{ \hat{r}_{b2}},
\qquad
\frac{\mathcal{N}_{ab} }{ \mathcal{N}_a  \mathcal{N}_b} \simeq -3.
\end{equation}
Thus, when $\hat{r}_{a1} \ll \hat{r}_{b2} \ll 1$ and $K_a \sim K_b$
where both $f_{\rm NL}$ and $n_{f_{\rm NL}}$ can be large, the terms
originating solely from $a$ curvaton dominate in
\eqref{eq:two_cur_fNL} and \eqref{eq:two_cur_nfNL} (i.e., $K_a^2
(\mathcal{N}_{aa} / \mathcal{N}_a^2)$ and $K_a (\mathcal{N}_{aa} /
\mathcal{N}_a^2)$ in the expressions of $f_{\rm NL}$ and $n_{f_{\rm
    NL}}$, respectively).  In this case, we can make a rough estimate
of the values of $f_{\rm NL}$ and $n_{f_{\rm NL}}$ as
\begin{equation}
\label{eq:fNL_ra}
f_{\rm NL} \simeq \frac53 K_a^2 \frac{1}{ \hat{r}_{a1}},
\qquad
n_{f_{\rm NL}} \simeq 4 (1-K_a) (\eta_a - \eta_b).
\end{equation}
In fact, for the case with $(\hat{r}_{a1},\hat{r}_{b2}) \simeq
(10^{-4}, 10^{-2})$, $n_{f_{\rm NL}}$ takes its maximum value at
around $K_a \sim 0.1 ~(K_b \sim 0.9)$.  Since now we are assuming that
$\hat{r}_{a1} \ll 1$, large $f_{\rm NL}$ is possible even with a
relatively small value of $K_a$.  Furthermore, although $\eta_a$
  gives a positive contribution to $n_s$ (see \eqref{eq:two_cur_ns}),
  small $K_a$ together with $\eta_b \simeq 0$ imply that the spectral
  index can be red-tilted when assuming a large field inflation model with $
  \dot{H}_\ast / \dot{H}_\ast^2 \sim - \mathcal{O}(0.01)$.  Thus, in
  this model, we can have large $f_{\rm NL}$ and $n_{f_{\rm NL}}$ with
  red-tilted power spectrum for some parameter region.  

\begin{figure}[htbp]
 \begin{center}
  \includegraphics[width=.48\linewidth]{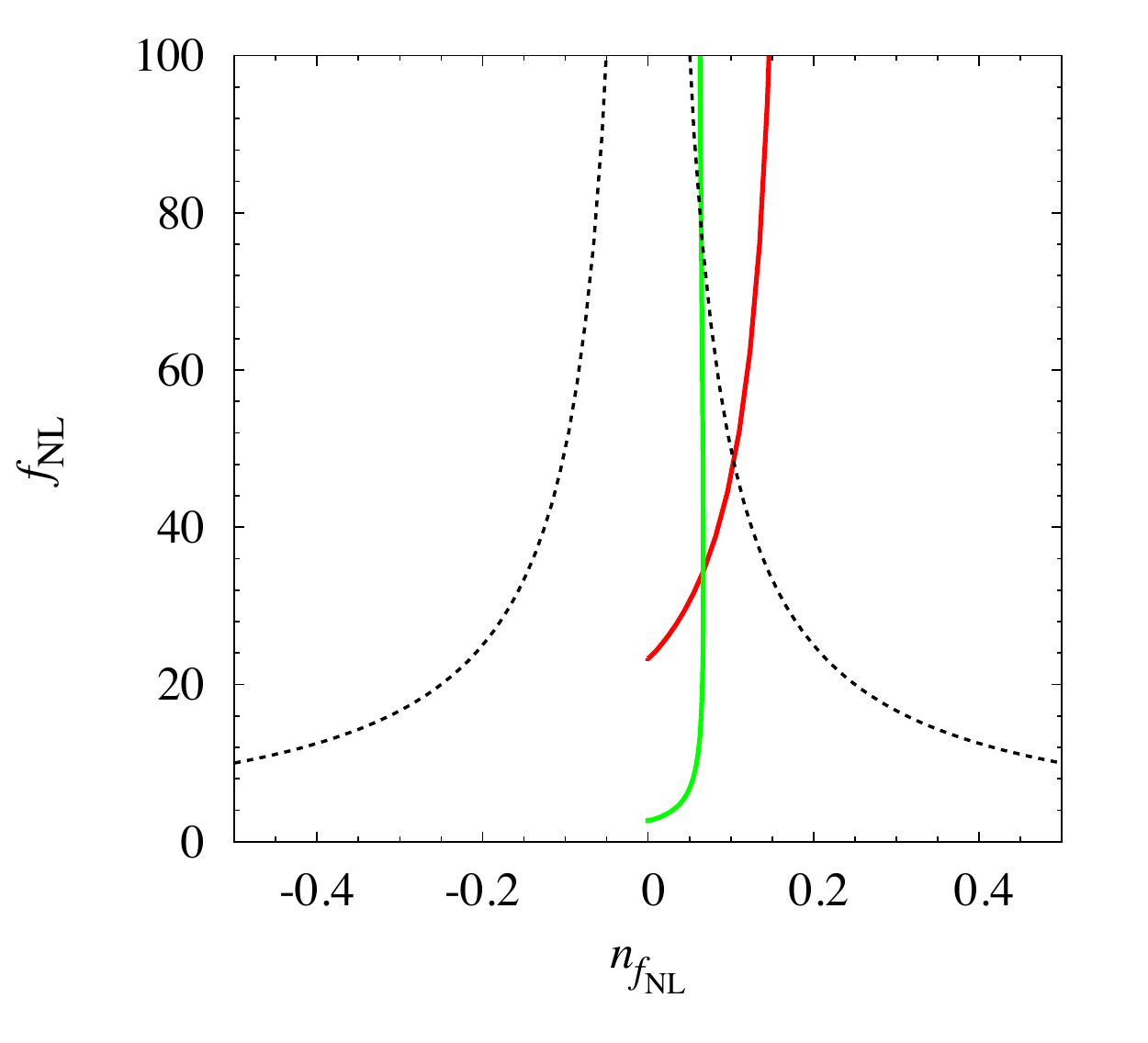}
  \caption{The same cases as Figure~\ref{fig:subdom_two_lam} are
    plotted in the $f_{\rm NL}$--$n_{f_{\rm NL}}$ plane.  The expected
    observational sensitivity of Planck is also shown (black
    dashed). }
  \label{fig:subdom_two_nfNL_fNL}
 \end{center}
\end{figure}

As emphasized several times in the previous sections, the
detectability of $n_{f_{\rm NL}}$ also depends on the size of $f_{\rm
  NL}$, thus we also show a plot in the $n_{f_{\rm NL}}$--$f_{\rm NL}$
plane in Figure~\ref{fig:subdom_two_nfNL_fNL}.  For reference, the
detection limit lines for Planck $|n_{f_{\rm NL}} f_{\rm NL} | = 5$
are also shown.  In the figure, the value of $\lambda$ is varied as in
Figure~\ref{fig:subdom_two_lam}.  In some region, the predictions of
the model for $f_{\rm NL}$ and $n_{f_{\rm NL}}$ are beyond the
detectable sensitivity line, thus if $n_{f_{\rm NL}}$ is detected in
future experiments, this model may also be a target of serious study.

\vspace{\baselineskip}

We should also remark that in the case discussed above, a detectable
level of $n_{f_{\rm NL}}$ can be obtained at the expense of
fine-tuning of the parameters.  Since here we are assuming that both
curvatons begin their oscillations during the radiation dominated
epoch after the inflaton reheating, the ratio of the energy densities
between $a$ and $b$ curvatons at the time of $a$ curvaton decay is
given by (independently of whether $m_a \lessgtr m_b$)
\begin{equation}
\label{6.17}
\frac{\hat{r}_{\rm a1}}{\hat{r}_{\rm b1}} = 
\left. \frac{\rho_a}{\rho_b} \right|_{t=t_{\rm adec}} 
\sim
\frac{1}{\lambda^2} \left(\frac{\eta_a}{\eta_b}  \right)^{1/4}, 
\end{equation}
where for simplicity we have taken $ b_{\rm osc} / a_{\rm osc} \sim
b_* / a_*$.  Therefore one sees that the parameter sets chosen in
Figure~\ref{fig:subdom_two_lam} suppose the mass~$m_b$ to be extremely
suppressed, much smaller than~$m_a$. See also discussions in
Footnote~\ref{foot18}.

\subsection{Both curvaton dominant at their decays}

Now we consider the case where both curvatons are dominant at the time
of their decays, i.e.,
\begin{equation}
 \hat{r}_{a1}, \, \hat{r}_{b2} \gg 1 , \qquad
\hat{r}_{a1} \gg \hat{r}_{b1}. 
\label{ra1rb2rb1}
\end{equation}
In this case, one obtains
\begin{equation}
 N_a \simeq \frac{8 }{9 \hat{r}_{b2}} \frac{1}{ a_*}, \quad
N_b \simeq \frac{2}{3 } \frac{1}{b_*}, \quad
N_{aa} \simeq \frac{40}{27 \hat{r}_{b2}} \frac{1}{ a_*^2}, \quad
N_{ab} \simeq -\frac{64}{27  \hat{r}_{b2}} \frac{1}{ a_* b_*}, 
\quad 
N_{bb} \simeq -\frac{2}{3 } \frac{1}{b_*^2}.
\label{eq:dom_Na}
\end{equation}
It should be noted here that $\zeta$ depends on $\hat{r}_{b2}$ but not
on $\hat{r}_{a1}$ nor $\hat{r}_{b1}$ at leading order as far as we
assume the condition~(\ref{ra1rb2rb1}).  Also one can easily see that
\begin{equation}
\frac{\mathcal{N}_{aa}}{\mathcal{N}_a^2} \simeq \frac{15 \hat{r}_{b2}}{
 8},  
\qquad
\frac{\mathcal{N}_{bb}}{\mathcal{N}_b^2} \simeq  -\frac{3}{2},
\qquad
\frac{\mathcal{N}_{ab}}{\mathcal{N}_a \mathcal{N}_b} \simeq -4. 
\end{equation}
Hence, when the assumption of \eqref{ra1rb2rb1} holds and $K_a \sim
K_b$ where $n_{f_{\rm NL}}$ can be sizable, $f_{\rm NL}$ and
$n_{f_{\rm NL}}$ are roughly estimated as
\begin{equation}
\label{6.20}
f_{\rm NL} \simeq \frac{25}{16}K_a^2 \hat{r}_{b2}, 
\qquad
n_{f_{\rm NL}} \simeq 4 (1-K_a) (\eta_a - \eta_b).
\end{equation}

\begin{figure}[htbp]
 \begin{center}
  \resizebox{180mm}{!}{
  \includegraphics{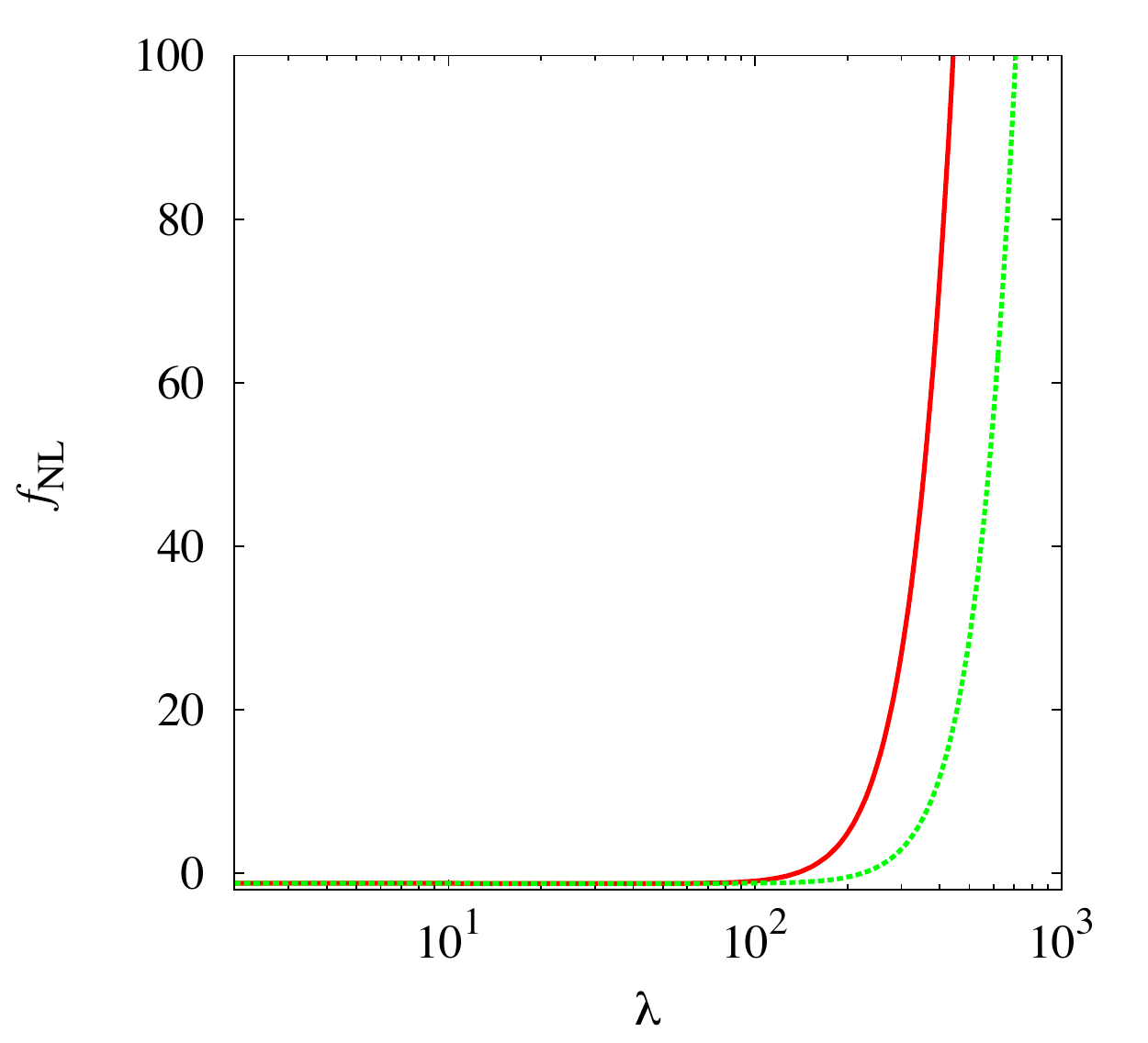}
   \includegraphics{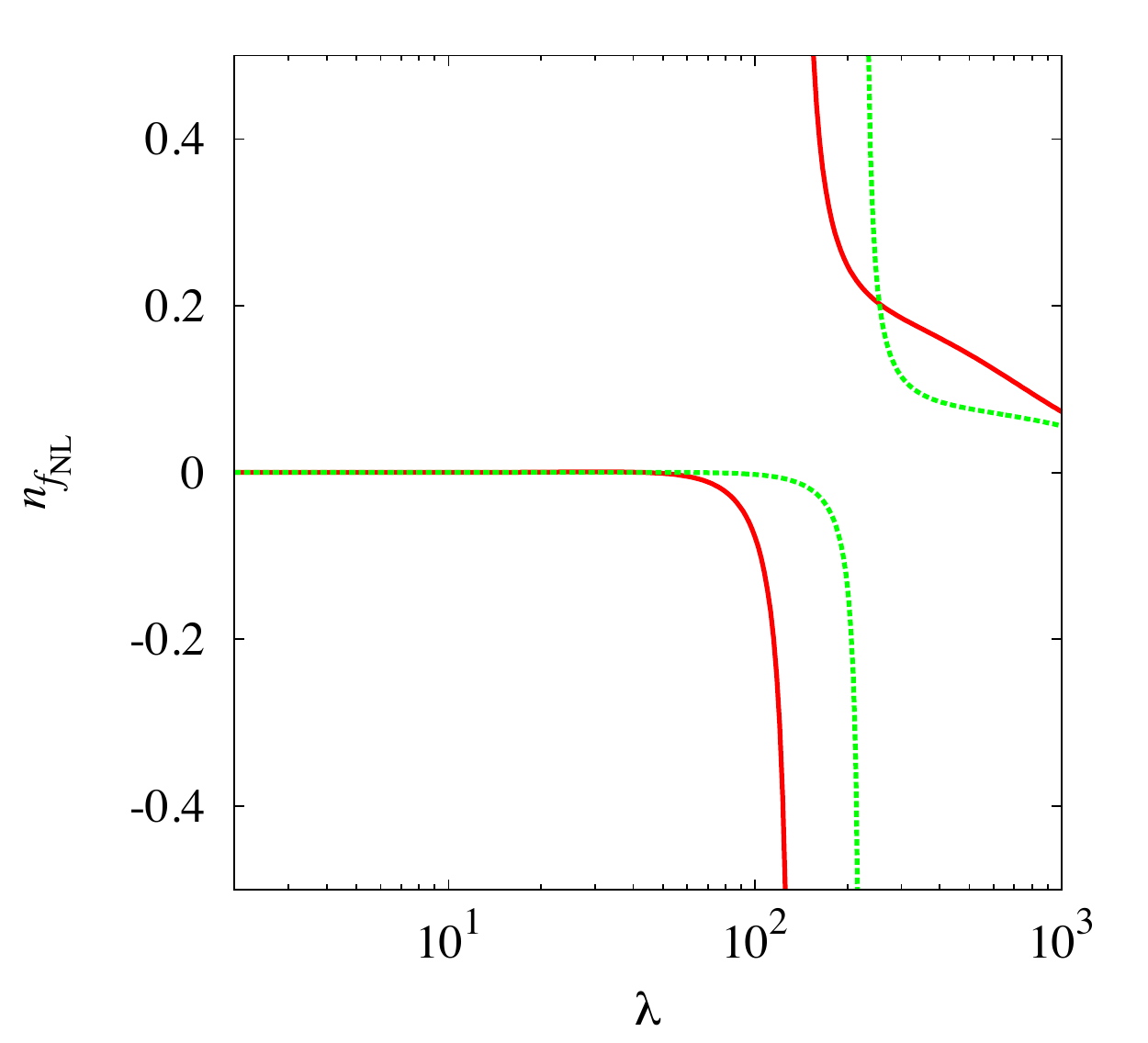}
   }
   \caption{Plots of $f_{\rm NL}$ (left panel) and $n_{f_{\rm NL}}$
     (right panel) as a function of $\lambda$ for the cases with both
     curvatons being {\it dominant} at their decays.  Here we fix
       the parameters as $(\hat{r}_{a1}, \hat{r}_{b1}, \hat{r}_{b2},
       \eta_a)=(10^3, 10^2, 10^{3}, 0.05)$ (red) and $(10^3, 10^2,
       2\times 10^{3}, 0.02)$ (green), while $ \eta_b$ takes values
       much smaller than~$\eta_a$ (cf. Footnote~\ref{foot18}).  }
  \label{fig:dom_two_lam}
  \end{center}
 \end{figure}

\begin{figure}[htbp]
 \begin{center}
  \includegraphics[width=.48\linewidth]{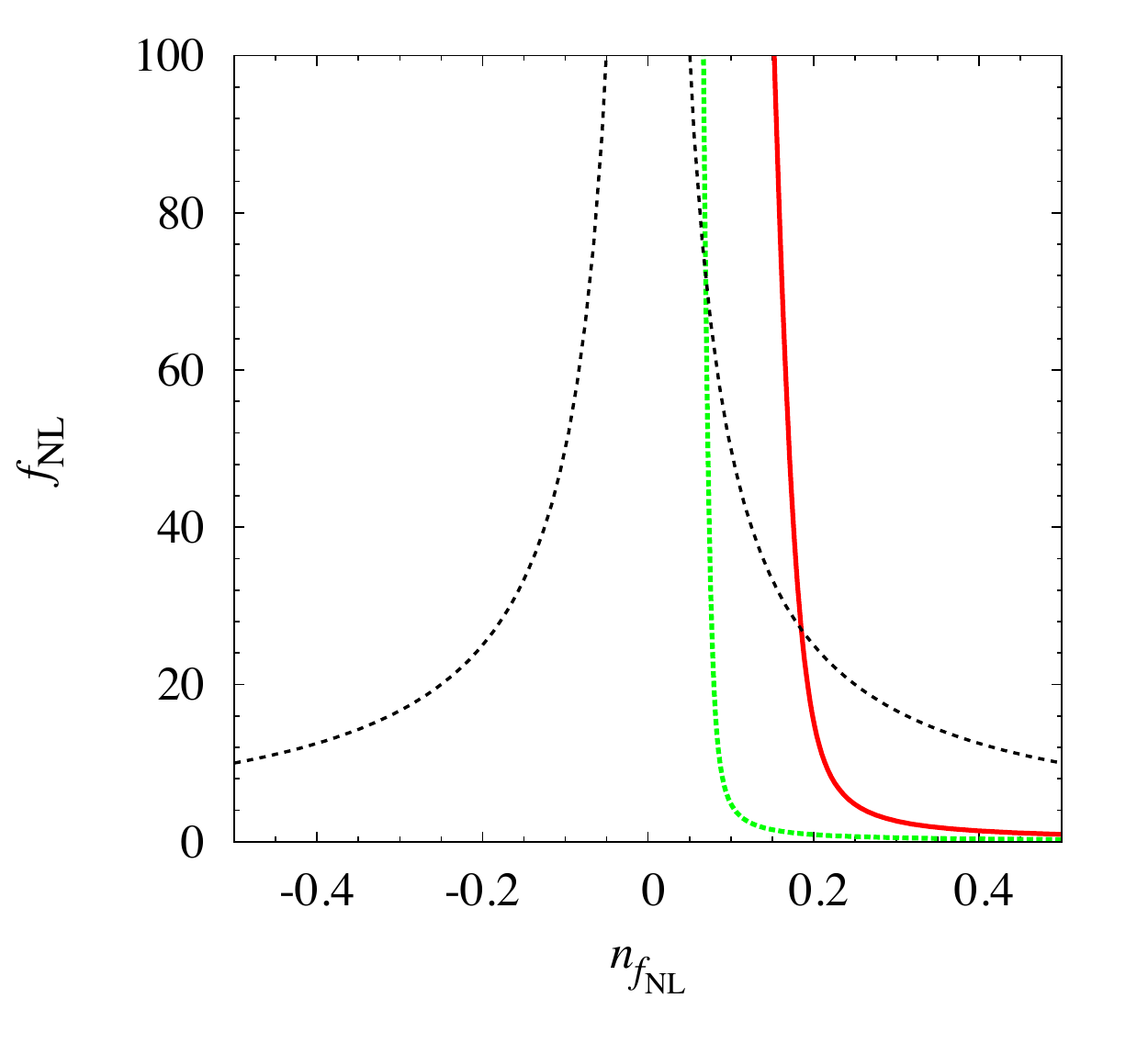}
  \caption{The same cases as Figure~\ref{fig:dom_two_lam} are plotted
    in the $f_{\rm NL}$--$n_{f_{\rm NL}}$ plane. The expected
    observational sensitivity of Planck is also shown (black dashed).}
   \label{fig:dom_two_nfNL_fNL}
   \end{center}
\end{figure}

In Figure~\ref{fig:dom_two_lam}, we plot $f_{\rm NL}$ and $n_{f_{\rm
    NL}}$ as functions of~$\lambda$ for several parameter sets of
$(\hat{r}_{a1}$, $\hat{r}_{b1}$, $\hat{r}_{b2}$, $\eta_a$, $\eta_b)$.
As mentioned above, as long as we assume (\ref{ra1rb2rb1}), the
expressions for $\mathcal{N}_a, \mathcal{N}_{aa}$ and so on are given
as ~\eqref{eq:dom_Na}. Thus the assumption on the values of
$\hat{r}_{a1} $ and $\hat{r}_{b1}$ have little effects on $f_{\rm NL}$
and $n_{f_{\rm NL}}$.  Under a fixed~$\hat{r}_{b2}$, the
  fraction~$K_a$ increases along with~$\lambda$.  It should be
noticed that $f_{\rm NL}$ becomes negative when $\lambda$ is small,
i.e. $K_a \ll 1$, due to the negative values of $\mathcal{N}_{bb} /
\mathcal{N}_b^2$.  However, as $\lambda$ becomes larger ($K_a$ is
larger), $f_{\rm NL}$ crosses zero and increases to large positive
values as seen from Figure~\ref{fig:dom_two_lam}.  Hence $n_{f_{\rm
    NL}}$ blows up at around the region where $f_{\rm NL} \sim 0$
similarly to the case of the self-interacting curvaton. 
Further increasing~$\lambda$, then $K_a$ becomes comparable to $K_b$
and (\ref{6.20}) is satisfied. For even larger~$\lambda$,
$n_{f_{\mathrm{NL}}}$ again becomes small.
To see the detectability of $n_{f_{\rm NL}}$ in this case, we again show
the plot in the $n_{f_{\rm NL}}$--$f_{\rm NL}$ plane in
Figure~\ref{fig:dom_two_nfNL_fNL}, from which we can clearly see that
$n_{f_{\rm NL}}$ can be detectable in some parameter region as in the
{\it subdominant} case discussed in the previous subsection.  We
remark that $n_s$ can also be red-tilted at around the region where
$n_{f_{\rm NL}}$ becomes large. In fact, with the parameter sets
assumed in
Figures~\ref{fig:dom_two_lam}~and~\ref{fig:dom_two_nfNL_fNL}, the
contribution to the power spectrum from the $a$ curvaton is somewhat
small as $K_a \sim \mathcal{O}(0.01)$ when $n_{f_{\rm NL}}$ blows
  up.  There the positive contribution of $K_a \eta_a$ in $n_s$
becomes small, in which the spectral index can be almost given by $n_s
- 1 \simeq 2 \dot{H}_\ast /H_\ast^2$ and be red-tilted for large field
inflation models.

We also note that (\ref{6.17}) requires the mass~$m_b$ to be highly
suppressed in the above case, as in the previous subsection.  Here,
especially since the $a$ curvaton is dominant at its decay, which
means that $\rho_a > \rho_b$, the constraint~(\ref{6.17}) indicates
$\sqrt{m_a/m_b} > \lambda^2$.  As seen from
Figure~\ref{fig:dom_two_lam}, the product of $f_{\rm NL} n_{f_{\rm
    NL}}$ can be large when $\lambda \sim \mathcal{O}(10^2)$, which
requires a large hierarchy between $m_a$ and $m_b$.

\section{Conclusions}
\label{sec:conc}

In this paper we have investigated the scale-dependence of linear and
second order density perturbations produced in the curvaton mechanism.
In particular, we focused on the runnings such as $\alpha$ and
$n_{f_{\rm NL}}$.  Since models of the curvaton have several variants,
we discussed the representative models: curvatons with non-quadratic
potentials, and multiple source cases.  For the later model, we
discussed multi-curvaton and mixed curvaton and inflaton ones.

Non-quadratic curvaton potentials give scale-dependence to the
non-Gaussianity, as well as to the spectral index of the linear
perturbation spectrum.  We have especially shown that the local-type
$f_{\mathrm{NL}}$ produced from curvatons can strongly depend on the
scale, even when the linear order perturbations are nearly
scale-invariant.  We analytically computed the running
$n_{f_{\mathrm{NL}}} \equiv d \ln |f_{\mathrm{NL}}| / d \ln k$ for
curvatons with general energy potentials in (\ref{nfNLfNL}), and
obtained conditions (\ref{cond1}) and (\ref{cond2}) under which
$f_{\mathrm{NL}}$ can be strongly scale-dependent.

The two conditions (\ref{cond1}) and (\ref{cond2}) can be satisfied
respectively by curvaton potentials that flatten and steepen compared
to a quadratic one.  As an example of the former case, we looked into
pseudo-Nambu-Goldstone curvatons with cosine-type potentials, for
which it was shown that $n_{f_{\mathrm{NL}}}$ is directly related to
the amplitude of $\tilde{\alpha}$ and thus strictly constrained by
current observational bounds on running spectral index, unless the
inflationary mechanism realized a largely time-varying~$H$ such that
cancelled out the contribution of $\tilde{\alpha}$ to the running
spectral index.  For the latter case, we studied self-interacting
curvatons in detail and saw that the steep potential wipes out the
initial field fluctuations~$\delta \sigma_*$, suppressing the resulting
density perturbations. This lead to the production of a strongly
scale-dependent $f_{\mathrm{NL}}$ even for a suppressed running
spectral index, when $\sigma_*$ was away from, but not too away from
the potential minimum. Moreover, the scale-dependence of
$f_{\mathrm{NL}}$ from self-interacting curvatons could be as large as
to be detected by upcoming CMB experiments.  

In Sections~\ref{sec:mixed} and \ref{sec:multi-cur}, we further
discussed the mixed curvaton and inflaton, and multi-curvaton models.
In these models, due to the multiple source nature, large values of
$n_{f_{\mathrm{NL}}}$ can also arise in a different way from the
conditions discussed for the curvaton only case.  We have made
quantitative discussion for the running and shown in what cases
$n_{f_{\rm NL}}$ and $f_{\rm NL}$ can be both large enough to be
detected in future cosmological observations.

The analytic methods used in this paper clarifies the intriguing
behaviors of the density perturbations produced from non-quadratic
and/or multiple curvatons, providing us with a systematic framework for
studying the curvaton scenario in general.  
While for pseudo-Nambu-Goldstone curvatons a running $f_{\mathrm{NL}}$
was accompanied by a large~$\alpha$ and thus was constrained,
the steep potential of self-interacting curvatons allowed strongly
scale-dependent $f_{\mathrm{NL}}$. 
We expect that in terms of the $f_{\mathrm{NL}}$
scale-dependence, these two examples respectively
give typical behaviors of curvatons whose potentials flatten/steepen
compared to a quadratic.
Multiple source scenarios could also produce large~$n_{f_{\mathrm{NL}}}$.
The systematic approach we presented will be helpful for probing the
physics of curvatons when combined with upcoming data.

\section*{Acknowledgements}

T.K. would like to thank Neal Dalal, Teruaki Suyama, Fuminobu
Takahashi, and Jun'ichi Yokoyama for helpful conversations.  The work
of T.T.  is supported in part by JSPS Grant-in-Aid for Scientific
Research No.~23740195 and also by Saga University Dean's Grant 2011
For Promising Young Researchers.

\appendix

\section{Density Perturbations from Curvatons with Non-Sinusoidal Oscillations}
\label{app:B}

The analytic expressions in Section~\ref{sec:GF} can be generalized to
include a period of non-sinusoidal oscillations, as was discussed in
Appendix~B of~\cite{Kawasaki:2011pd}. Here we suppose that the energy
density of the oscillating curvaton initially redshifts as
$\rho_\sigma \propto a^{-n}$ with a constant~$n$, then when its energy
approaches a certain value~$\rho_{\sigma \mathrm{sin}}$, the curvaton
suddenly switches to a sinusoidal oscillation and then redshifts as
$\rho_\sigma \propto a^{-3}$.  We take $\rho_{\sigma \mathrm{sin}}$ as
a constant (i.e. independent of~$\sigma_*$), considering for e.g.,
self-interacting curvatons whose oscillations are determined by the
curvaton field value.  The curvaton energy density is assumed to be
negligibly small until the inflaton decay or the onset of the curvaton
oscillation, whichever is later.

Then the linear perturbation amplitude becomes
\begin{equation}
 \mathcal{P}_\zeta = \left(\frac{\partial \mathcal{N}}{\partial \sigma_*} 
  \frac{H_*}{2 \pi} \right)^2,
\end{equation}
with 
\begin{equation}
 \frac{\partial \mathcal{N}}{\partial \sigma_*} = \frac{\hat{r}}{4+3 \hat{r}}
  \left(1 - X(\sigma_{\mathrm{osc}})\right)^{-1}
 \left\{\frac{3}{n}\frac{V'(\sigma_{\mathrm{osc}})}{V(\sigma_{\mathrm{osc}})} -
  \frac{3 X(\sigma_{\mathrm{osc}})}{\sigma_{\mathrm{osc}}}  \right\}
 \frac{V'(\sigma_{\mathrm{osc}})}{V'(\sigma_*)},
\end{equation}
where $X(\sigma_{\mathrm{osc}})$ is defined in (\ref{Xdef}).  The
spectral index $n_s$ and its running~$\alpha$ are the same as in
(\ref{eq:ns}) and (\ref{runningalpha}), respectively.  Furthermore,
the non-linearity parameter is
\begin{equation}
\begin{split}
 f_{\mathrm{NL}} 
 & =
  \frac{40 (1+\hat{r})}{3 \hat{r} (4+3 \hat{r})}  
 + \frac{5 (4+3 \hat{r})}{6 \hat{r}}
 \left\{\frac{3}{n}\frac{V'(\sigma_{\mathrm{osc}})}{V(\sigma_{\mathrm{osc}})}-
 \frac{3 X(\sigma_{\mathrm{osc}})}{\sigma_{\mathrm{osc}}}  \right\}^{-1}  
  \Biggl[(1-X(\sigma_{\mathrm{osc}}))^{-1} X'(\sigma_{\mathrm{osc}})    \\
 & 
  + \left\{\frac{3}{n}  \frac{V'(\sigma_{\mathrm{osc}})}{V(\sigma_{\mathrm{osc}})} -
 \frac{3X(\sigma_{\mathrm{osc}})}{\sigma_{\mathrm{osc}}}  \right\}^{-1} 
 \left\{\frac{3}{n}\frac{V''(\sigma_{\mathrm{osc}})}{V(\sigma_{\mathrm{osc}})} -
 \frac{3}{n}\frac{V'(\sigma_{\mathrm{osc}})^2}{V(\sigma_{\mathrm{osc}})^2} -
 \frac{ 3 X'(\sigma_{\mathrm{osc}})}{\sigma_{\mathrm{osc}}} + \frac{3
 X(\sigma_{\mathrm{osc}})}{\sigma_{\mathrm{osc}}^2} 
 \right\} \\
 & \qquad \qquad \qquad \qquad \qquad \qquad \qquad \qquad \qquad
  + \frac{V''(\sigma_{\mathrm{osc}})}{V'(\sigma_{\mathrm{osc}})} -
 (1-X(\sigma_{\mathrm{osc}}))
 \frac{V''(\sigma_*)}{V'(\sigma_{\mathrm{osc}})} 
\Biggr],
\end{split}
\end{equation}
and its running takes the form
\begin{equation}
 n_{f_\mathrm{NL}}  \simeq
 \frac{1}{f_{\mathrm{NL}}}
 \frac{5 (4+3\hat{r})}{18 \hat{r}} 
 \left\{\frac{3}{n} \frac{V'(\sigma_{\mathrm{osc}})}{V(\sigma_{\mathrm{osc}})}-
 \frac{3 X(\sigma_{\mathrm{osc}})}{\sigma_{\mathrm{osc}}}  \right\}^{-1}  
 \left( 1 - X(\sigma_{\mathrm{osc}})\right)
 \frac{V'(\sigma_*)}{ V'(\sigma_{\mathrm{osc}})}
 \frac{V'''(\sigma_*) }{ H_*^2 }.
\end{equation}
Of course, the above equations reproduce (\ref{NfuncX}),
(\ref{fNLfuncX}), and (\ref{nfNLfNL}) for purely sinusoidal
oscillations, i.e. $n=3$. 

For example, for a self-interacting curvaton~(\ref{SIpot}) with $m=4$,
then when $\sigma_{\mathrm{osc}} $ is larger than $f$, we need to use
the expressions in this appendix with $n =4$.


\end{document}